\documentclass[11pt]{article}
\usepackage{amsfonts}
\usepackage{graphicx}
\usepackage{amssymb}
\usepackage{authblk}
\usepackage{amsmath}
\usepackage{mathtools}
\usepackage{mathrsfs}
 \usepackage{setspace}
\usepackage{algorithm}
\usepackage{algpseudocode}
\usepackage{placeins}

\def \vC{\mbox{\boldmath $C$ \unboldmath}\!\!}
\def \vZ{\mbox{\boldmath $Z$ \unboldmath}\!\!}

\def \vI{\mbox{\boldmath $I$ \unboldmath}\!\!}

\def \vSigma{\mbox{\boldmath $\Sigma$ \unboldmath}\!\!}
\def \vTheta{\mbox{\boldmath $\Theta$ \unboldmath}\!\!}

\def \valpha{\mbox{\boldmath $\alpha$ \unboldmath}\!\!}

\def \vepsilon{\mbox{\boldmath $\epsilon$ \unboldmath}\!\!}

\def \vs{\mbox{\boldmath $s$ \unboldmath}\!\!}

\def \vZ{\mbox{\boldmath $Z$ \unboldmath}\!\!}
\def \vH{\mbox{\boldmath $H$ \unboldmath}\!\!}

\def \vM{\mbox{\boldmath $M$ \unboldmath}\!\!}
\def \vE{\mbox{\boldmath $E$ \unboldmath}\!\!}

\def \vTheta{\mbox{\boldmath $\Theta$ \unboldmath}\!\!}

\def \vmu{\mbox{\boldmath $\mu$ \unboldmath}\!\!}

\def \vS{\mbox{\boldmath $S$ \unboldmath}\!\!}

\def \vH{\mbox{\boldmath $H$ \unboldmath}\!\!}

\def \vs{\mbox{\boldmath $s$ \unboldmath}\!\!}
\def \vS{\mbox{\boldmath $S$ \unboldmath}\!\!}

\def \vX{\mbox{\boldmath $X$ \unboldmath}\!\!}
\def \vZ{\mbox{\boldmath $Z$ \unboldmath}\!\!}

\def \vI{\mbox{\boldmath $I$ \unboldmath}\!\!}

\def \vepsilon{\mbox{\boldmath $\epsilon$ \unboldmath}\!\!}

\def \vmu{\mbox{\boldmath $\mu$ \unboldmath}\!\!}

\def \vSigma{\mbox{\boldmath $\Sigma$ \unboldmath}\!\!}
\def \vTheta{\mbox{\boldmath $\Theta$ \unboldmath}\!\!}

\usepackage[default]{jasa_harvard}    
\usepackage{JASA_manu}

\begin{document}

\title{A Potts-Mixture Spatiotemporal Joint Model for Combined MEG and EEG Data}
\author[1]{Yin Song\thanks{This work is based on Yin Song's PhD thesis supervised by F.S. Nathoo and is supported by the Natural Sciences and Engineering Research Council of Canada and the Canadian Statistical Sciences Institute.}} 
\author[1]{Farouk S. Nathoo\thanks{Corresponding Author: nathoo@uvic.ca. The authors wish it to be known that the first two authors
should be regarded as joint First Authors.}} 
\author[2]{Arif Babul}

\affil[1]{Department of Mathematics and Statistics, University of Victoria}
\affil[2]{Department of Physics and Astronomy, University of Victoria}

\maketitle

\begin{center}
\textbf{Abstract}
\end{center}
We develop a new methodology for determining the location and dynamics of brain activity from combined magnetoencephalography (MEG) and electroencephalography (EEG) data. The resulting inverse problem is ill-posed and is one of the most difficult problems in neuroimaging data analysis. In our development we propose a solution that combines the data from three different modalities, MRI, MEG, and EEG, together. We propose a new Bayesian spatial finite mixture model that builds on the mesostate-space model developed by Daunizeau and Friston (2007). Our new model incorporates two major extensions: (i) We combine EEG and MEG data together and formulate a joint model for dealing with the two modalities simultaneously; (ii) we incorporate the Potts model to represent the spatial dependence in an allocation process that partitions the cortical surface into a small number of latent states termed mesostates. The cortical surface is obtained from MRI. We formulate the new spatiotemporal model and derive an efficient procedure for simultaneous point estimation and model selection based on the iterated conditional modes algorithm combined with local polynomial smoothing. The proposed method results in a novel estimator for the number of mixture components and is able to select active brain regions which correspond to active variables in a high-dimensional dynamic linear model. The methodology is investigated using synthetic data and simulation studies and then demonstrated on an application examining the neural response to the perception of scrambled faces. R software implementing the methodology along with several sample datasets are available at the following GitHub repository \emph{https://github.com/v2south/PottsMix}.

\noindent\textsc{Keywords}: {Bayesian Mixture Model, Electromagnetic Inverse Problem, Iterated Conditional Modes, Maxwell's Equations, Potts Model, Spatiotemporal Model}

\newpage

\section{Introduction}
Magnetoencephalography (MEG) and electroencephalography (EEG) are neuroimaging modalities that have been widely used to study the function of the brain non-invasively using an array of sensors placed on (EEG) or above the scalp (MEG).  These sensor arrays can be used to capture the time-varying electromagnetic field that exists around the head as a result of electrical neural activity within the brain. At a given position of the array, each sensor records a scalar-valued time series representing either the electric potential (EEG) or the magnetic field (MEG) at that position. While the magnetic field is a vector field in space, most MEG machines measure only one component of this field (curl-free tangential), so both EEG and MEG measure scalar fields around the scalp. 
The entire array thus produces a multivariate time series such as the data depicted in Figure 1, panel (a), an example of an MEG, and Figure 1, panel (c), an example of an  EEG. Viewed from a spatial perspective, at a given time point, each array records a spatial field such as that depicted in Figure 1, panel (b), which shows the MEG spatial field at a particular time point, and Figure 1, panel (d), which shows the EEG spatial field at the same time point.

\begin{figure}[htbp]
\centering
\begin{tabular}{cc}
\includegraphics[scale=0.32]{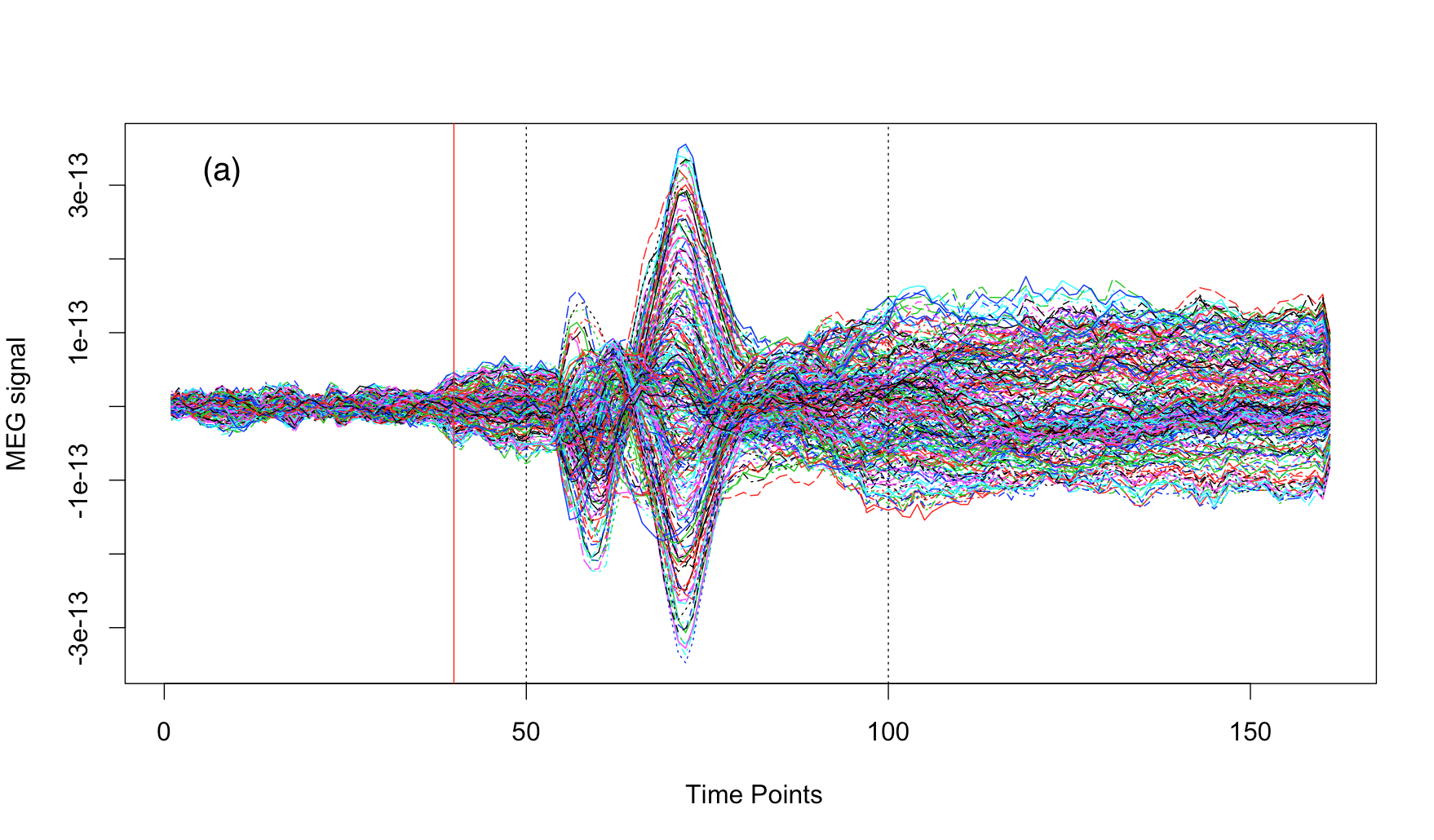} &
\hspace{-2em} 
\includegraphics[scale=0.32]{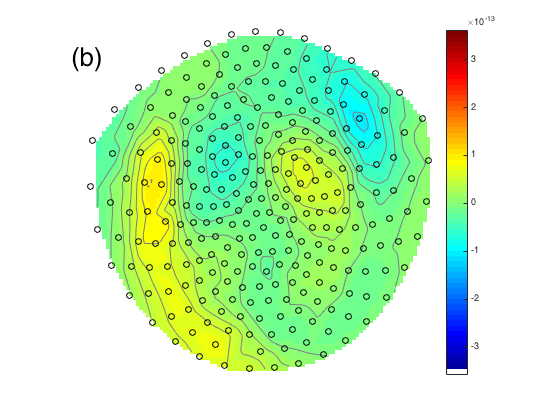}  \\ 
\includegraphics[scale=0.32]{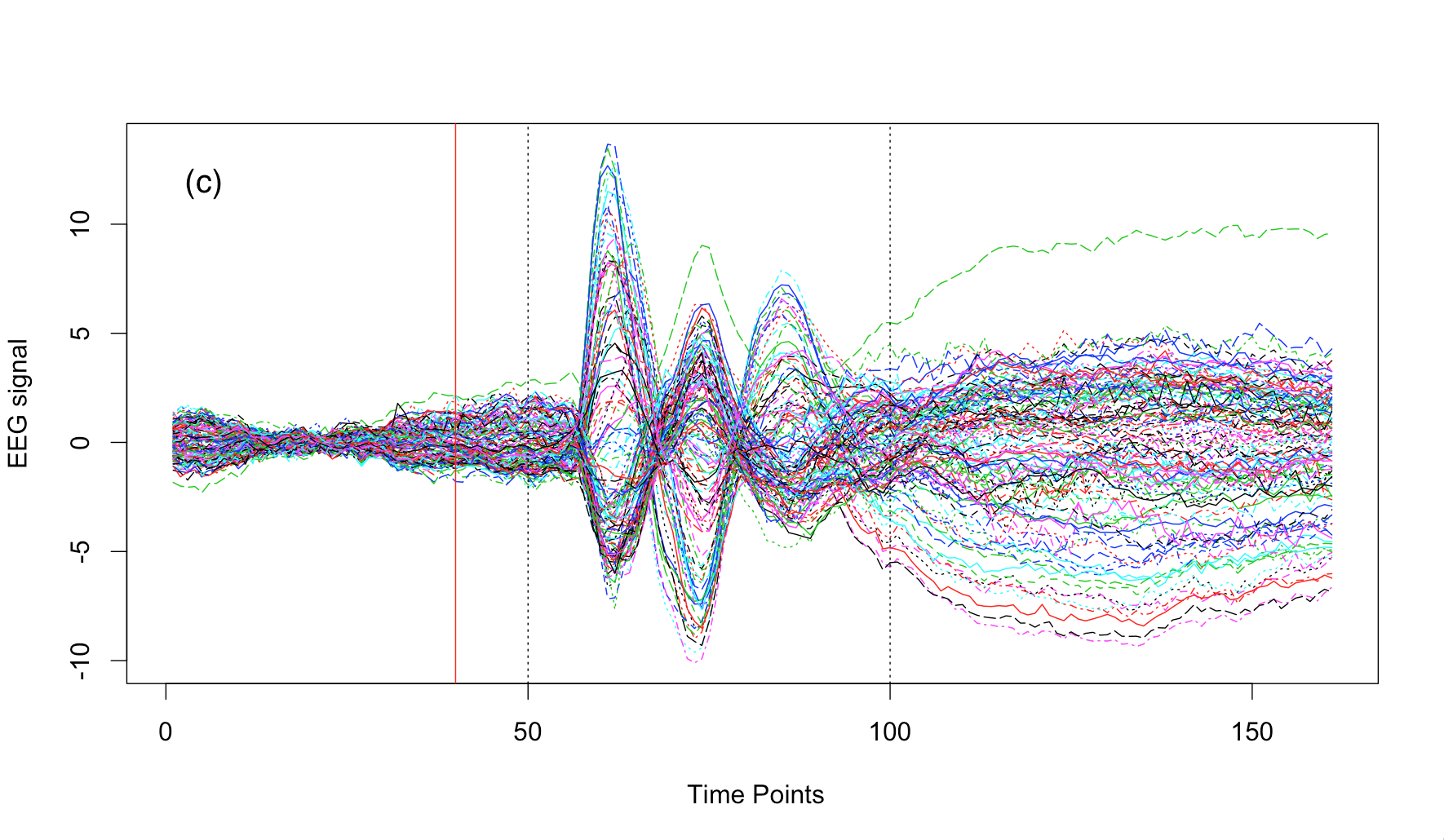}&
\hspace{-2em}
\includegraphics[scale=0.32]{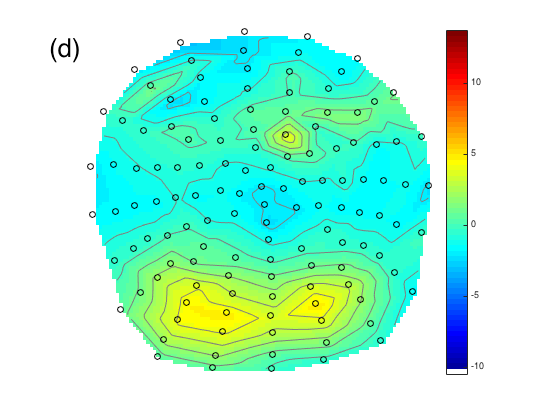} \\ 
\end{tabular}
\caption{The MEG and EEG data considered in the face perception study: panels (a) and (c) show the time series observed at each MEG sensor and EEG sensor, respectively; panels (b) and (d) depict the spatially interpolated values of the MEG data and the EEG data, respectively, each observed at $t=80$, roughly 200ms after presentation of the stimulus. In panels (b) and (d) the black circles correspond to the sensor locations after projecting these locations onto a 2-dimensional grid (for presentation). The MEG and EEG data represent averages over 336 and 344 independent and identically distributed trials respectively.}
\end{figure}

While both EEG and MEG are generated by neural activity, each modality measures this activity indirectly in a different way, through the associated electric field and magnetic field respectively. In typical studies, EEG and MEG are used separately to study brain activity that is evoked by a particular stimulus, or to study the brain while it is at rest. Our emphasis in this article is the development of methodology for the analysis of such data in the former case when brain activity is evoked with a particular stimulus. The focus is on situations where the MEG and EEG data are collected simultaneously or the data are collected sequentially in a situation where the data mimic a simultaneous recording paradigm. MEG and EEG pick up currents from mostly disjoint, though contiguous sections of the cortex. Therefore, technically the sources of the EEG signal will contribute only modestly to the MEG signal, and vice versa, although if the activity is continuous over large regions of the cortex, the sources will be spatially correlated. Thus a joint spatial model for combined MEG and EEG data should lead to improved source estimation. 

While EEG and MEG data can and often are analyzed directly at the sensor level (see, e.g. Ismail et al., 2013), our objective in the analysis of these data is to localize the sources of neural activity within the brain. That is, we want to take the data collected over the sensor arrays and map these data back to the brain. In doing so we want to combine the MEG and EEG data together as both datasets are generated from the neural response of interest. Our proposed methodology is applicable to settings where it is believed that the neural activity is generated by a small number (e.g., 2 to 9) of latent states, with each state having its own dynamics. Thus two primary challenges we are faced with when considering this problem are: (i) a combined analysis of MEG and EEG data, and (ii) estimation of low-dimensional latent structure.

In considering our objective of mapping the sensor array data back to the brain we must consider the relationship between the observed data and the unknown neural activity within the brain. This relationship is governed by the theory of electrodynamics, which is described by Maxwell's equations. This theory will thus play a role in our solution to the inverse problem and will be incorporated into an associated statistical model. 

Generally, an inverse problem is said to be well-posed if a solution exists, the solution is unique, and if the solution's behavior changes continuously with the initial conditions (Hadamard, 1902). A problem that is not well-posed is said to be ill-posed. Within the setting of electromagnetic data it has been shown theoretically (von Helmholtz, 1853) that the problem of finding the sources of electromagnetic fields outside a volume conductor has an infinite number of solutions. 

Outside of the statistical literature this inverse problem has received a great deal of attention in the field of neuroimaging. Many existing solutions are based on regularization, often within the context of penalized likelihood estimation. Methods based on either an $L_{2}$ penalty (Pascual-Marqui, Michel, and Lehmann, 1994) or an $L_{1}$ penalty (Matsuura and Okabe, 1995) have been proposed, as have more general approaches, such as the solution proposed by Tian and Li (2011), where a group elastic net (Zou and Hastie, 2005) for MEG source reconstruction is developed. Bayesian approaches have also been developed by a number of authors including Friston et al. (2008) and Wipf and Nagarajan (2009), who consider Gaussian scale mixtures incorporating a potentially large number of covariance components representing spatial patterns of neural activity, and Henson et al. (2009b) who extend this approach for combined MEG and EEG data, while Henson et al. (2010) extend this approach for combined EEG/MEG and fMRI data. Long et al. (2011) account for the dynamic nature of the problem by using the Kalman filter with implementation on a network of high performance computers, while Calvetti et al. (2015), Vivaldi and Sorrentino (2016), and Sorrentino and Piana (2017) consider Bayesian smoothing approaches. 

Zakharova et al. (2017) discuss the inverse problem and develop a solution within the context of noninvasive preoperative methods for the localization of sources which can guide the outcome of brain surgery. Aydin et al. (2017) develop an approach for source localization based on combined MEG and EEG data where the source localization is guided by a detailed and calibrated finite element head model that considers the variation of individual skull conductivities and white matter anisotropy. Lim et al. (2017) develop a method for sparse EEG/MEG source estimation based on the group lasso that has been adapted to take advantage of functionally-defined regions of interest for the definition of physiologically meaningful groups within a functionally-based common space. Belaoucha and Papadopoulo (2017) use spatial information based on diffusion MRI to solve the MEG/EEG inverse problem, while Nathoo et al. (2014) use spatial spike-and-slab priors to solve the EEG/MEG inverse problem while incorporating fMRI data. 

In all of the approaches mentioned above, the unknown neural activity is restricted to the cortical surface, and the solution to the inverse problem is informed by the anatomy of the brain using structural MRI. This has the advantage of using an anatomical constraint that is realistic since it is widely believed that the primary generators of the MEG/EEG signal are restricted to the cortex though it does have the disadvantage of excluding sources deeper in the brain. 


While all of the above mentioned techniques can be applied generally to evoked response data, our methodology will focus specifically on settings where it is believed a priori that the neural response is generated by a small set of hidden states so that the continuous-current distribution based approaches will not accurately reflect this prior information on the structure of neural activity. In this case a finite mixture model seems more appropriate. Such a model, known as the mesostate-space model (MSM) has been developed by Daunizeau and Friston (2007), based on the following four assumptions, taken directly from Daunizeau and Friston (2007)
\begin{enumerate}
\item \emph{bioelectric activity is generated by a set of distributed sources}
\item \emph{the dynamics of these sources can be modelled as random fluctuations about a small number of mesostates}
\item \emph{mesostates evolve in a temporally structured way and are functionally connected (i.e. influence each other)}
\item \emph{the number of mesostates engaged by a cognitive task is small (e.g. between one and a few).}
\end{enumerate}

While it is well suited for the settings considered here, the MSM is developed for either EEG or MEG data only, and it is not directly applicable for combined EEG and MEG data. Equally important, the MSM assigns each location on the cortical surface to one of a small number of mixture components using a simple multinomial labelling process, where the corresponding mixture allocation variables are assumed independent and identically distributed. More realistically, we expect these discrete allocation variables to be spatially correlated across brain locations and it is therefore important to incorporate this prior information into the structure of the mixture model.

Motivated by the issues discussed above, we develop a new Bayesian model that builds on the MSM in two ways. First, we formulate the model for combined MEG and EEG data. Second, we relax the assumption of independent mixture allocation variables and instead model these variables using the Potts model. The Potts model contains a hyperparameter that controls the degree of spatial correlation and we assign a hyperprior to this parameter that accounts for the phase transition point of the Potts model, so that unrealistic mixture allocations are avoided. 

For our new model formulation we propose an approach for simultaneous point estimation and model selection based on the iterated conditional modes (ICM) algorithm (Besag, 1986) combined with a pseudolikelihood approximation to the normalizing constant of the Potts model, and local polynomial smoothing. By model selection, we mean choosing the number of mixture components for the latent Gaussian process and our ICM algorithm results in a very simple and novel estimator that appears to have reasonable properties, while being computationally efficient. 

The paper is structured as follows. Our Bayesian model specification is presented in Section 2. In Section 3 we discuss our computational algorithm and the resulting estimator for the number of mixture components. In Section 4 we present two simulation studies. Section 5 presents a demonstration of the methodology on a real dataset. The paper concludes with a discussion in Section 6.

\section{Spatiotemporal Mixture Model}

We let $\vM(t) = (M_1(t), M_2(t), . . . , M_{n_M}(t))'$ and $\vE(t) = (E_1(t), E_2(t), . . . , E_{n_E}(t))' $ denote the MEG and EEG, respectively, at time $t$, $t = 1, \dots, T$; where $n_{M}$ and $n_{E}$ denote the number of MEG and EEG sensors outside the head. We assume that the sensor locations for the two modalities have been co-registered to the same space through an appropriately defined transformation of coordinates (see, e.g., Penny et al., 2011). Correspondingly, we let $\vX_M$ and $ \vX_E$ denote $n_{M} \times P$ and $n_{E} \times P$ design matrices, respectively, which we refer to as the forward operators. In this case, $P$ represents a large number of point sources of potential neural activity within the brain corresponding to known locations $\vs_{1},\dots,\vs_{P}$ covering the cortical surface, and it is typical that the value of $P$ is assumed large enough so that $P>>n_{E}$ and $P>>n_{M}$. As cortical neurons with their large dendritic trunks locally oriented in parallel, and pointing perpendicularly to the cortical surface are believed to be the main generators of MEG and EEG, the orientations of the point sources are assumed normal to the cortical surface. 

The computation of the forward operators is based on a solution to Maxwell's equations under the quasi-static approximation (Sarvas, 1987). A detailed theoretical treatment of the forward problem can be found in Penny et al. (2011). Within the current context it is sufficient to note that $\vX_{M_{ij}}$ represents the noise free MEG measurement that would be observed at the $i^{th}$ MEG sensor given a unit current source at location $\vs_{j}$ in the brain. The elements of $\vX_E$ have a similar interpretation for EEG. Under the quasi-static approximation to Maxwell's equations the MEG/EEG measurement at a given time point, denoted as \vM/$\vE$, is related to the unknown neural activity at the same time point $\vS = (S_{1},\dots,S_{P})'$ through a linear relationship

\begin{equation}
\label{EEG_linear}
	      \vM = \vX_{M}\vS, \,\,\vE = \vX_{E}\vS.
\end{equation}

We assume that the data have been transformed as described in Section 1 of the Supplementary Material and specify a model that is applicable to situations where the number of states is reasonably low or the primary activity can be approximated by a low dimensional process. We further assume that the $P$ cortical locations are embedded in a 3D regular grid composed of $N_v$ voxels. This assumption allows us to better formulate the hyper-prior for the Potts model and also facilitates a more efficient computational implementation as described below. Given this grid of voxels, we define the mapping $v: \{1,...., P\}  \rightarrow \{1,...N_v\}$ such that $v(j)$ is the index of the voxel containing the  $j^{th}$ cortical location. Beginning with equation (\ref{EEG_linear}) we add the temporal dimension and incorporate Gaussian measurement error leading to the following specification 
 \begin{align*}
	 \vM(t) & =  \vX_M \vS(t) + \vepsilon_M(t),\,\,\,\,\,\, \vepsilon_M(t) | \sigma_{M}^{2} \stackrel{iid}{\sim} MVN(\mathbf{0},\sigma_{M}^{2}\vH_{M}), \,\,\, t = 1,\dots,T   \\   
	\vE(t) & =  \vX_E \vS(t) + \vepsilon_E(t) ,\,\,\,\,\,\, \vepsilon_E(t) | \sigma_{E}^{2} \stackrel{iid}{\sim} MVN(\mathbf{0},\sigma_{E}^{2}\vH_{E}), \,\,\, t = 1,\dots,T,   
\end{align*} 
where $\vH_M$ and $\vH_E$ are known $n_M \times n_M$ and $n_E \times n_E$ matrices which can be obtained from auxiliary data providing information on the covariance structure of EEG and MEG sensor noise, or we can simply set $\vH_M = \vI_M $ and $\vH_E = \vI_E$. The latter is assumed hereafter. 

At the second level of the model we assume that the unknown neural activity arises from a Gaussian mixture model
	\begin{equation}
	\label{mixture_prior}
	S_j(t)| \pmb{\mu}(t),\valpha,  \vZ \stackrel{ind}{\sim} \prod_{l=1}^{K}N(\mu_l(t), \alpha_l)^{Z_{v(j)l}},  
	\end{equation}
$j = 1,\dots,P, \,\, t = 1,\dots,T$; where $\pmb {Z} = (Z_1^{'},Z_2^{'}, ..., Z_{N_v}^{'})' $ is  a labelling process defined over the 3D regular grid of voxels such that for each $v\in \{1,. . . , N_v\}$, $ \pmb Z_v^{'} =  (Z_{v1}, Z_{v2}, . . . , Z_{vK}) $ with $Z_{vl} \in\{0,1\}$ and $\sum_{l=1}^{K} Z_{vl} = 1$;  $\pmb{\mu}(t) = (\mu_1(t), \mu_2(t), . . . , \mu_K(t))' = (\mu_1(t), \pmb \mu^A(t) ')'$, where $\pmb \mu^A(t) = (\mu_2(t), . . . , \mu_K(t))'$ denotes the mean of the 'active' states and $\mu_1(t) = 0 \;\;\text{for all} \;t$, so that the first mixture component corresponds to an 'inactive' state. The variability of the $l^{th}$ mixture component about its mean $\mu_{l}(t)$ is represented by $\alpha_{l}, l = 1,\dots,K$.

The $j^{th}$ location on the cortex is allocated to one of $K$ states through  $\pmb Z_{v(j)}$, and we assume that the labelling process follows a Potts model so that
\begin{equation*}
		P(\pmb{Z}|\beta)  = \frac{\exp\{ \beta \sum_{h\sim j} \delta(\pmb{Z_j, Z_h})\} } {G(\beta)},\,\,\,\,\, \delta(\pmb{Z_j, Z_h})  = 2 \pmb{Z_j^{'}Z_h} - 1,
\end{equation*}
where $G(\beta)$ is the normalizing constant for this probability mass function, $\beta \ge 0$ is a hyper-parameter which governs the strength of spatial cohesion,  and $i \sim j$ indicates that voxels $i$ and $j$ are neighbours, with a first-order neighbourhood structure over the 3D regular grid of voxels.

We assume that the mean temporal dynamics follow a first-order vector autoregressive process:
	$
	\pmb \mu^A(t)  = \pmb A \pmb \mu^A(t-1) + \mathbf{a}(t), \,\,\, \mathbf{a}(t) |\sigma^2 _a \overset{i.i.d}{\sim} MVN(\pmb 0 , \sigma^2 _a \mathbf{I}) 
	$
	$t = 2, \dots, T$, $\pmb \mu^A(1) \sim MVN(\pmb 0, \sigma^2_{\mu_1}\pmb I)$, with $\sigma^2_{\mu_1}$ known, but $\sigma^2 _a$ unknown. The hyper-parameter for the Potts model is assigned a uniform prior $\beta \sim \text{Unif}[0,\beta_{crit}]$, where $\beta_{crit}$ is an approximation of the phase transition point of the $K$-state Potts model on a 3-dimensional regular lattice  (Moores et al., 2009), 
		 $\beta_{crit} = \frac{2}{3}\log\{\frac{1}{2}[\sqrt{2} + \sqrt{4K-2}]\}.$
Additional priors completing the model specification are as follows:
		 $\alpha_l \overset{i.i.d}{\sim} \text{Inverse-Gamma}(a_\alpha, b_\alpha ),\,  l= 1,2 , . . . K$,
	$A_{ij} \overset{i.i.d}{\sim} N(0, \sigma^2_A), i=1, . . . , K-1, j=1, . . . , K-1$, 
	$\sigma^2_q \sim  \text{Inverse-Gamma}(a_q, b_q), \,\,\, q \in \{a, M, E\}$.  The model parameters are then:
$
	\vTheta  = \{\pmb Z, \{ \pmb \mu^A(1), \pmb \mu^A(2), ... , \pmb \mu^A(T)\} ,\\ \{ \alpha_1, \alpha_2, . . . ,\alpha_k\}, \sigma^2_E, \sigma^2_M, \{ S_j(t), t= 1,2, . . . , T, j = 1,2,. . . , P\},\\
	\beta, \pmb A,
	 \sigma^2_a\}
$
with prior distributions fully determined after specification of $a_E,b_E, a_M, b_M, a_\alpha, b_\alpha, \sigma^2_A, a_a, b_a, \sigma^2_{\mu_1}$, and these hyper-parameters are chosen so that the resulting priors are somewhat diffuse.

The posterior distribution takes the form $P(\vTheta | \vE, \vM) = P(\vTheta, \vE, \vM)/P(\vE, \vM)$ where 
	\begin{align*}
	& P(\vTheta, \vE, \vM) = P(\vE,\vM|\vTheta)P(\vTheta) = P(\vE|\vTheta)P(\vM|\vTheta)P(\vTheta) \\ 
	& = \prod_{t=1}^{T}MVN(\vE(t);  \vX_E\vS(t), \sigma^2_E\vH_E)\; \times MVN(\vM(t); \vX_M\vS(t), \sigma^2_M \vH_M)  \\
	& \times IG(\sigma^2_E; a_E, b_E) \times IG(\sigma^2_M;  a_M, b_M) \times [\prod_{j=1}^{p}\prod_{t=1}^{T}\prod_{l =1}^{K}N(S_j(t);  \mu_l(t), \alpha_l)^{Z_{v(j)l}}] \\
	&\times [\prod_{t=2}^{T}MVN(\pmb \mu^A(t); \pmb A\mu^A(t-1) ,  \sigma^2_a\pmb I)] \times MVN(\pmb \mu^A(1);  \pmb 0 , \sigma^2_{\mu_1} \pmb I) \\
	& \times \text{Potts}(\vZ;  \beta) \times \prod_{l =1}^{K} IG(\alpha_l;  a_\alpha, b_\alpha) \times \text{Unif}(\beta; 0, \beta_{crit}) \\
	& \times [\prod_{i=1}^{K-1}\prod_{j=1}^{K-1}N(A_{ij};   0, \sigma^2_A)] \times IG(\sigma^2_a;  a_a, b_a)
	\end{align*}
Where $MVN(\mathbf{x}; \vmu,\vSigma)$ denotes the density of the $\text{dim}(\mathbf{x})$-dimensional multivariate normal distribution with mean $\mathbf{\vmu}$ and covariance $\vSigma$ evaluated at $\mathbf{x}$; $IG(x; a,b)$ denotes the density of the inverse-gamma distribution with parameters $a$ and $b$ evaluated at $x$; $N(x; \mu,\sigma^{2})$ denotes the density of the normal distribution with mean $\mu$ and variance $\sigma^{2}$ evaluated at $x$; $\text{Potts}(\vZ; \beta)$ is the joint probability mass function of the Potts model with parameter $\beta$ evaluated at $\vZ$; $\text{Unif}(x; a, b)$ is the density of the uniform distribution on $[a,b]$ evaluated at $x$. 

With $T = 161$ time points and $P = 8,196$ brain locations selected on the cortex, the dimension of $\vS = (\vS(1)', \dots, \vS(T)')'$ is $1,319,556$ and this high-dimensional parameter space poses challenges for Bayesian computation. In addition, the parameter space includes the discrete-valued mixture allocation variables $\vZ$ and a large number of such variables will cause problems for standard MCMC sampling algorithms typically used for the required computation. We must therefore consider some approximations, and in the following section we discuss simultaneous point estimation for $\vTheta$ and model selection for the number of mixture components in the latent process using an algorithm that makes the required computation feasible and relatively efficient.

\section{Computation and Estimating the Number of Mixture Components}
The iterated conditional modes (ICM) algorithm is a approach for computing a point estimate for $\vTheta$ and has a long history of application to image restoration. The properties of the algorithm within this context were first described by Besag (1983). The algorithm proceeds by partitioning $\vTheta$ into a set of convenient blocks $\vTheta = \{\vTheta_{1}, \dots, \vTheta_{M} \}$, and is iterative, where, given an initial value $\vTheta^{(0)}$, the $i^{th}$ iteration proceeds by cycling through each of the blocks $\vTheta_{1}, \dots, \vTheta_{M}$ in turn, and updating block $\vTheta_{j}$ by maximizing the corresponding full conditional distribution. That is, the $j^{th}$ sub-step of the $i^{th}$ iteration is based on the following equation
$$
\vTheta_{j}^{(i)} = \operatorname*{argmax}_{\vTheta_{j}}P(\vTheta_{j}|\vE,\vM,\vTheta_{1}^{(i)},\dots,\vTheta_{j-1}^{(i)}, \vTheta_{j+1}^{(i-1)}, \vTheta_{M}^{(i-1)})
$$
where the objective function is the probability mass/density function of the corresponding full conditional distribution $[\vTheta_{j}|\vE,\vM, \vTheta_{1},\dots,\vTheta_{j-1}, \vTheta_{j+1}, \vTheta_{M}]$. 

Within our ICM algorithm the update for the spatial cohesion parameter $\beta$ requires repeated evaluation of the normalizing constant $G(\beta)$ in the Potts model. It is well known that evaluating this normalizing constant requires fairly extensive computation and this computation is often approached using thermodynamic integration (Johnson et al., 2013). We avoid thermodynamic integration by using the pseudolikelihood approximation $$\text{Potts}(\vZ;  \beta) \approx \prod_{i=1}^{N_v}P(\pmb Z_i|\pmb Z_{-i}, \beta) = \prod_{i=1}^{N_v}\frac{\exp \big( 2\beta \sum_{j=1}^{k}Z_{ij}\sum_{l\in\delta_i}Z_{lj}{}\big)}{\sum_{q=1}^{k} \exp (2\beta \sum_{l \in \delta_i}{}Z_{lq})},$$ where $\delta_{i}$ denotes the set of indices corresponding to the neighbours of voxel $i$. While this approximation will incur some bias, it allows us to avoid lengthy computations.

Within the framework of the ICM algorithm there are several options for updating the labelling process $\vZ$. In the simplest case, the variable associated with each individual voxel $\vZ_{j}$ is updated one-by-one in sequence, $j=1,\dots,n_{v}$. We adopt a more efficient approach that begins with partitioning $\vZ$ into two blocks $\vZ = \{\vZ_{W}, \vZ_{B}\}$ according to a 3-dimensional chequerboard arrangement, where $\vZ_{W}$ corresponds to the 'white' voxels and $\vZ_{B}$ corresponds to the 'black' voxels. Under the Markov random field prior with a first-order neighbourhood structure, the elements of $\vZ_{W}$ are conditionally independent given $\vZ_{B}$, the remaining parameters, and the data $\vE$, $\vM$. This allows us to update $\vZ_{W}$ in a single step which involves simultaneously updating its elements from their full conditional distributions, and this updating can be implemented using multiple cores. Even without multiple cores, this scheme will typically result in faster convergence when compared with the sequential one-by-one updates. The variables $\vZ_{B}$ are updated in the same way. This idea was recently employed by Ge et al. (2014) within the context of a Metropolis-Hastings algorithm. 

As described in Besag (1986), the algorithm converges rapidly to a point estimate; however, this estimate and even convergence of the algorithm will depend on the initial value of $\vTheta$. A careful choice for the initial value is thus important. To obtain the initial values for $\vS$ we use regularized least squares with either a ridge or lasso penalty. The estimates for $\vS_{j} = (S_{j}(1), \dots, S_{j}(T))',\,\, j=1,\dots,P$ are then clustered into $K$ groups using a K-means algorithm and these groups are then used to assign the initial values for $\vZ$. Initial values for $\mu_{j}(t)$ are then obtained by taking the average of the initial $\vS$ values assigned to each of the mixture components. The initial value for $\beta$ is drawn from its prior distribution, and the initial values for the remaining parameters are set according to the mode of the associated prior. 

To improve the quality of the final estimates we have found it useful to apply smoothing to the estimated source time series $\hat{S}_{j}(t), \,\, t = 1,\dots,T$ at each location. This is accomplished by using local polynomial regression implemented via the \emph{loess} function in the R programming language. 

To reduce the dimension of the parameter space and as a result the required computation time, we use the K-means algorithm to cluster the $P$ locations on the cortex into a smaller number of $J \le P$ clusters, and assume that $S_{j}(t) = S_{l}(t)$ for cortical locations $l,j$ belonging to the same cluster. Typical values for $J$ are $J=250, 500, 1000$, and these choices are investigated in our simulation studies. 

While the value of $K$, the number of mixture components in equation (\ref{mixture_prior}), is fixed with no prior assigned to it, it will typically be the case that the number of mixture components will not be known beforehand. For an evoked response study, our model specification assumes that the number of mixture components is small, no more than 10, but likely less. From our ICM algorithm we can obtain a simple estimate for the number of mixture components based on the estimated allocation variables $\hat{\vZ}$. In particular, we run the algorithm with a value of $K$ that is considerably larger than the expected number of mixture components. For example, the value of $K$ can be set as $K=15$ when running the algorithm. The $j^{th}$ location on the cortex is allocated to one of the mixture components based on the value of $\hat{\vZ}_{v(j)}$, where $\hat{\pmb{Z}}_{v(j)} = (\hat{Z}_{v(j)_{1}}, \hat{Z}_{v(j)_{2}},\dots,\hat{Z}_{v(j)_{K}})'$ and $\hat{Z}_{v(j)_{l}} = 1$ if location $j$ is allocated to component $l \in \{1,\dots,K\}$. When the algorithm is run with a value of $K$ that is larger than necessary, there will exist empty mixture components that have not been assigned any locations under  $\hat{\vZ}$. In a sense these empty components have been automatically pruned out as redundant. The estimated number of mixture components is then obtained as follows:
\begin{equation}
\label{K_hat}
\hat{K}_{ICM} = \sum_{l=1}^{K}I\{\sum_{v=1}^{n_{v}} \hat{Z}_{v_{l}} \ne 0\}.
\end{equation}
This estimator is very simple and only requires us to run the ICM algorithm once for a single value of $K$. Multiple runs of the algorithm with different values of $K$ are avoided altogether. Properties of the estimator $\hat{K}_{ICM}$ are investigated in Section 4.2. 

The overall estimation procedure is presented in Algorithm 1 and the corresponding derivations are presented in Section 2 of the Supplementary Material. Convergence of the algorithm is monitored by examining the relative change of the Frobenius norm of the estimated sources on consecutive iterations. Section 3 of the Supplementary Material presents a sequence of examples looking at synthetic data simulated from a number of known source configurations and these examples demonstrate that the algorithm is able to recover the source configurations from corresponding simulated data.

\begin{algorithm}
\caption{ - ICM  Algorithm for Potts Mixture Model }\label{euclid}
\begin{algorithmic}[1]
\small
\State $ \Theta\gets \text{Initial Value}$
\State $ \text{Converged} \gets 0$
\While{$\text{Converged} = 0$} 
        \State  $\sigma_M^2 \gets  \Big [ \sum_{t=1}^{T} \frac{1}{2} \big(\vM(t) - \vX_M\vS(t)\big)' \vH_M ^{-1} \big(\vM(t) - \vX_M\vS(t)\big)  +b_M \Big ] / \Big[  a_M+\frac{TN_M}{2} + 1\Big]$ 
        
        \State  $\sigma_E^2 \gets  \Big [ \sum_{t=1}^{T} \frac{1}{2} \big(\vE(t) - \vX_E\vS(t)\big)' \vH_E ^{-1} \big(\vE(t) - \vX_E\vS(t)\big)  +b_E \Big ] / \Big[  a_E+\frac{TN_E}{2} + 1\Big]$ 

         \State $\sigma_a^2 \gets  \Big[ \sum_{t=2}^{T}  \frac{1}{2}  (\pmb \mu^A(t) - \pmb {A\mu}^A(t-1)) ' (\pmb \mu^A(t) - \pmb {A\mu}^A(t-1))  + b_a\Big] / \Big[  a_a + \frac{(T-1)(K-1)}{2} +1 \Big]$
         
         \State  $vec(\pmb A) \gets \bigg( \frac{1}{\sigma^2_a} \Big( \sum_{t=2}^{T} \pmb \mu^A(t)' \pmb {Kr_t} \Big) \times \pmb C^{-1}_1\bigg)'$,  \text{where}   $\vC_1 = \frac{1}{\sigma^2_A} \pmb I_{(K-1)^2} + \frac{1}{\sigma^2_a} \bigg( \sum_{t=2}^{T} \pmb{Kr_t}'\pmb{Kr_t}\bigg)$, \newline \text{and}  $\pmb {Kr_t} = \Big(\pmb \mu^A(t-1) '\otimes \pmb I_{K-1} \Big)$

	\For{ $l = 1,..., K$}
        \State  $ \alpha_l \gets \Big[ \frac{ \sum\limits_{j=1}^{P}\sum\limits_{t=1}^{T}Z_{v(j)l} (S_j(t) - \mu_l(t))^2}{2 }  + b_\alpha\Big] / \Big[\frac{T\sum\limits_{j=1}^{P}Z_{v(j)l}}{2} +a_\alpha + 1 \Big]$
      \EndFor
      \State  $\pmb\mu(1) \gets   \bigg(  \Big(\sum_{j=1}^{P}(S_j(1)\vec{I}_{K-1})' \pmb D_j  +\frac{1}{\sigma^2_a} \pmb\mu^A(2)' \pmb A\Big) \times \pmb B^{-1}_1
\bigg) '$,  \text{where}  $\pmb B_1  = \sum_{j=1}^{P} \pmb D_j  + \frac{1}{\sigma^2_a} \pmb{A'A}  + \frac{1}{\sigma^2_{\mu_1}}\pmb I_{K-1},  \;  \pmb D_j =\text{Diag}(\frac{ Z_{v(j)l}}{\alpha_l}, l = 2, ... ,K), \; \vec{I}_{K-1} = (1,1,\dots, 1)' \; \text{with dim} \;( \vec{I}_{K-1}) = K-1 $
         
         \For { $t = 2,..., T-1$}
         
  \State  $\pmb\mu(t) \gets  \Bigg( \bigg( \sum_{j=1}^{P}(S_j(t)\vec{I}_{K-1})' \pmb D_j +\frac{1}{\sigma^2_a} (\pmb\mu^A(t+1))' \pmb A + \frac{1}{\sigma^2_a}(\pmb\mu^A(t-1)'\pmb A' )\bigg) \times \pmb B^{-1}_2\Bigg)'$ \newline  \text{where} $\pmb B_2 = \sum_{j=1}^{P} \pmb D_j  + \frac{1}{\sigma^2_a} (\pmb{A'A}  + \pmb I_{K-1})$
  \EndFor
  
                      \State  $\pmb\mu(T) \gets \Bigg( \bigg( \sum_{j=1}^{P}(S_j(T)\vec{I}_{K-1})' \pmb D_j  + \frac{1}{\sigma^2_a}(\pmb\mu^A(T-1)'\pmb A' )\bigg) \times \pmb B^{-1}_3\Bigg)'$ 
                      \newline  \text{where} $\pmb B_3  = \sum_{j=1}^{P} \pmb D_j  + \frac{1}{\sigma^2_a}  \pmb I_{k-1}$
                      
		\For{ $j = 1,..., P$}
		\State  $\pmb S_{j} \gets -\frac{1}{2}\pmb \Sigma_{S_j} \pmb W_{2j}$ \Comment{$\pmb S_{j}  = ( S_{j}(1), S_{j}(2), ...., S_{j}(T))'$}
		\newline $\pmb \Sigma^{-1}_{S_j}  = W_{1j} \vI_{T}$, $\pmb W_{2j}' =  (W_{2j}(1), W_{2j}(2), . . . , W_{2j}(T))$
		\newline \text{where} $W_{1j} = \frac{1}{\sigma^2_M} \Big( \vX_M[,j]  '\vH_M^{-1}\vX_M[,j] \Big)+\frac{1}{\sigma^2_E} \Big( \vX_E[,j]  '\vH_E^{-1}\vX_E[,j]  \Big) + \sum_{l=1}^{K}\frac{Z_{v(j)l}}{\alpha_l}$
		\newline $W_{2j}(t)= \frac{1}{\sigma^2_M} \Big(  - 2\vM(t)' \vH_M^{-1} \vX_M[,j] + 2(\sum_{v\neq j}^{}\vX_M[,v] S_v(t))'\vH_M^{-1}\vX_M[,j]   \Big) \newline+ \frac{1}{\sigma^2_E} \Big(  - 2\vE(t)' \vH_E^{-1} \vX_E[,j] + 2(\sum_{v\neq j}^{}\vX_E[,v] S_v(t))'\vH_E^{-1}\vX_E[,j] \Big)  - 2 \sum_{l=1}^{K}\frac{\mu_l(t)}{\alpha_l}$
		\newline $\vX_M[,j], \; \vX_E[,j]$ denote the $jth$ column of $\vX_E$ and $\vX_M$
		\EndFor

		 \algstore{myalg}
    \end{algorithmic}
    \end{algorithm}
    \begin{algorithm}                     
    \begin{algorithmic} [1]              
    \algrestore{myalg}
\small
                 \State Let $\mathbb{B}$ denote the indices for 'black' voxels and $\mathbb{W}$ denote the indices for 'white' voxels. 
	    	\For{$\kappa \in \mathbb{B}\; \text{\emph{simultaneously}}$}
		\State $Z_{\kappa q} \gets 1$ and $Z_{\kappa l} \gets 0,\,\, \forall l \ne q$ \newline
		\text{where} $q = \operatorname*{argmax}_{h \in \{1,\dots, K\}}P(h)$, and
		\vspace{1em}
		\State  $P(h) = \frac{ \alpha_{h}^{-TN_{j\kappa}/2 } \times \exp \big(  -\frac{1}{2} \sum_{j|v(j) = \kappa}{} \alpha_h^{-1}\sum_{t=1}^{T} ( S_j(t) - \mu_{h}(t))^2 +  2\beta\sum_{v \in \delta_\kappa}{}Z_{vh}\big) }{	\sum_{ l=1}^{K}\alpha_{l}^{-TN_{j\kappa}/2 } \times \exp \big(  -\frac{1}{2} \sum_{j|v(j) = \kappa}{} \alpha_l^{-1}\sum_{t=1}^{T} ( S_j(t) - \mu_{l}(t))^2 +  2\beta\sum_{v \in \delta_\kappa}{}Z_{vl}\big) 
}
$	
		\vspace{1em}
		\newline where $N_{j\kappa}$ is the number of cortical locations contained in voxel $\kappa$.
		\EndFor

		\For{$\kappa \in \mathbb{W}\; \text{\emph{simultaneously}}$}
		\State $Z_{\kappa q} \gets 1$ and $Z_{\kappa l} \gets 0,\,\, \forall l \ne q$ \newline
		\text{where} $q = \operatorname*{argmax}_{h \in \{1,\dots, K\}}P(h)$, and
		\vspace{1em}
		\State  $P(h) = \frac{ \alpha_{h}^{-TN_{j\kappa}/2 } \times \exp \big(  -\frac{1}{2} \sum_{j|v(j) = \kappa}{} \alpha_h^{-1}\sum_{t=1}^{T} ( S_j(t) - \mu_{h}(t))^2 +  2\beta\sum_{v \in \delta_\kappa}{}Z_{vh}\big) }{	\sum_{ l=1}^{K}\alpha_{l}^{-TN_{j\kappa}/2 } \times \exp \big(  -\frac{1}{2} \sum_{j|v(j) = \kappa}{} \alpha_l^{-1}\sum_{t=1}^{T} ( S_j(t) - \mu_{l}(t))^2 +  2\beta\sum_{v \in \delta_\kappa}{}Z_{vl}\big) 
}
$	
		\vspace{1em}
		\newline where $N_{j\kappa}$ is the number of cortical locations contained in voxel $\kappa$.
		\EndFor

       \State  $\hat \beta \ \gets \operatorname*{argmax}_{\beta \in [ 0, \;\beta_{crit}]} H(\beta)$, where
	$$ H(\beta) =  2\beta \sum_{i=1}^{N_v}\sum_{j=1}^{k}Z_{ij}\sum_{l \in \delta_i}{}Z_{lj} - \sum_{i=1}^{N_v}\log\{ \sum_{q=1}^{k} \exp(2\beta\sum_{l\in \delta_i}{} Z_{lq})\}$$		
       \State  Check for convergence.  Set Converged  = 1 if so.
       
      \EndWhile
            
\end{algorithmic}
\end{algorithm}

\section{Simulation Studies}
\subsection{Evaluation of Neural Source Estimation}
Our first simulation study evaluates the quality of the source estimates as the number of activated brain regions change, and we make comparisons between our approach with and without smoothing, and the mesostate-space model (MSM) of Daunizeau and Friston (2007) applied to either EEG or MEG data. We generate both MEG and EEG data based on neural activity at  8,196 locations on the cortex, and this activity is projected onto the sensor arrays using the forward operators $\vX_M$ and $\vX_E$, with Gaussian noise added at each sensor. A detailed description of simulation parameter settings is provided in Section 3 of the Supplementary Material.

For this study we set the number of mixture components $K$ to be the true number of latent states (either two, three, four, or nine) in both our model as well as the MSM, so that fixing $K$ in this study does not give either approach an advantage over the other. In the next section we present another simulation study where we focus on estimating the number of mixture components and evaluate the sampling distribution of $\hat{K}_{ICM}$. In the current study we consider four scenarios with two, three, four, and nine latent states, and in each case one of these states is inactive, while the other states have activity generated by Gaussian signals. In the simplest case we have only a single activated region, and this region is depicted in Supplementary Figure 1. The temporal signal arising from locations contained in the activated region for this case is depicted in Supplementary Figure 2, panel (a). The other two cases have two, three, and eight activated regions, and these regions are depicted in Supplementary Figure 3, panels (a) and (c), for the case of two and three active regions, and Supplementary Figure 5, for the case of eight active regions, while the corresponding temporal signals associated with the activated regions are depicted in Supplementary Figure 4, panels (a), (c) and (g). 

In each case we simulate 500 replicate datasets and each of the four approaches is applied to each dataset. For each replicate we compute the correlation between the estimated sources and the true sources $\text{Corr}[(\vS(1)', \vS(2)', \dots, \vS(T)'), (\hat{\vS}(1)', \hat{\vS}(2)', \dots, \hat{\vS}(T)')]$ as a measure of agreement, and this measure is averaged over the 500 replicate datasets. For each simulated dataset we apply our algorithm with $J= 250, 500, 1000$ clusters so as to evaluate how the performance varies as this tuning parameter changes. In addition, we make comparisons between these methods based on the mean-squared error of $\hat{S}_{j}(t)$. In particular, for each brain location $j$ and time point $t$ we estimate the mean-squared error (MSE) of the estimator $\hat{S}_{j}(t)$ based on the  $R=500$ simulation replicates. These MSE's are then totalled over brain locations and time points in order to obtain the Total MSE (TMSE). This total is obtained separately for locations in active regions and then for the inactive region.

The average correlation between the estimated values and the truth for each of the distinct settings in our study is presented in Table 1. Examining Table 1, we see that for most of the cases considered our Potts-mixture model, both with and without smoothing, yields a higher average correlation than the MSM with either EEG or MEG. With respect to average correlation, we also see that smoothing does not improve the correlation obtained when using the Potts-mixture model. With respect to the number of clusters $J$, we find that using a lower number of clusters results in a higher average correlation, even in the case where $K=9$.

The Total MSE's for the same distinct cases are depicted in Table 2. From the results in this table, several key observations are made:
\begin{enumerate}
	\item When we consider active regions and our approach with a differing number of clusters, we observe that using $J=250$ clusters yields the lowest TMSE compared with the alternatives of $J=500$ or $J=1000$ clusters. In addition, for the best case corresponding to $J=250$, we see that incorporating temporal smoothing after the ICM algorithm yields optimal TMSE values among those settings considered for our approach. We also notice that these optimal TMSE values obtained from our method are uniformly lower than the TMSE obtained from MSM-EEG and MSM-MEG for all values of $K$. 
	\item When we consider inactive regions, it is clear that the MSM with either modality has lower total mean-squared error than the Potts-mixture model in all cases. 
	\item When  we consider inactive regions for the specific case where $K=9$ our approach with $J=250$ clusters yields a TMSE that is very large. In this case, the TMSE drops significantly when the number of clusters is increased from 250 to 500 or 1000 and we see here an advantage to using a larger number of clusters. 
\end{enumerate}

The results for the case where $K=9$ in Table 1 (Average Correlation) may at first seem contradictory to those  for $K=9$ in Table 2 (TMSE), where we see our approach outperforming MSM in the former and underperforming in the latter. This contradiction can be explained, however, when considering scale. In particular, our approach appears to estimate the patterns of spatiotemporal activation more accurately than MSM in this most complicated setting leading to the higher correlation measure, while the scale of the true sources appears to be better estimated by MSM leading to the improved TMSE. It is our view that the patterns of activation are far more important than the scale. 


Finally, our separation of TMSE into active and inactive regions has lead a reviewer to suggest that we examine the false discovery rate of the methods being compared in this simulation study. To be specific, let $FP$ denote the number of vertices that are declared active by the model that are in fact truly inactive. Let $TP$ denote the number of vertices that are declared active by the model that are in fact truly active. Then $FP + TP$ is the total number of vertices declared active. For a given dataset, the false positive rate is $p_{FP} = FP/(FP + TP)$. We have computed this quantity for each simulation replicate and then taken the average false positive rate over all simulation replicates. In a similar way, let $FN$ denote the number of vertices declared inactive by the model that are in fact active. Let $TN$ denote the number of vertices declared inactive by the model that are in fact inactive. The false negative rate, for a given dataset, is then $p_{FN} = FN/(FN+TN)$ and we compute the average of this quantity over simulation replicates. The results are presented in Table 3.

Examining Table 3, we first note that the false positive rate associated with MSM is in general unacceptably high, with over 60\% (and in many cases much higher) of the vertices labelled as active by MSM being inactive in all cases. The extreme false positive rate found for MSM is in and of itself a result that we find particularly interesting. We note that the average false positive rate of $0.641$ observed for MSM (MEG) when $K=3$ is lower than the other values observed for MSM (MEG) and MSM (EEG) because the average is pulled down by a number of very small values in this case. The performance of our approach with respect to false discoveries is considerably better than that of MSM. With respect to the false negative rate, broadly speaking, the approaches perform equally well, though with MSM (EEG) having the best performance when the number of active regions is low. Overall, these results demonstrate a significant improvement obtained from our methodology when considering false discoveries and roughly equal performance when considering false negative rates.

\begin{table}[htbp]
\small
\begin{center}
\caption{Simulation study I - Neural Source Estimation. Average (Ave.) correlation between the neural source estimates and the true values for the Potts-Mixture model, the Potts-Mixture model without local polynomial smoothing, the Mesostate-space model with MEG data, and the Mesostate-space model with EEG data. The simulation study is based on $R=500$ simulation replicates where each replicate involves the simulation of MEG and EEG data based on a known configuration of the neural activity. For each replicate we compute the correlation between the estimated sources and the true sources as a measure of agreement and this correlation is then averaged over the $R=500$ simulation replicates in order to obtain the Ave. Correlation. "NS" refers to no smoothing for Potts model.} \vspace*{5mm}
\begin{tabular}{lcccccc}
\hline
&\multicolumn{1}{c}{} &\multicolumn{1}{c}{$K = 2$} & \multicolumn{1}{c}{$K = 3$} & \multicolumn{1}{c}{$K = 4$} & \multicolumn{1}{c}{$K = 9$}  \\
Method & Clusters &Ave. Corr.  & Ave. Corr.  & Ave. Corr. & Ave. Corr. \\
\hline
Potts-Mixture & 250&0.59  & 0.62 & 0.61 & 0.50\\
Potts-Mixture (NS)& 250 &  0.59& 0.62& 0.59 &0.50\\
Potts-Mixture & 500&0.51&0.50&0.43&0.49\\
Potts-Mixture (NS)& 500&0.51&0.50&0.42&0.48\\
Potts-Mixture & 1000&0.38&0.40&0.39&0.41\\
Potts-Mixture (NS)& 1000 &0.37&0.40&0.37&0.40\\
MSM (MEG) & NA & 0.20 &  0.47  &  0.37 &0.34\\
MSM (EEG)  & NA&  0.24 & 0.41 & 0.41 & 0.28\\
\hline
\end{tabular}
\end{center}
\normalsize
\end{table}

%
%
\begin{table}[htbp]
\footnotesize
\begin{center}
\caption{Simulation study I - Neural Source Estimation. Total mean-squared error of the neural source estimators for the Potts-Mixture model, the Potts-Mixture model without local polynomial smoothing, the Mesostate-space model with MEG data, and the Mesostate-space model with EEG data. The simulation study is based on $R=500$ simulation replicates where each replicate involves the simulation of MEG and EEG data based on a known configuration of the neural activity. For each brain location $j$ and time point $t$ we obtain (estimate) the mean-squared error (MSE) of the estimator of $S_{j}(t)$ based on the  $R=500$ simulation replicates. These MSE's are then totalled over brain locations and time points in order to obtain the Total MSE indicated in the table. "NS" refers to no smoothing for Potts model.} \vspace*{5mm}
\begin{tabular}{lcccccccccc}
\hline
 & \multicolumn{1}{c}{}& \multicolumn{2}{c}{$K = 2$} & \multicolumn{2}{c}{$K = 3$} & \multicolumn{2}{c}{$K = 4$}  &\multicolumn{2}{c}{$K = 9$}\\
Method & Clusters& Active  & Inactive  & Active & Inactive & Active & Inactive  & Active & Inactive \\
\hline
Potts-Mixture &250&86   &139  &557 &479 &802 &662 &2216 & 16900 \\
Potts-Mixture (NS) & 250&90   & 138 & 609&471 &874 &702  &5318&16488\\
Potts-Mixture &500&207&200 & 847&356&1451&573 &2346	&1001\\
Potts-Mixture (NS) & 500&209&201 & 1289&536&1669&711 &2734&1340\\
Potts-Mixture &1000&267&235 &1141	&398&2170&676&3050&1017\\
Potts-Mixture (NS)  &1000&274&239 &1819 &615&2448&793&3605&1337\\

MSM (MEG)  &NA&380   &19  &842 &55 &1297 &65  &2345 &357\\
MSM (EEG)  &NA&325   &71  &795 &172 &1094 &237  &2499&665\\
\hline
\end{tabular}
\end{center}
\normalsize
\end{table}

\begin{table}[htbp]
\small
\begin{center}
\caption{Simulation study I - Neural Source Estimation. False Positive Rate ($p_{FP}$) and False Negative Rate($p_{NP}$) . "NS" refers to no smoothing for Potts model.} \vspace*{5mm}
\begin{tabular}{lcccccccccc}
\hline
 & \multicolumn{1}{c}{}& \multicolumn{2}{c}{$K = 2$} & \multicolumn{2}{c}{$K = 3$} & \multicolumn{2}{c}{$K = 4$}  &\multicolumn{2}{c}{$K = 9$}\\
Method & Clusters& $p_{FP}$  & $p_{NP}$  & $p_{FP}$  & $p_{NP}$ & $p_{FP}$  & $p_{NP}$  &$p_{FP}$ & $p_{NP}$  \\
\hline
Potts-Mixture &250&0.245&0.008 &0.361& 0.016&	0.444&	0.023&	0.537&0.050   \\
Potts-Mixture (NS) &250&0.245&0.008&0.401&	0.014&	0.446&0.022&	0.534&	0.050\\
Potts-Mixture &500&0.291&	0.010&	0.322&	0.023&	0.385&	0.035&	0.426&	0.060\\
Potts-Mixture (NS) &500&0.290&	0.010&	0.338&	0.021&0.379&	0.034&0.426&	0.060\\
Potts-Mixture &1000&0.340&	0.011&	0.329&	0.027&	0.417&	0.040&	0.425&	0.062&\\
Potts-Mixture (NS)&1000&0.340&	0.011&	0.341&	0.025&0.418&	0.040&	0.422&	0.062&\\

MSM (MEG) &NA&0.922   &0.003   &0.641  &0.019  &0.879  &0.011  &0.914  &0.025\\
MSM (EEG)  &NA& 0.966  &0.000   &0.922  &0.005  &0.898  &0.020  &0.917  &0.033\\
\hline
\end{tabular}
\end{center}
\normalsize
\end{table}

\subsection{Evaluation of Mixture Component Estimation}

In this section we conduct simulation studies to evaluate the sampling distribution of $\hat{K}_{ICM}$. In simulating the data we consider the following scenarios:
\begin{enumerate}
	\item Two latent states as depicted in Supplementary Figure 1 with the active regions having a Gaussian signal as depicted in Supplementary Figure 2, panel (a).
	\item Three latent states as depicted in Supplementary Figure 3, panel (a) with the active regions having Gaussian signals as depicted in Supplementary Figure 4, panel (a).
	\item Four latent states as depicted in  Supplementary Figure 3, panel (c) with the active regions having Gaussian signals as depicted in Supplementary Figure 4, panel (c).
	\item Four latent states as depicted in Supplementary Figure 3, panel (e) with the active regions having Gaussian signals and a sinusoid signal as depicted in Supplementary Figure 4, panel (e).
	\item Nine latent states as depicted in Supplementary Figure 5 with the active regions having Gaussian signals as depicted in Supplementary Figure 4, panel (g).
\end{enumerate}
For each of the five cases the data are simulated with 5\% Gaussian noise added at the sensors as in the previous section, with 1000 replicate datasets used in each case. The ICM algorithm is run with $K=10$ for each of the 5000 simulated datasets with the other settings for the model and algorithm set as in previous sections. For each dataset we compute the value of the estimator (\ref{K_hat}) and histograms illustrating the sampling distribution of $\hat{K}_{ICM}$ are presented in Figure 2, panels (a) - (e) for each of the five cases above. 

For the first the four cases we see that the mode of the sampling distribution corresponds exactly to the true number of latent states. In the final case where we have a larger number (nine) of latent states we see that $\hat{K}_{ICM}$ is biased and under-estimates the number of latent states. This is not unexpected with a mixture model having a large number of components that are not well separated as the estimation method will merge two adjacent components (where the signals peak at similar time points) into one estimated component leading to under-estimation. 

We now repeat the simulation studies for the five cases above but make the estimation problem more difficult by reducing the separation of the Gaussian signals. The modified temporal signals are depicted in Supplementary Figure 2. The sampling distribution of  $\hat{K}_{ICM}$ for each of the five cases under the less well-separated setup are depicted in Figure 2, panels (f) - (j). In this case the mode of the sampling distribution does not correspond to the true number of latent states in any of the cases and we see that $\hat{K}_{ICM}$ tends to under-estimate the true number of states, with the mode of the sampling distribution being one-less than the true value in the first four cases and we see fairly substantial under-estimation in the fifth case. 

Overall, given the complexity of latent Gaussian spatial mixture model, the inherent difficulty of the problem of estimating the number of components in a high-dimensional latent mixture model, the simplicity of the proposed estimator, and its computational efficiency, we find the performance of $\hat{K}_{ICM}$ satisfactory, though we acknowledge the tendency of the estimator to under-estimate the true number of mixture components in some settings. We note further that the idea of simultaneous point estimation and model selection developed in our specific context has considerable potential for further development in latent mixture models more generally.  

\begin{figure}[htbp]
\centering
\begin{tabular}{l}
\includegraphics[scale=0.19]{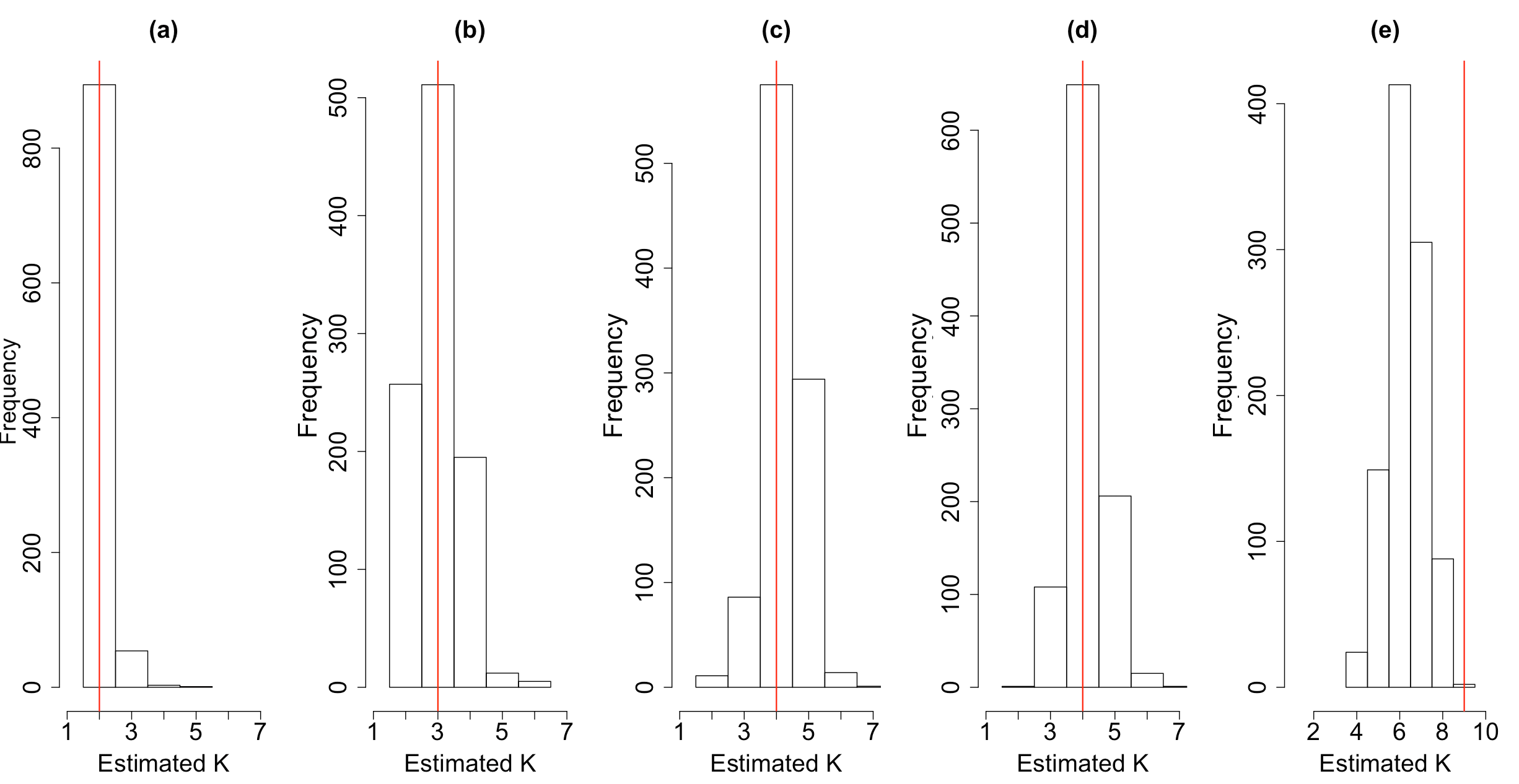}  \\ 
\includegraphics[scale=0.19]{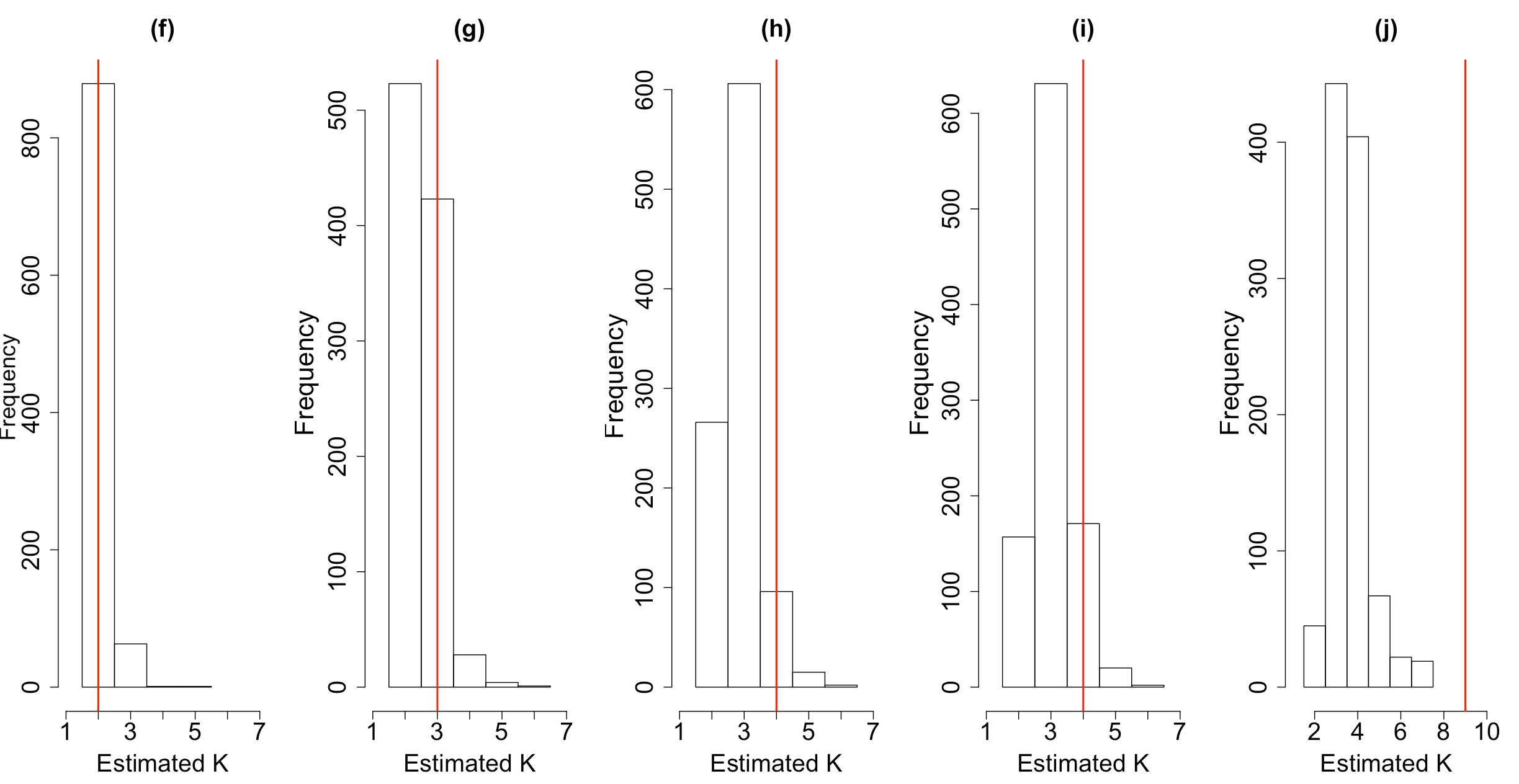}  \\ 
\hspace{-2.75em}
\end{tabular}
\caption{Histograms illustrating the sampling distribution of $\hat{K}_{ICM}$ obtained in the simulation studies of Section 4.2.  The first row corresponds to the case where the true signals are well-separated; panel (a), $K=2$; panel (b), $K=3$; panel (c), $K=4$ with three Gaussian signals; panel (d), $K=4$ with two Gaussian signals and one sinusoid; panel (e), $K=9$ with eight Gaussian signals. The second row corresponds to the case where the true signals are less well-separated. In each case the vertical red line indicates the true number of latent states underlying the simulated data.}
\end{figure}

\section{Electromagnetic Brain Mapping of Scrambled Faces}

We apply our algorithm to the MEG and EEG data presented in Figure 1, to reconstruct the associated neural activity which is constrained to lie on the cortical surface. The data are from an experiment where a single subject is repeatedly presented with pictures of scrambled faces while required to make a symmetry judgement. The experiment and related analyses are described in detail in Henson et al. (2003, 2007, 2009a, 2009b, 2010). Beginning with pictures of faces, each scrambled face is created from a single picture by 2D Fourier transformation, random phase permutation, inverse transformation and outline-masking of each face. Thus the scrambled faces are closely matched with the corresponding faces for low-level visual properties. 

The experiment involves a sequence of trials each lasting 1800ms, where in each trial the subject is presented with one of the pictures for a period of 600ms while being required to make a four-way, left-right symmetry judgment while brain activity is recorded over the array. Both scrambled faces and unscrambled faces are presented to the subject; however, our analysis will focus only on trials involving scrambled faces. This produces a multivariate time series for each trial, and the trial-specific time series are then averaged across trials to create a single multivariate time series. In the case of EEG data, the average evoked response is an average of 344 trials while for MEG data, the average evoked response is an average of 336 trials. The degree of inter-trial variability is quite low. This experiment is conducted while EEG data are recorded, and then again on the same subject while MEG data are recorded. The EEG data are acquired on a 128-sensors ActiveTwo system, sampled at 2048Hz and subsequently downsampled to 200 Hz. The resulting average evoked response to scrambled faces is depicted in Figure 1, panel (c). The MEG data are acquired on 274 sensors with a CTF/VSM system, sampled at 480 Hz and subsequently downsampled to 200 HZ. The resulting average evoked response to scrambled faces is depicted in Figure 1, panel (a). In total, each average evoked response covers roughly 805ms leading to $T = 161$ time points, where the $40^{th}$ time point $t = 40$ corresponds to the time at which the stimulus is presented (the red vertical line in Figure 1). Averaging across trials is considered standard practice to increase the signal to noise ratio, but is based on the assumption that the process of interest is stationary across trials. This averaging also increases the viability of the Gaussian assumption on the data via the central limit theorem.

In the EEG data we see three peaks after the presentation of the stimulus at roughly $t = 60$ (100ms after stimulus), $t = 75$ (175ms after stimulus), and $t=85$ (225ms after stimulus). In the MEG data we see two peaks after the presentation of the stimulus at roughly $t = 60$ (100ms after stimulus) and $t = 70$ (150ms after stimulus). In order to capture the actual neural response of interest, that is, the response to the observation of scrambled faces while the subject makes a symmetry judgment, we use a temporal segment of the data from time point $t=50$ to $t = 100$, indicated by the vertical dashed lines in Figure 1, panel (a) and panel (c). The algorithm is run with $K=10$, $n_{v} = 560$ voxels, and $J=250$ clusters with initial values selected as described in Section 3. 

From the algorithm output we find the estimated number of states to be $\hat{K}_{ICM} = 3$, indicating that there are two active states. For each of the two active states, we determine the location at which the corresponding source activity $\hat{S}_{j}(t)$ has the highest power. The estimated curves for each of these two locations is depicted in Figure 3. In the first active state we see a large negative peak, while in the second active state we see two peaks including a large positive peak. In both states the largest peak (at these two high power locations) occurs within the vicinity of 175ms after the presentation of the stimulus. The large peak in the signal from the first state occurs slightly before the large peak in the signal from the second state.

\begin{figure}[htbp]
\centering
\includegraphics[scale=0.35]{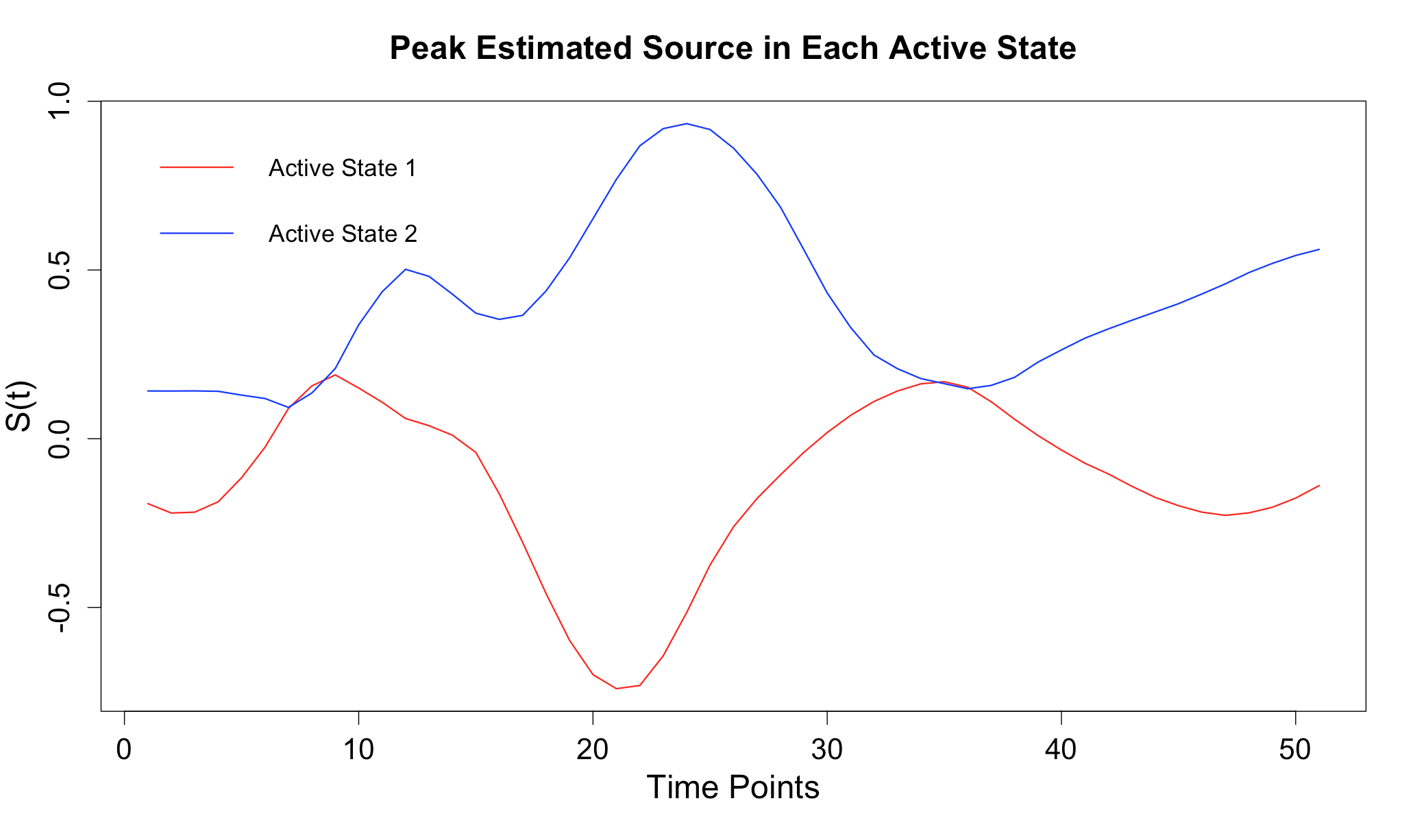}
\caption{Brain Activation for Scrambled Faces - Peak source $\hat{S}_{j}(t)$ in each of the two active states.}
\end{figure}

The spatial patterns of the estimated neural sources $\hat{\vS}(t)$ are of primary interest. The overall power ($\sum_{t=1}^{T}\hat{S}_{j}(t)^{2}$) of these estimated sources obtained from our model at each brain location $j$ is mapped onto the cortex in the first row of Figure 4. Examining these results we see that the greatest power occurs on the bilateral ventral occipital cortex and the occipital cortex. More specifically, the highest power signals on both hemispheres seem to arise within Brodman areas 18 and 19 which are visual association areas. Brodman area 18 is responsible for the interpretation of images while Brodman area 19 has feature-extracting, shape recognition, attentional, and multimodal integrating functions. In general, the power map seems to represent regions that would be expected to show scrambled face-related activity. For example, Daunizeau and Friston (2007) analyze the EEG response to scrambled faces for a single subject and also find regions with high probability of being active on the bilateral ventral occipital cortex and the occipital cortex. These authors also find a region with high probability of being active in the right frontal lobe; our analysis based on the combined MEG and EEG data does not detect high power in this region. 

For comparison, we also apply MSM to the MEG data only, and then apply MSM to the EEG data only. The corresponding results obtained from MSM-MEG are depicted in the second row of Figure 4. Broadly speaking, MSM-MEG seems to indicate similar results to those obtained from our model, in particular with respect to activation on the bilateral ventral occipital cortex, Brodman areas 18 and 19. Interestingly, the results from MSM-EEG, depicted in third row of Figure 4, differ strongly when compared with results of MSM-MEG and our model. In particular, the spatial spread of the high power regions on the ventral occipital cortex, Brodman area 18, is considerably smaller and Brodman area 19 is not indicated. Importantly, MSM-EEG also detects high power in a region on the right frontal lobe, Brodman area 8, which is involved in the management of uncertainty. Recall, that the experimental paradigm requires that the subject make a symmetry judgment when presented with a scrambled face, and there may be uncertainty associated with this judgement. Relating this back to our first simulation study, we observed that MSM-EEG tends to outperform MSM-MEG in the active regions of the brain. From this observation, in addition to the previous results found in Daunizeau and Friston (2007), we suspect that the high power detected in the right frontal lobe by MSM-EEG is a true neural signal that has not been detected by our model. Ideally, our results would show some combination of the results found by MSM-MEG and MSM-EEG, but in this particular case our solution seems to align well only with MSM-MEG. 

We note that there are more MEG than EEG sensors so that there is more MEG data than EEG data in this case; however, the weighting of the two modalities will depend on the relative values chosen for the two standard deviation parameters $\sigma_{E}^{2}$ and $\sigma_{M}^{2}$. Our approach to tuning these parameters has been through maximizing the posterior distribution using the ICM algorithm though alternative approaches such as cross-validation could be used to estimate these parameters. 

\begin{figure}[htbp]
\centering
\begin{tabular}{lll}

\includegraphics[scale=0.17]{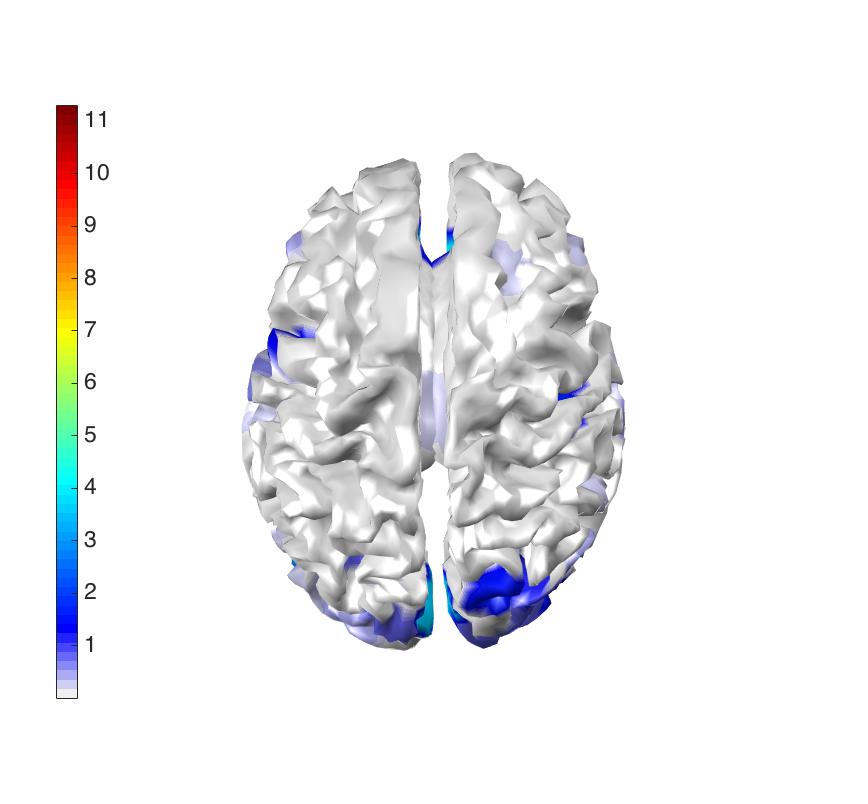} &
\includegraphics[scale=0.17]{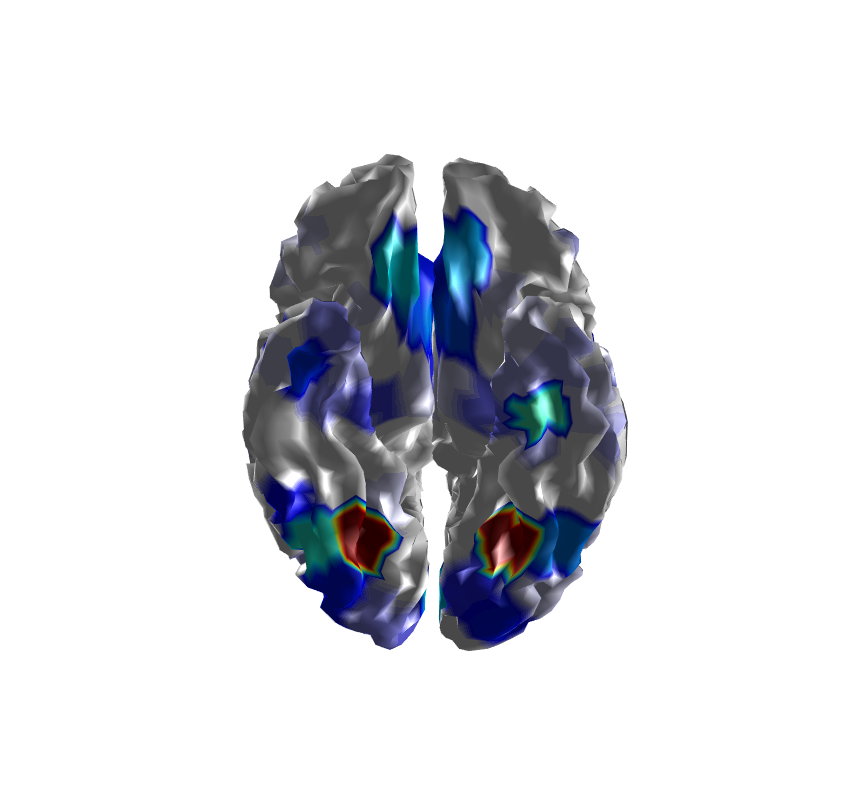} 
\includegraphics[scale=0.17]{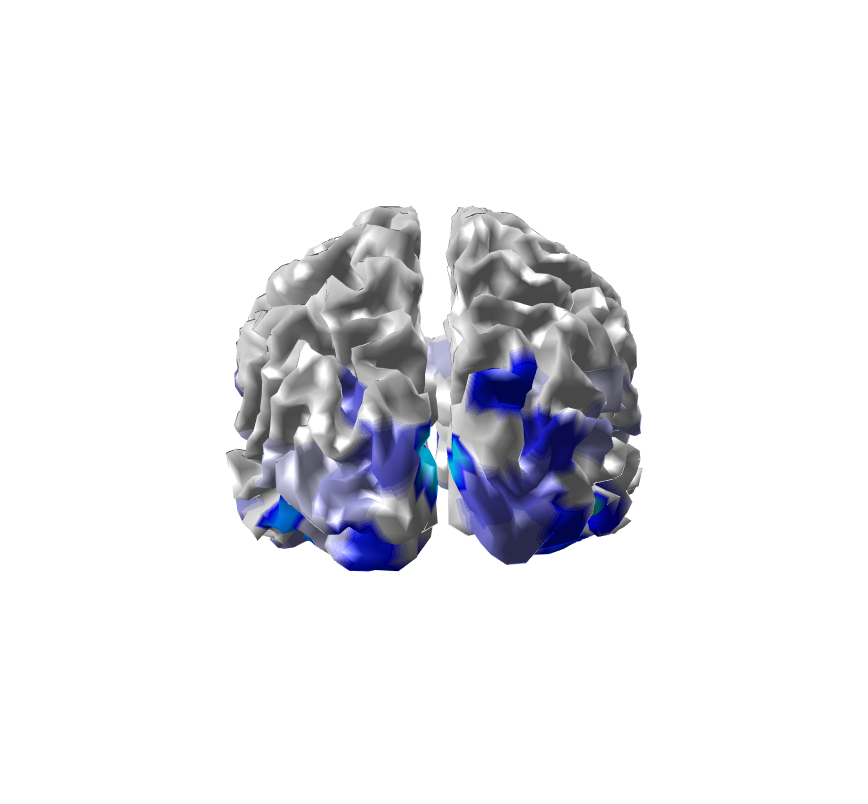} \\

\includegraphics[scale=0.17]{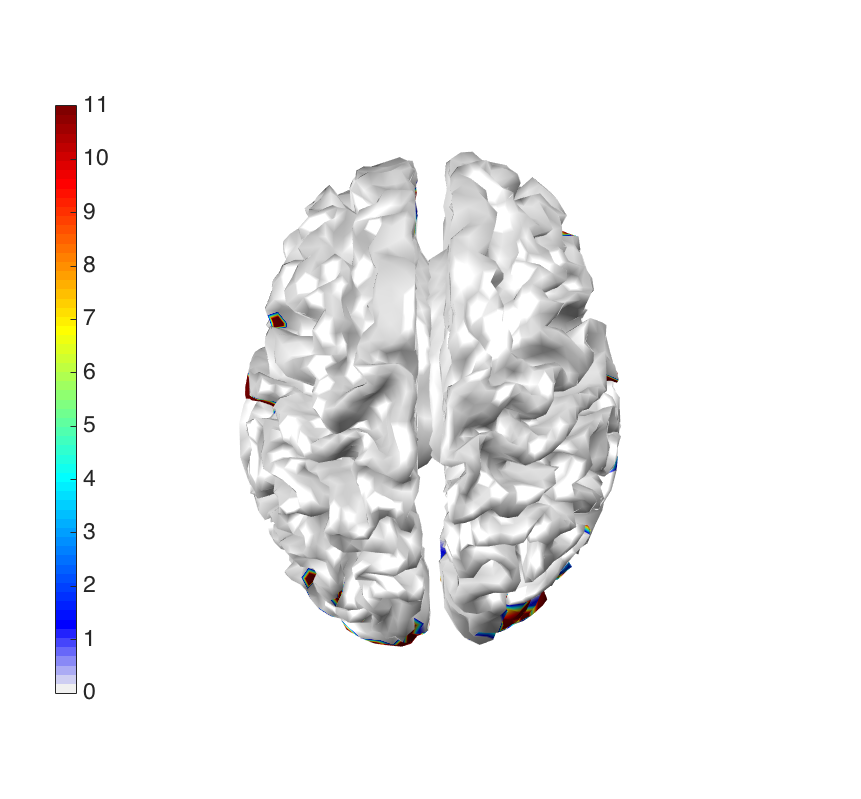} &
\includegraphics[scale=0.17]{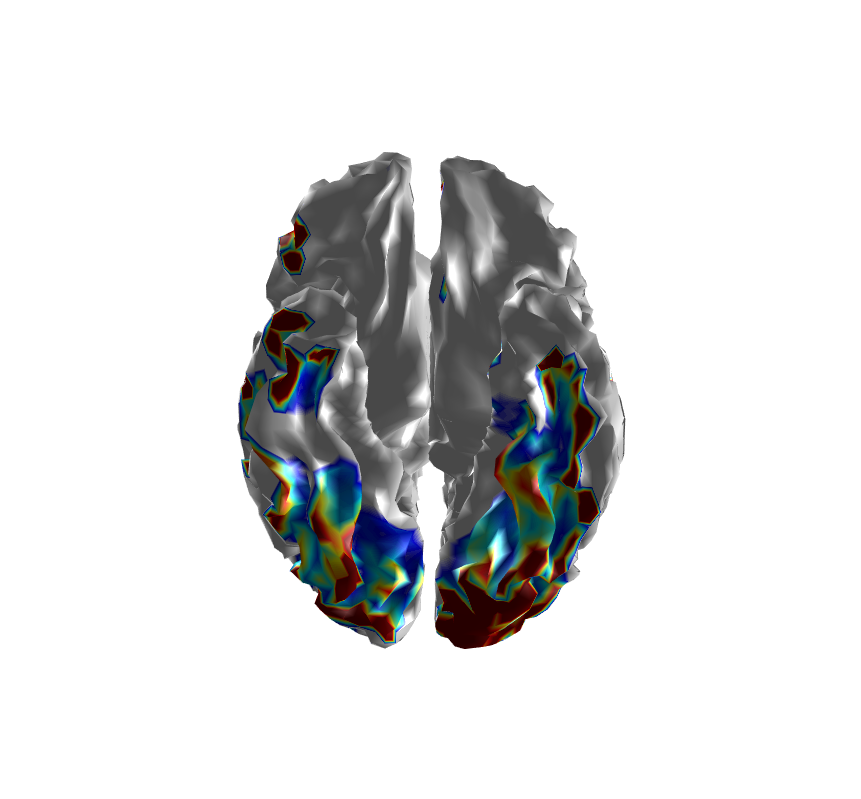} 
\includegraphics[scale=0.17]{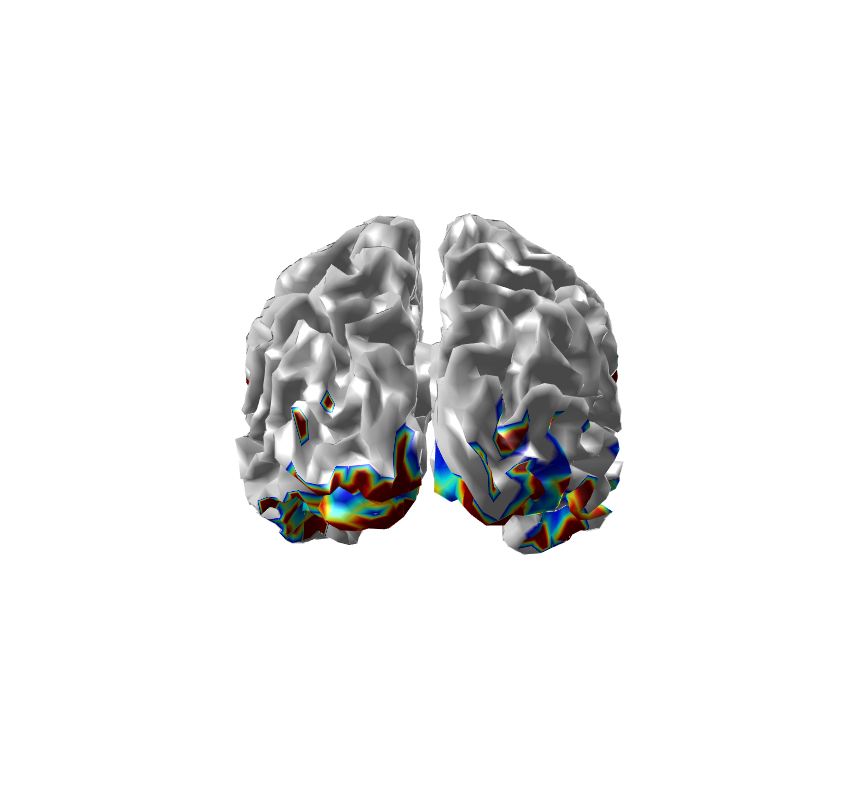} \\

\includegraphics[scale=0.17]{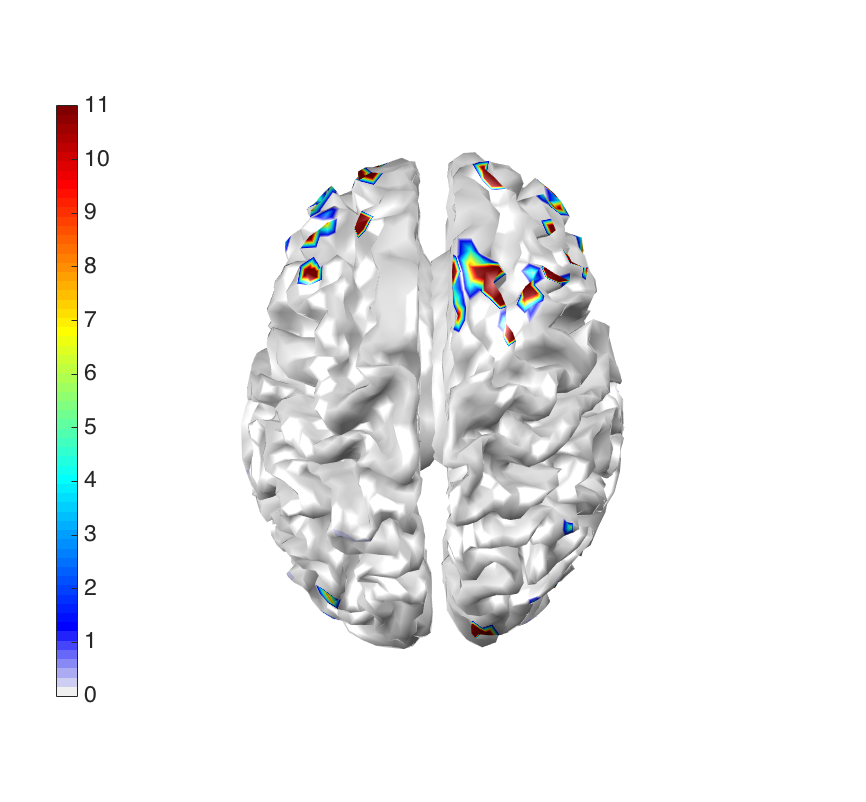} &
\includegraphics[scale=0.17]{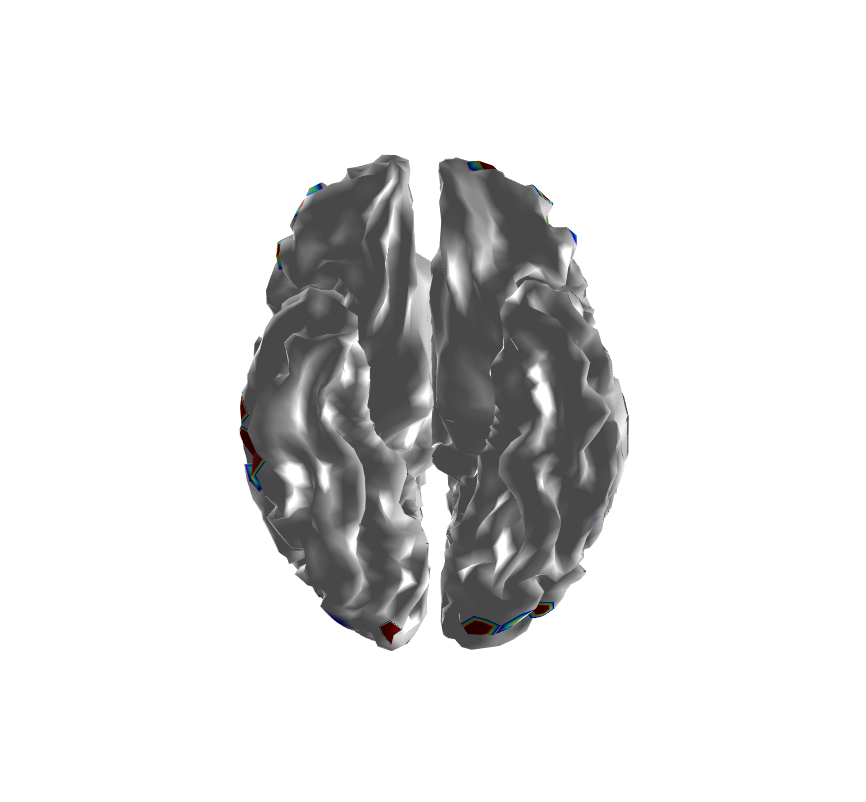} 
\includegraphics[scale=0.17]{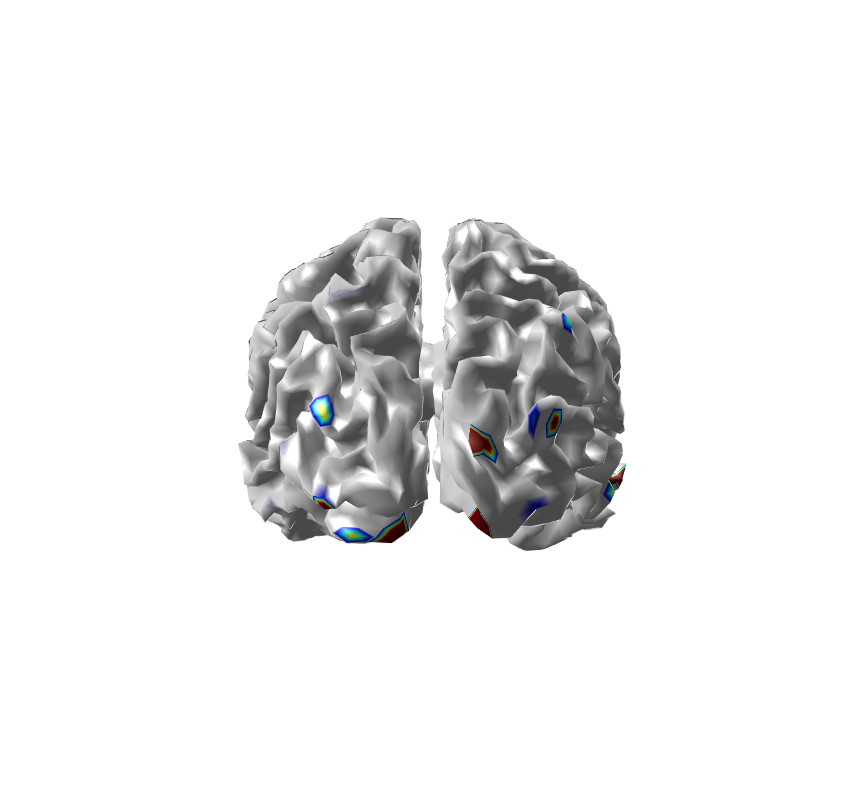} \\

\end{tabular}
\caption{Brain Activation for Scrambled Faces - The power of the estimated source activity $\sum_{t=1}^{T}\hat{S}_{j}(t)^{2}$ at each location $j$ of the cortical surface. Row 1 displays results from our proposed method applied to the combined MEG and EEG data; row 2 displays results from MSM applied to the MEG data; row 3 displays results from MSM applied to the EEG data.}
\end{figure}

In Figure 5 we examine cortical maps of $|\hat{S}_{j}(t)|$ at three peak time points $t = 50 +10$ (100ms after presentation of the stimulus), $t = 50 +25$ (175ms after presentation of the stimulus), and $t = 50 + 35$ (225ms after presentation of the stimulus). For each of the three selected time points we see that highest activity is observed on the left ventral surface; more specifically at Brodman area 19 at the first two peaks, and then moving into the perirhinal cortex at the third peak. The perirhinal cortex is involved in visual perception. The activation on the right ventral surface is relatively consistent across the three time points with peak activation in Brodman areas 18 and 19. 

\begin{figure}[htbp]
\centering
\begin{tabular}{cccc}
\hspace{-5em}
\includegraphics[scale=0.25]{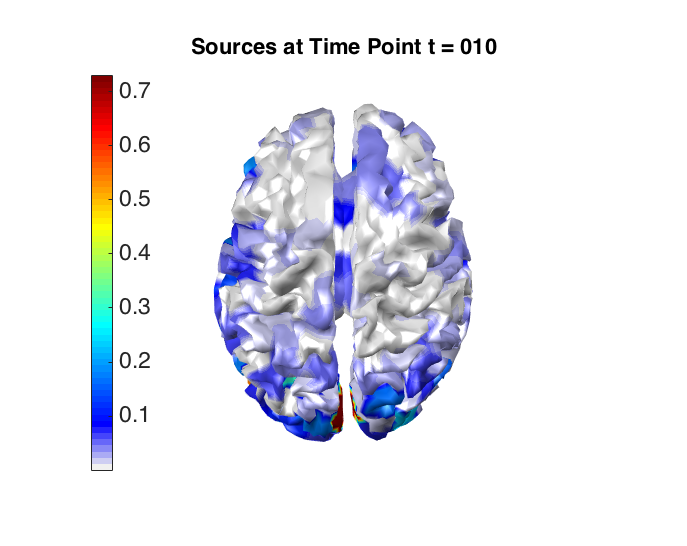} &
\hspace{-5.5em}
\includegraphics[scale=0.25]{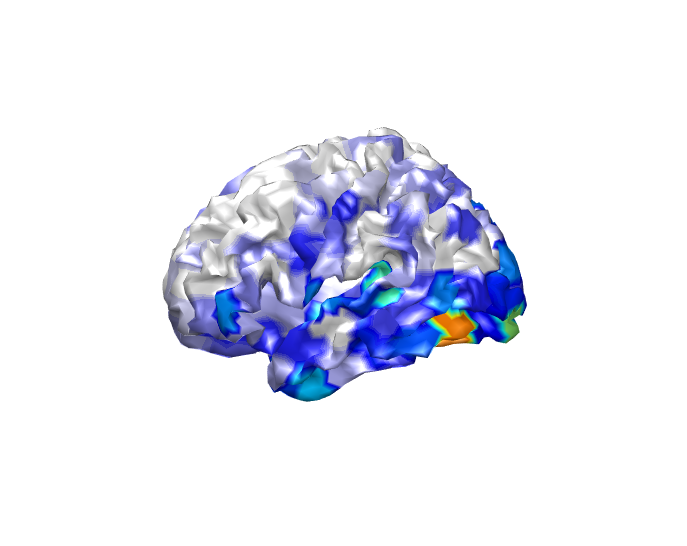} &
\hspace{-5.5em}
\includegraphics[scale=0.25]{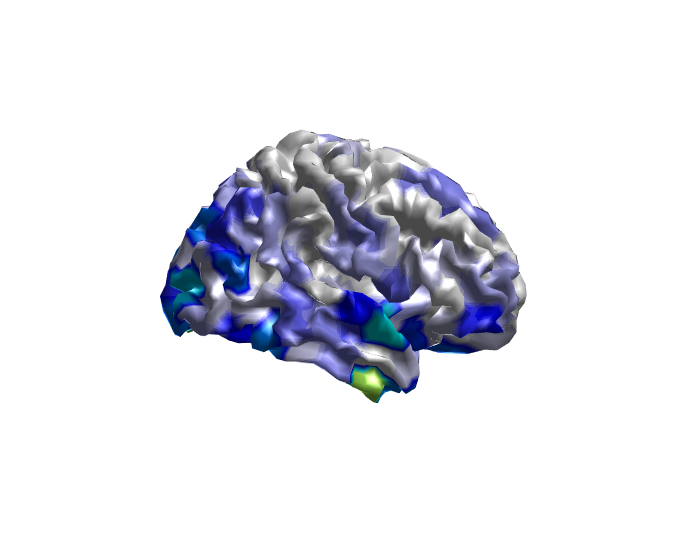} &
\hspace{-5.5em}
\includegraphics[scale=0.25]{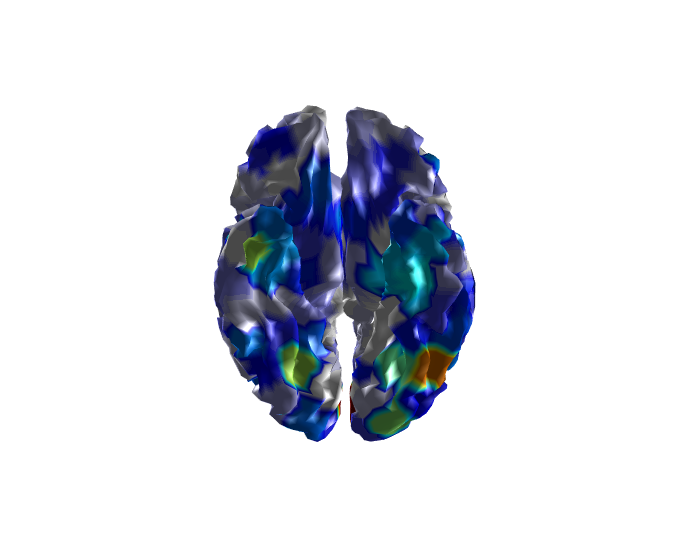} \\
\hspace{-5em}
\includegraphics[scale=0.25]{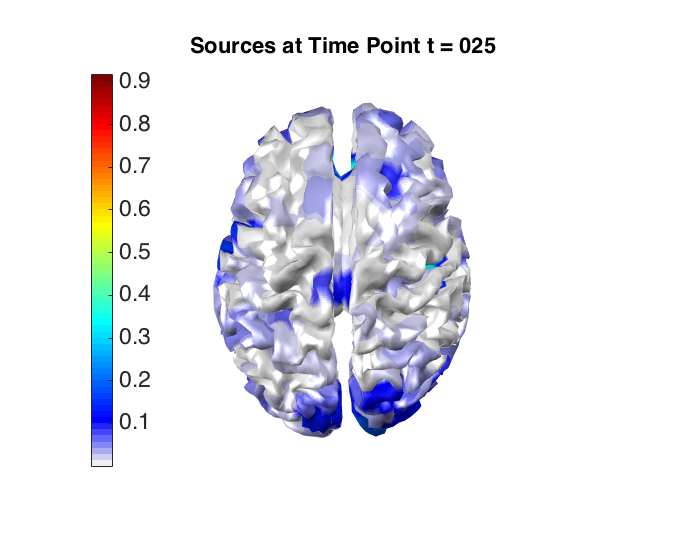} &
\hspace{-5.5em}
\includegraphics[scale=0.25]{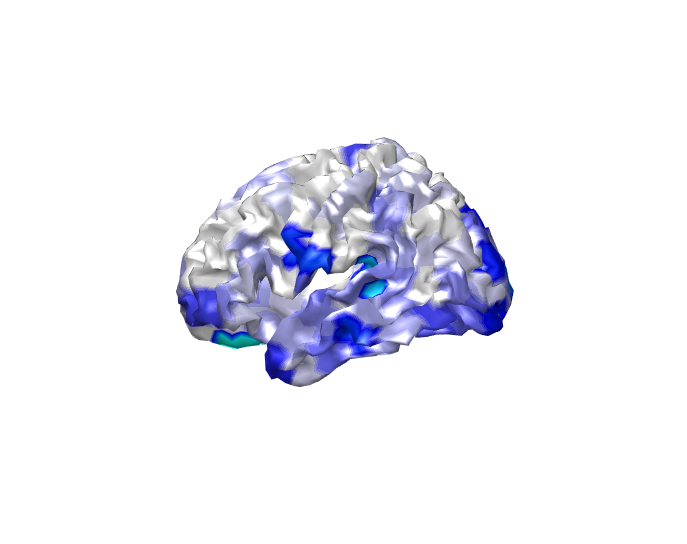} &
\hspace{-5.5em}
\includegraphics[scale=0.25]{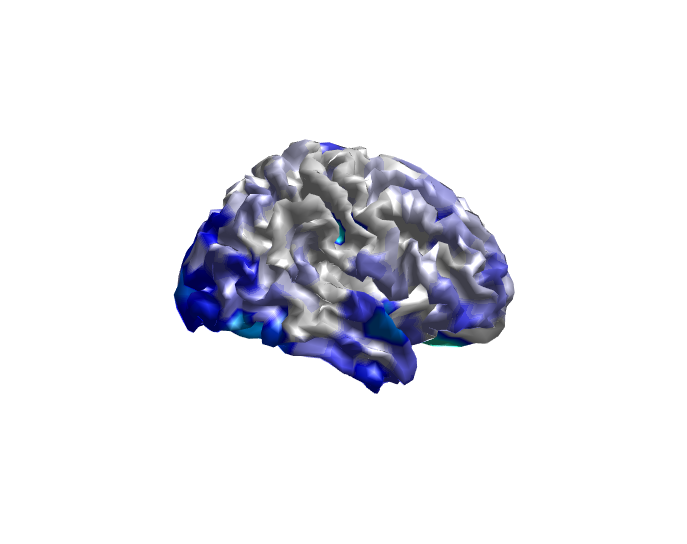} &
\hspace{-5.5em}
\includegraphics[scale=0.25]{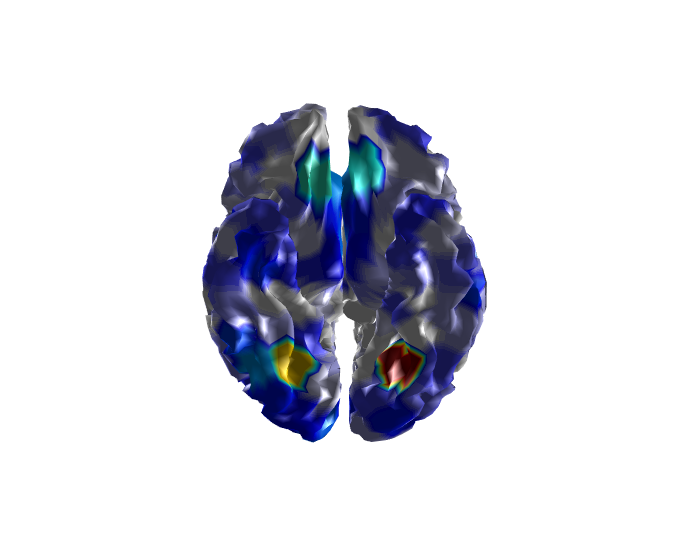} \\
\hspace{-5em}
\includegraphics[scale=0.25]{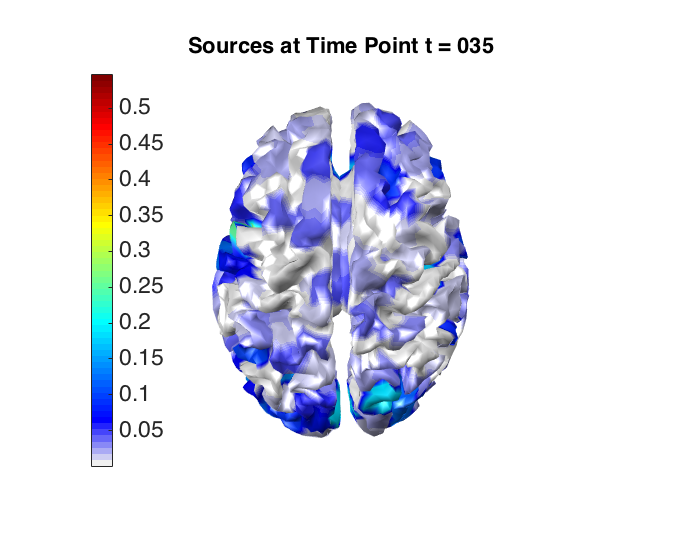} &
\hspace{-5.5em}
\includegraphics[scale=0.25]{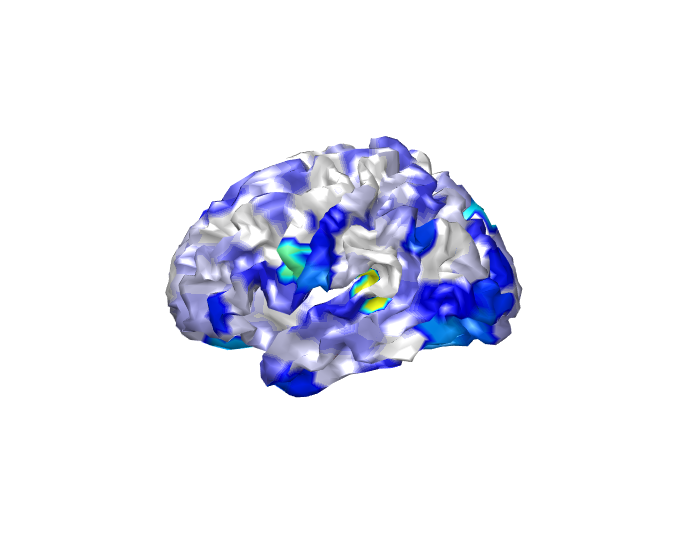} &
\hspace{-5.5em}
\includegraphics[scale=0.25]{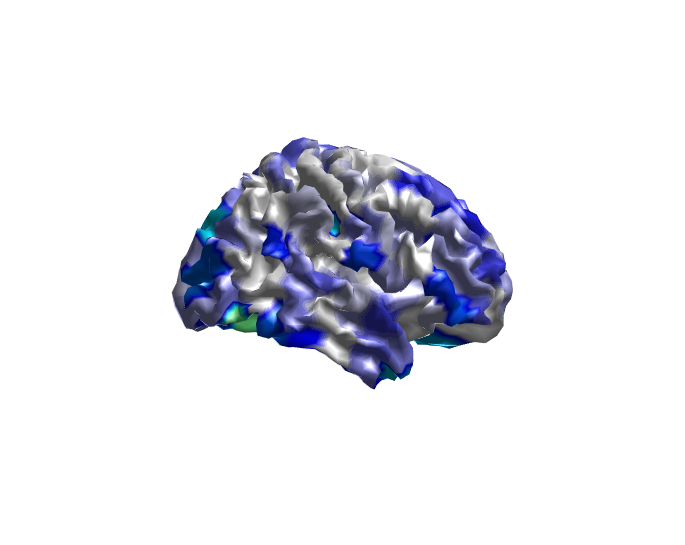} &
\hspace{-5.5em}
\includegraphics[scale=0.25]{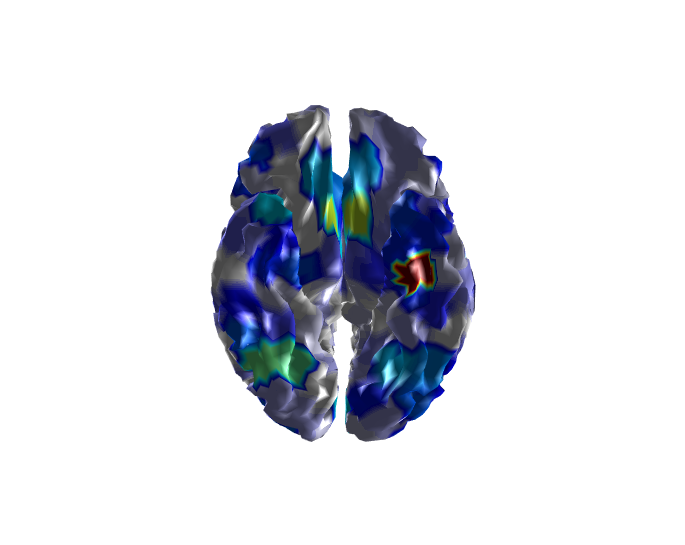} 
\end{tabular}
\caption{Brain Activation for Scrambled Faces - The magnitude of the estimated source activity $|\hat{S}_{j}(t)|$ at each location $j$ of the cortical surface and at three different time points, $t = 50 +10$ (Row 1; 100ms after presentation of the stimulus), $t = 50 +25$ (Row 2; 175ms after presentation of the stimulus), and $t = 50 + 35$ (Row 3; 225ms after presentation of the stimulus).}
\end{figure}

While it is not of specific interest to our brain mapping application we note that the spatial dependence parameter $\beta$ of the Potts model is estimated at $\hat{\beta} = 0.44$, which is right at the upper boundary of our restricted parameter space based on the approximate phase transition point of the Potts model. 


\subsection{Goodness-of-Fit to the Scrambled Faces MEG and EEG Data}

Finally, to examine the goodness-of-fit for the model we compute the residuals $ \hat{\vepsilon}_M(t) = \vM(t) -   \vX_M \hat{\vS}(t)$ and $\hat{\vepsilon}_E(t) = \vE(t) - \vX_E \hat{\vS}(t)$ for each time point $t = 1, \dots, T$. We will make here the assumption that they should be draws from a mean-zero Gaussian distribution if the assumed model generated the observed data. Figure 6, panels (a) and (b) show the time series plots of the residuals for EEG and MEG respectively. In the case of the EEG data, the model seems to have captured the signal at most of the sensors, though there are a few sensors where it appears that some part of the evoked signal has not been captured and remains in the residuals. The same holds true for the MEG data, and in addition, the residuals for the MEG data exhibit a periodic signal. This periodic signal is not part of the evoked response but is rather a property of the brain noise and so we are not overly concerned to see this pattern in the residuals. More concerning are the few sensors where large peaks still remain, which to us indicate parts of the evoked response that have not been captured adequately by the model. 

In Figure 6, panels (c) and (d) we show plots of the residuals versus fitted values for EEG and MEG respectively. For the EEG data there are no striking patterns, while for the MEG data we see higher values to the left of zero and lower values to the right of zero. These high and low values likely arise from the peaks of the periodic signal not captured by the fitted mean. 

Finally, Figure 6, panels (e) and (f) show normal quantile-quantile plots for the EEG and MEG residuals respectively. The Gaussian assumption seems somewhat tenable for the EEG data while this assumption does not seem reasonable for the MEG data, in particular, we see a strong deviation from normality in the left tail of the distribution. 

Overall, the residual analysis indicates a number of modelling assumptions that might be questionable for the data at hand. This is particularly true for the MEG data while the model yields a relatively better fit for the EEG data. Examining robustness to model misspecification and developing a more flexible model that can accommodate some of the features of the data not adequately captured by the current model (e.g., non-Gaussian distribution for the MEG data; temporally correlated residuals) is an important avenue for future work.

\begin{figure}[htbp]
\centering
\begin{tabular}{cc}
\vspace{-1em}
\hspace{-1.5em}
\includegraphics[scale=0.12]{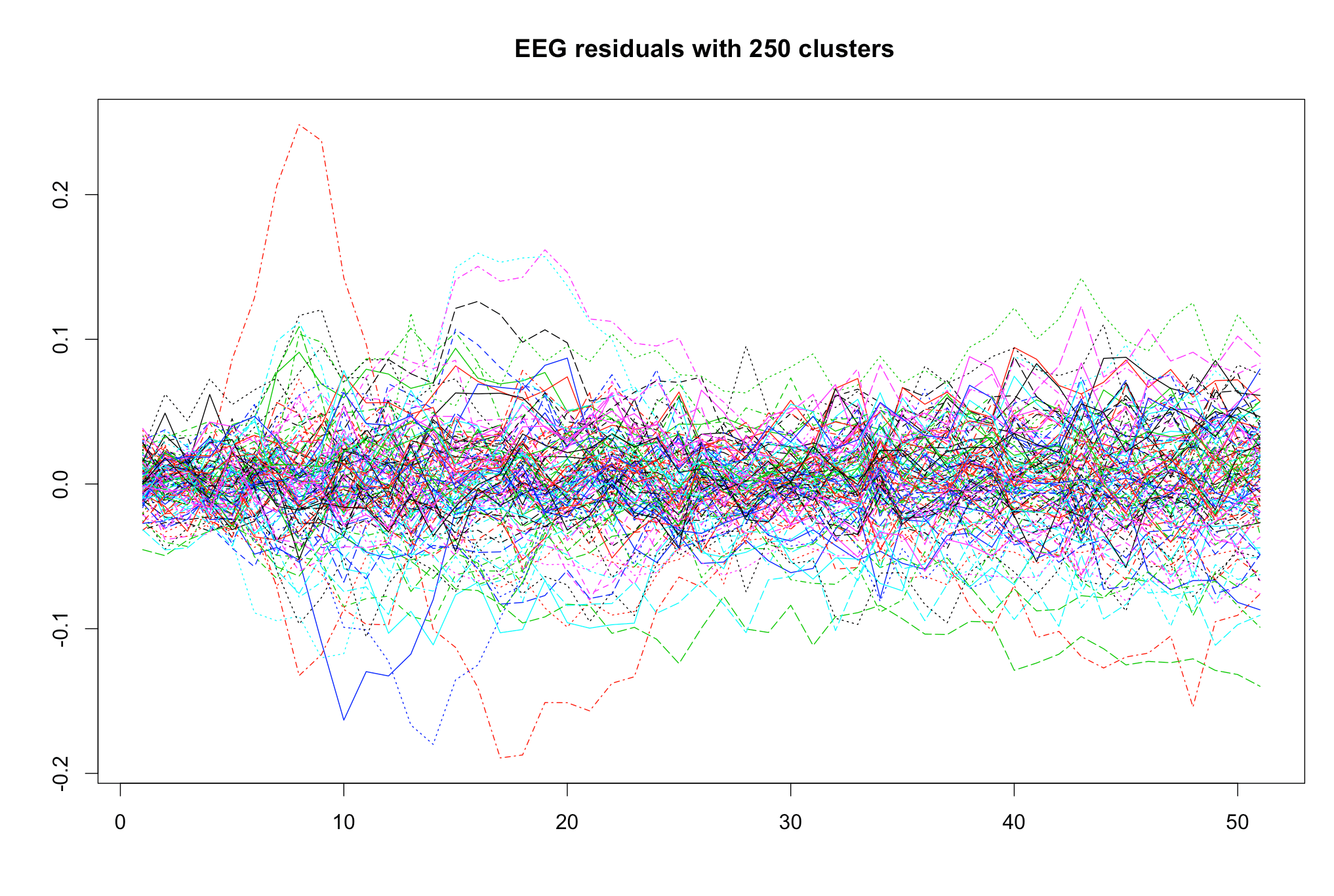} &
\hspace{-1.5em}
\includegraphics[scale=0.12]{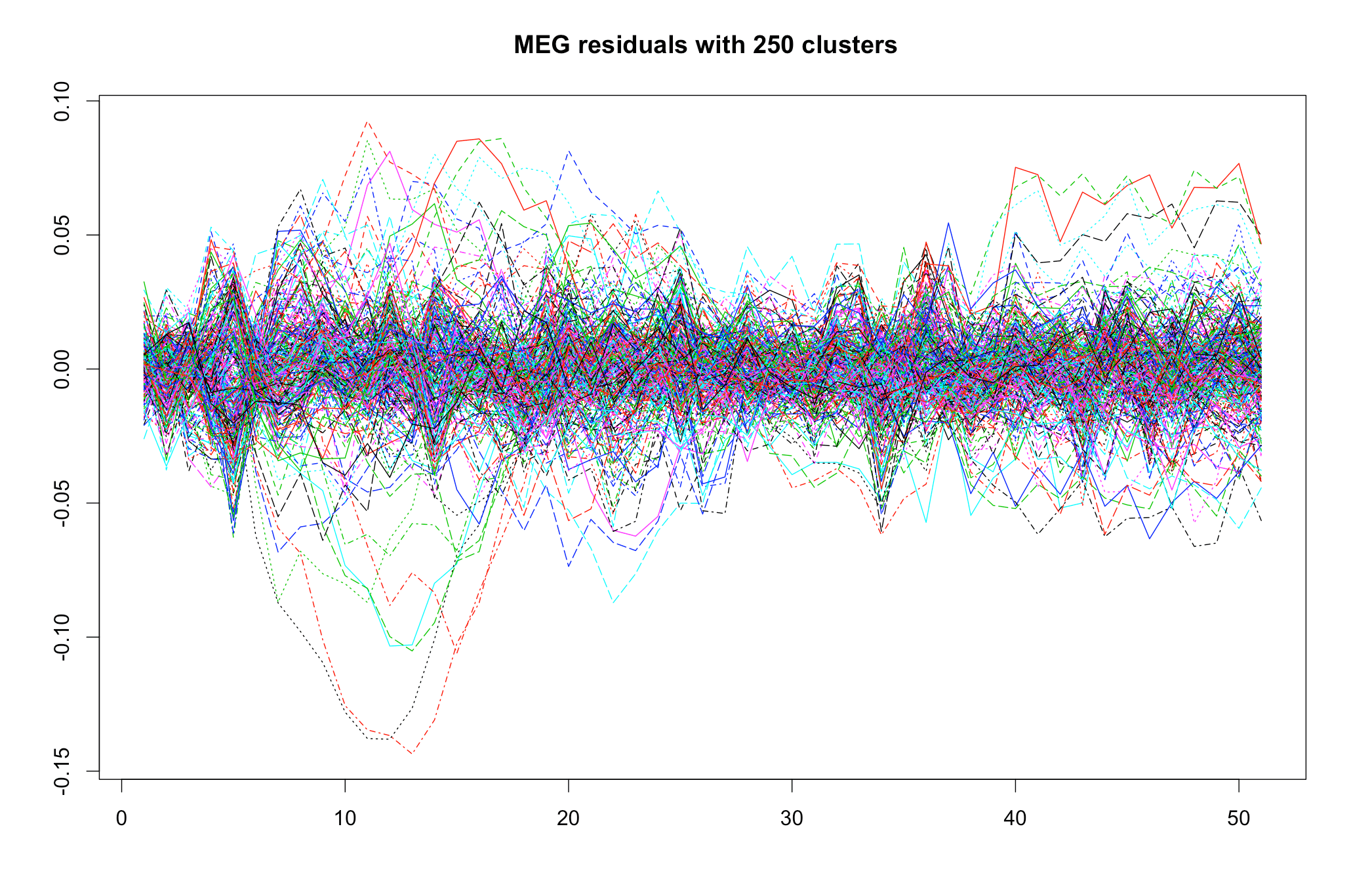}\\
(a) & \hspace{2em} (b)\\ \vspace{-0.5em}
\hspace{-1.5em}
\includegraphics[scale=0.235]{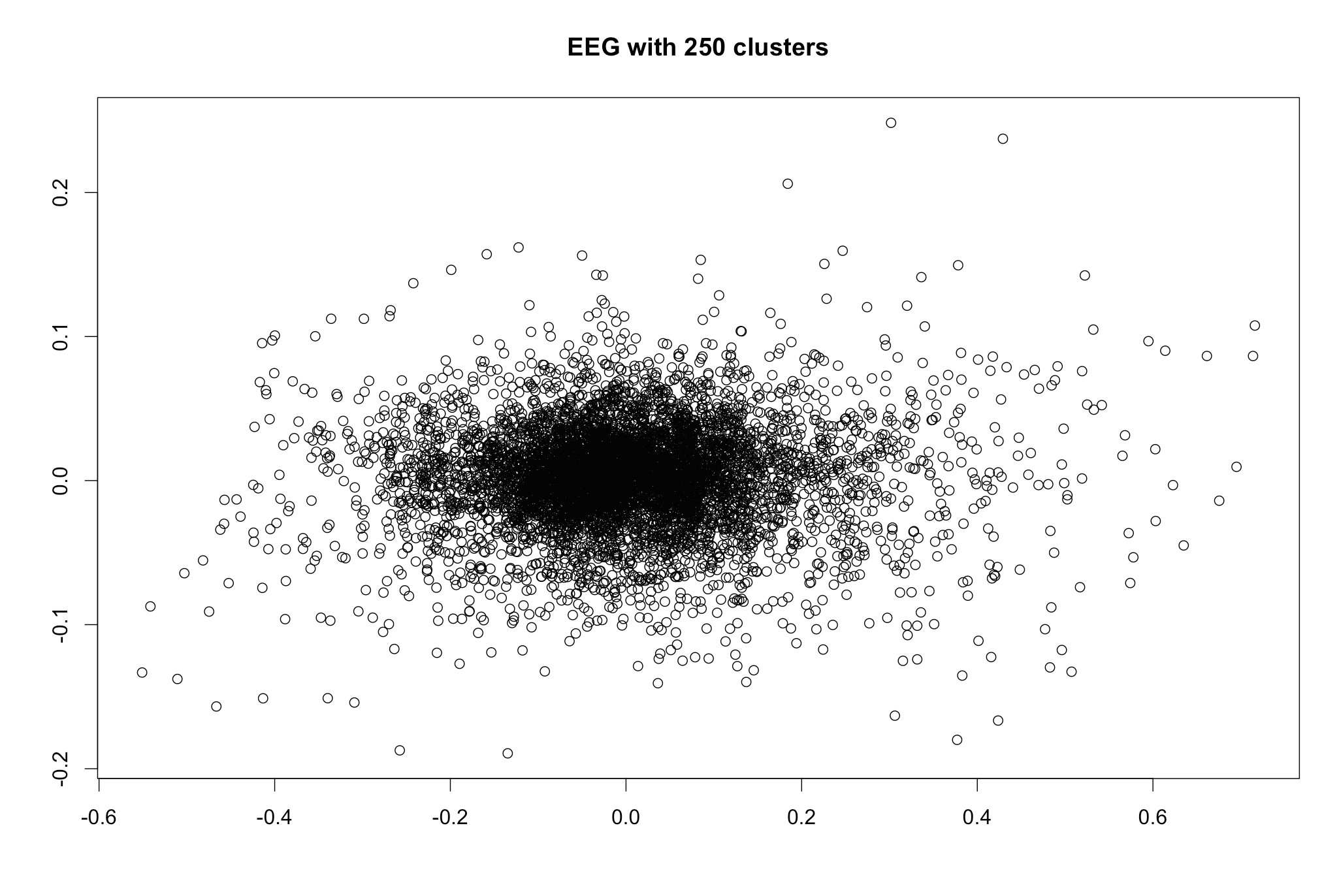} &
\hspace{-1.5em}
\includegraphics[scale=0.235]{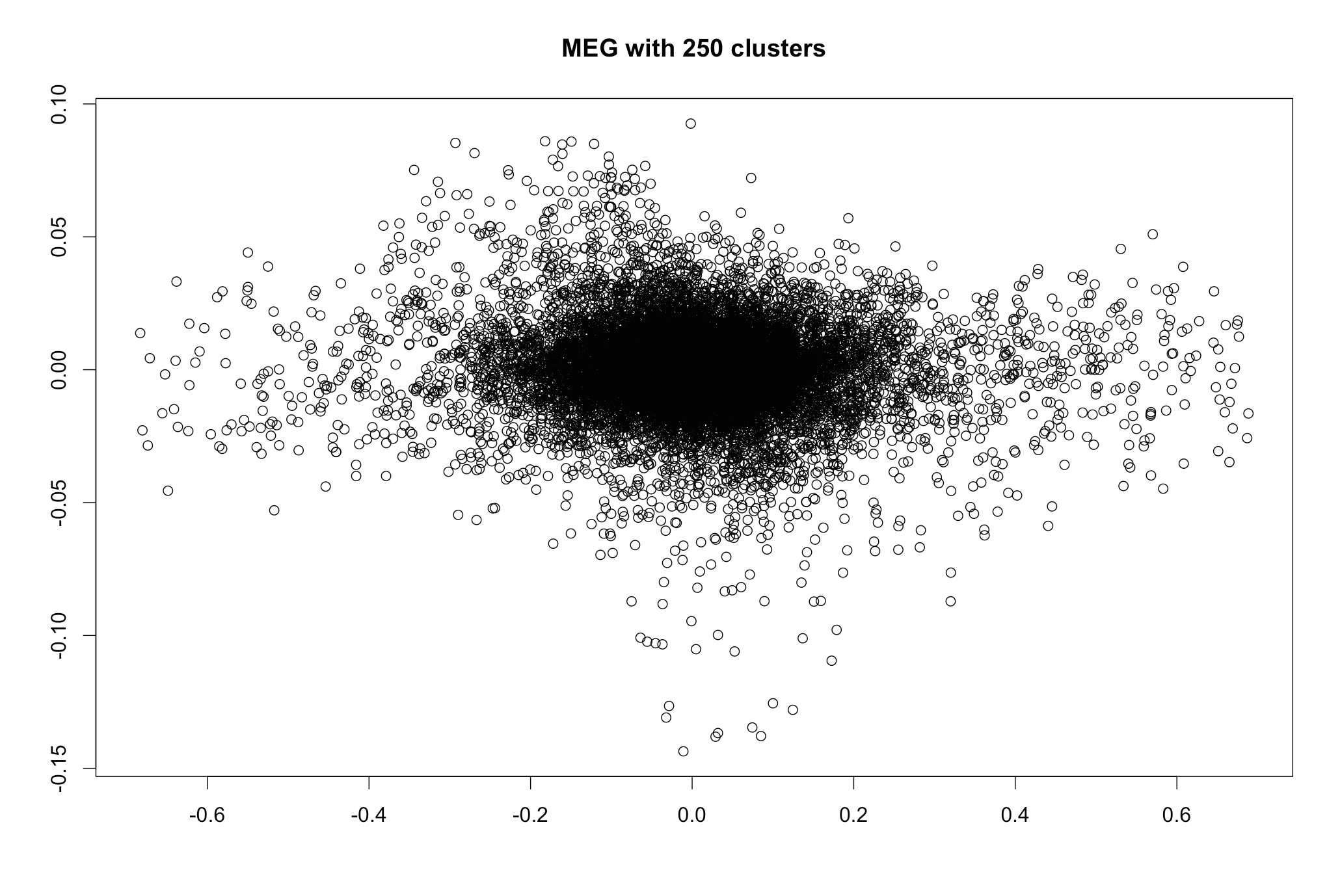} \\
(c) & \hspace{2em} (d)\\ \vspace{-0.5em}
\hspace{-1.5em}
\includegraphics[scale=0.235]{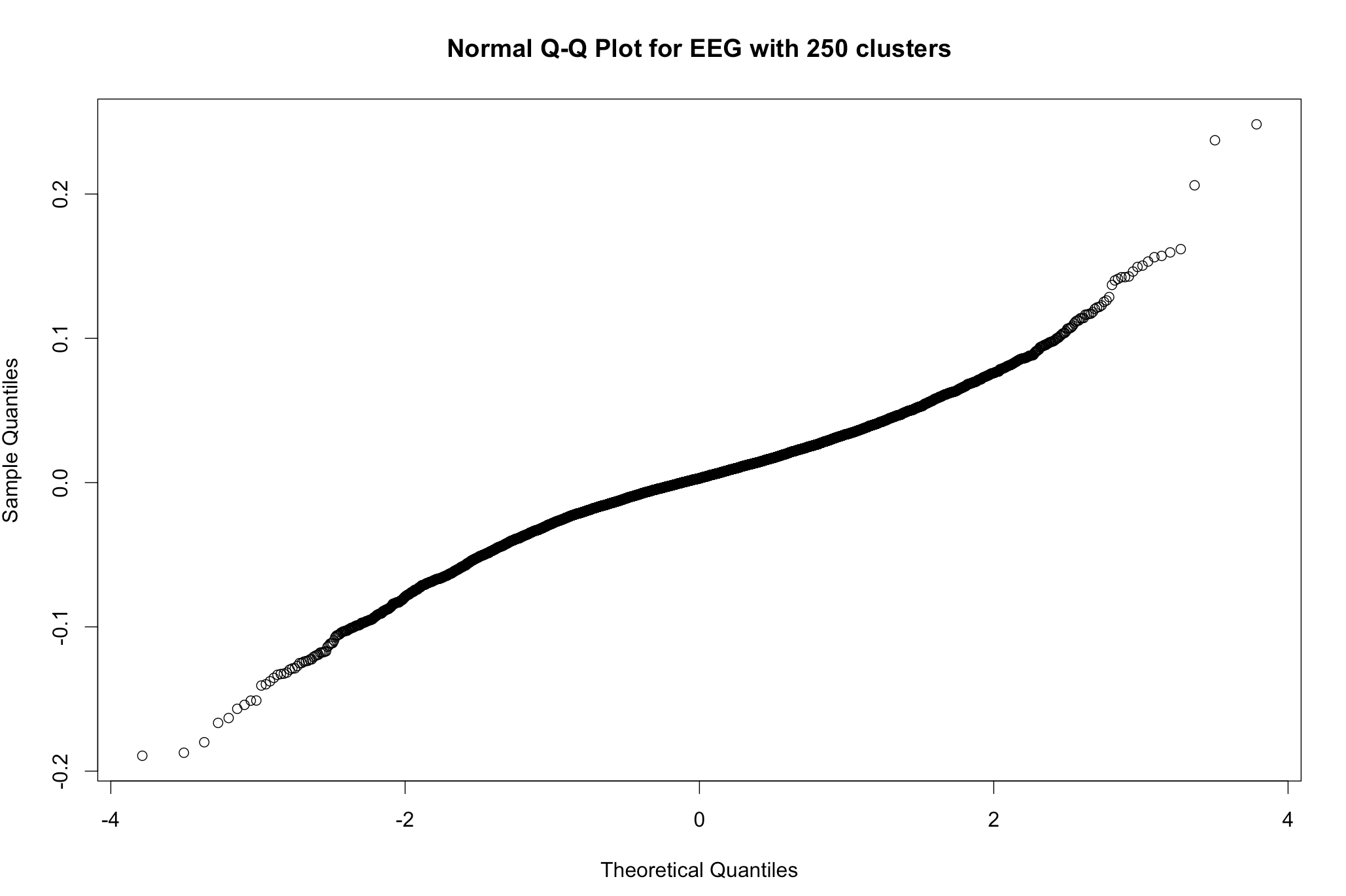} &
\hspace{-1.5em}
\includegraphics[scale=0.235]{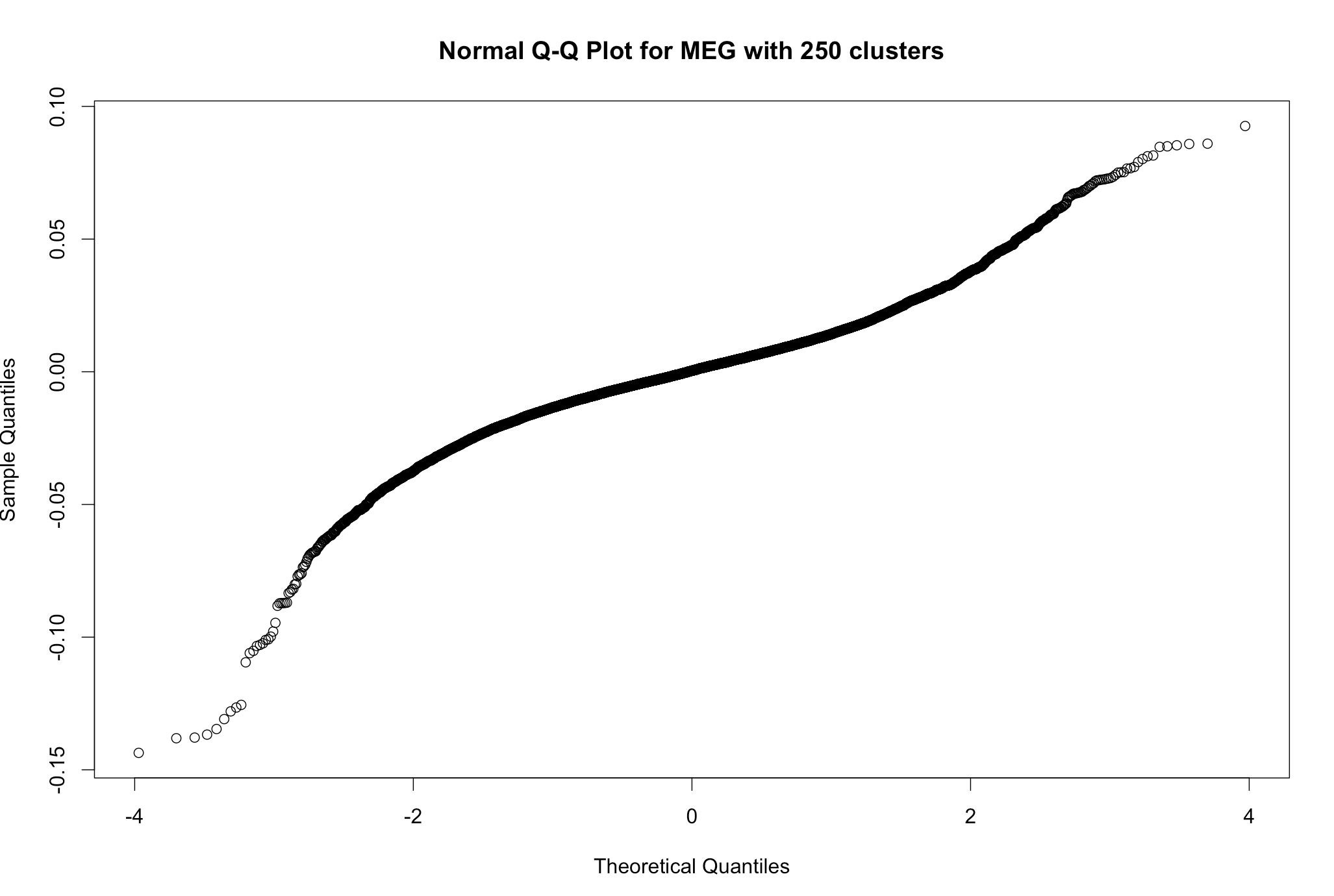}\\ 
(e) & \hspace{2em} (f)\\ 
\end{tabular}
\caption{Brain Activation for Scrambled Faces - Residual Diagnostics: Time series of residuals, (a) EEG, (b) MEG;  Residuals versus fitted values, (c) EEG, (d) MEG;  Residual normal quantile-quantile plots, (e) EEG,  (f) MEG.}
\end{figure}

\section{Discussion}

Motivated by a study examining the neural response to scrambled faces, we have developed a new approach for solving the inverse problem associated with combined EEG and MEG data. We view our methodology as primarily applicable both to situations where the MEG and EEG data are collected simultaneously and also to situations where the data are collected sequentially in a situation  where the data mimic a simultaneous recording paradigm. Our new model incorporates two simple ideas. First, it can be beneficial to combine complementary sources of information when estimating unknown parameters, in particular when in the high-dimensional setting. We thus develop a joint model that links together MEG, EEG, and MRI data. Second, the hidden states of the brain are spatially correlated and we incorporate this prior information into a model that estimates the states of the brain, the number of such states, and their dynamics. Combining these two ideas together we have developed an approach for source localization that appears to result in some improvements over and above the original MSM mixture model, for the settings considered.  It is worth noting that it is not only the models but also the model fitting algorithms that differ between the two approaches. Ours employs ICM whilst Daunizeau and Friston (2007) employ mean-field variational Bayes. 

Our methodology makes the very strong assumption that neural activity is generated by a small number of latent states. While many neuroscientists might hesitate to claim such low dimensions, there are certainly many situations where the bulk of neural activity may turn out to be closely approximated by a low dimensional process with linear interaction. On a coarser time scale, most fMRI studies of tasks find dozens of distinct small regions activated, although they may not be statistically independent. Extending our approach to accommodate a large number of sources is an open problem.

\emph{Initial Values for ICM}: It is well known that the ICM algorithm is sensitive to initial values and we have also found this to be the case with the ICM algorithm developed for our model. The solution obtained, and even the convergence of the algorithm will depend rather heavily on the initial values chosen. At present our approach to this problem is to make a 'sensible' choice for the initial values based on the heuristics described in Section 3. This approach for obtaining the initial values was used in the simulation examples, simulation studies, and in the data analysis in our application and seems to provide satisfactory results, in particular as observed in the simulation studies. Furthermore, the K-means algorithm is used as part of our approach for selecting the initial values for $\vZ$ and also for clustering locations on the cortex. As the K-means algorithm itself depends on randomly chosen initial values, this stochasticity will propagate into our solution. Generally, multiple runs of the algorithm lead to solutions that are similar in terms of the general spatial and temporal localizations, but the solutions that will vary from one run to another as a result of random initial values. This is less of a problem with simulated data but with real data where the signal is not as strong and where the model is likely misspecified to a greater extent than in our examples we do see an admittedly undesirable amount of variability in $\hat{\vZ}$ from one run to another. The estimated sources $\hat{\vS}(t)$ exhibit a lesser degree of variability and indicate generally the same spatial regions from one run to another, and the same temporal patterns. The map of total estimated power $\sum_{t=1}^{T}\hat{S}_{j}(t)^{2}$ is quite stable, as is $\hat{K}_{ICM}$, and the latter estimate usually converges after only a few iterations. One possible solution to the problem of dependence on initial values and the resulting variability may be through the use of some type of annealing approach or through ant colony system optimization. 

In our current work we are investigating the performance of ant colony optimization (ACO) in comparison to ICM and simulated annealing (SA). ACO is a population-based nature inspired global optimization algorithm that mimics the behaviour of real biological ants searching for food with pheromone-based communication. Our comparisons are being made within the context of spatial mixture models with Gaussian components and labelings based on the Potts model. While such a model is quite a bit simpler than the dynamic joint model developed in this paper, we have noticed in our current investigations that ACO substantially outperforms ICM and SA, both in terms of finding a larger value of the objective function and also in obtaining better pixel/voxel labelings when the ground truth is known. In a separate paper we will soon report on our results comparing ACO, ICM, and SA for spatial mixture models. We hope that a population-based algorithm if successfully developed and applied will make the results obtained from our model more stable. Currently the most reliable outputs are the map of total estimated power and $\hat{K}_{ICM}$.  It is worth noting that sensitivity and instability are intrinsic characteristics of the ill-posed inverse problem that we are dealing with; however, successful implementation of a global optimization algorithm such as ACO should help in alleviating the problem to some extent.

It should be noted that our methodology can also be applied to studies involving multiple subjects that are partitioned into groups (e.g., disease groups) where the model can be applied to each subject separately to reconstruct the sources of neural activity in a first stage analysis, and then the reconstructed neural activity can be compared across groups in a second stage analysis. In addition to looking at the reconstructed source activity, one interesting possibility in this case would be to compute $\hat{K}_{ICM}$ for each subject in the study and then to compare the estimated number of neural states across the different study groups. We suspect that such an analysis would be of great interest in a number of studies involving neuroimaging.

\nocite{*}
\bibliographystyle{ECA_jasa}
\bibliography{JASA_example}
\pagebreak

\begin{center}
\textbf{\large Supplementary Material for 'A Potts-Mixture Spatiotemporal Joint Model for Combined MEG and EEG Data'}
\end{center}

\setcounter{equation}{0}
\setcounter{figure}{0}
\setcounter{table}{0}
\setcounter{page}{1}
\setcounter{section}{0}
\makeatletter
\renewcommand{\theequation}{S\arabic{equation}}
\renewcommand{\thetable}{S\arabic{table}}

\section{Data Transformations and Supplementary Figures}

Given the original MEG and EEG data collected at the sensor arrays	
\begin{align*}
	\tilde \vM(t) & = (\tilde M_1(t), \tilde M_2(t), . . . , \tilde M_{n_M}(t))', \,\,\, t= 1,\dots,T \\
	\tilde \vE(t) & = (\tilde E_1(t), \tilde E_2(t), . . . , \tilde E_{n_E}(t))' , \,\,\, t= 1,\dots,T,
\end{align*}
we let $\tilde \vM$ and $\tilde \vE$ denote the corresponding data matrices of dimensions	$n_{M} \times T$ and $n_{E} \times T$, respectively. Similarly, we let $\tilde \vX_E$ and $ \tilde \vX_M$ denote the $n_E \times P$ and $n_M \times P$ EEG and MEG forward operators computed based on Maxwell's equations, the sensor array locations, the pre-specified locations on the cortex, and other assumptions on the conductivity of the fluids and tissues within the head.

Our model assumes that these data have been transformed as suggested by Henson et al. (2009b) as follows:
	\begin{align*}
	 \vM &= \frac{\tilde \vM}{\sqrt{\frac{1}{n_{M} }tr(\tilde \vM \tilde \vM^T)}}, &
	 \vE &= \frac{\tilde \vE}{\sqrt{\frac{1}{n_{E} }tr(\tilde  \vE \tilde \vE^T)}} \\
	 \vX_{M} &= \frac{\tilde \vX_{M}}{\sqrt{\frac{1}{n_{M} }tr(\tilde  \vX_{M} \tilde \vX_{M}^T)}}, &
	 \vX_{E} &= \frac{\tilde \vX_{E}}{\sqrt{\frac{1}{n_{E} }tr(\tilde  \vX_{E} \tilde \vX_{E}^T)}}, &
	\end{align*}
and the joint model is then specified for the transformed data as described in Section 3 of the paper.

\begin{figure}[htbp]
\centering
\begin{tabular}{c}
\includegraphics[scale=0.8]{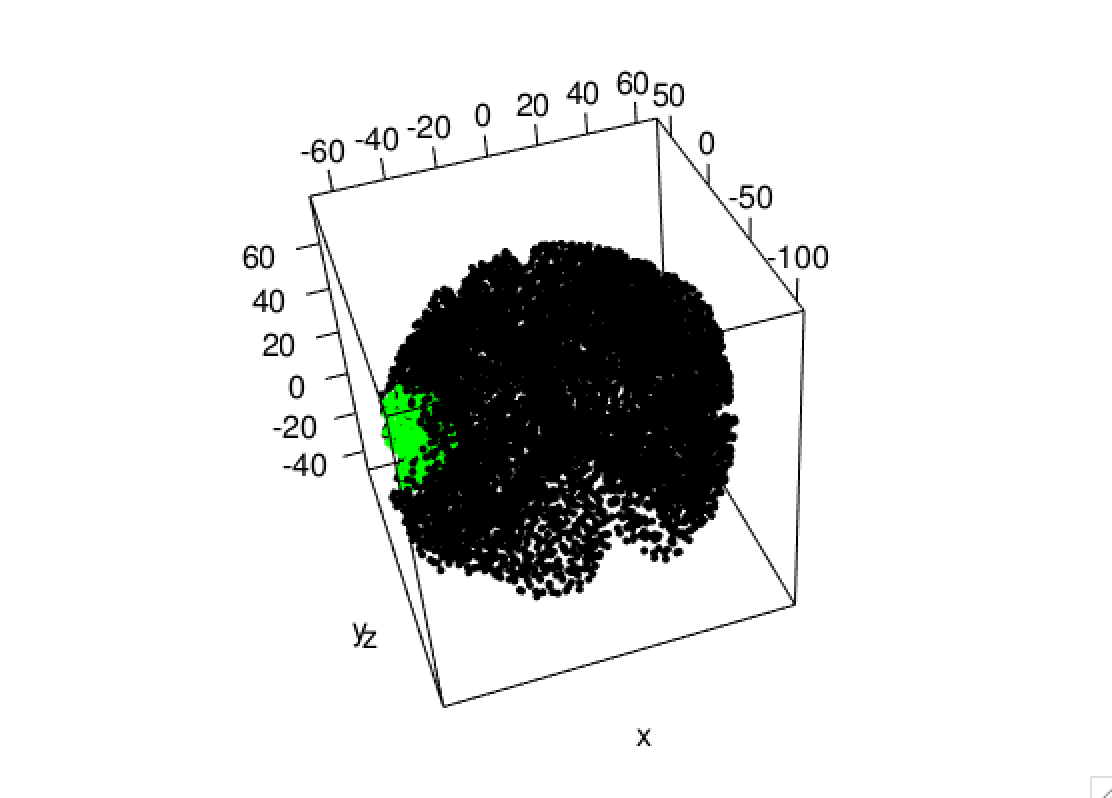} 
\end{tabular}
\caption{The true allocation of cortical locations to mixture components in the case where $K_{true} = 2$. Locations coloured green correspond to active locations while the other locations are inactive. The signal at active locations is based on the Gaussian curve depicted in supplementary Figure 2 panel (a). In total, there are 8,196 cortical locations used in this example.}
\end{figure}

\begin{figure}[htbp]
\centering
\begin{tabular}{cc}
\includegraphics[scale=0.25]{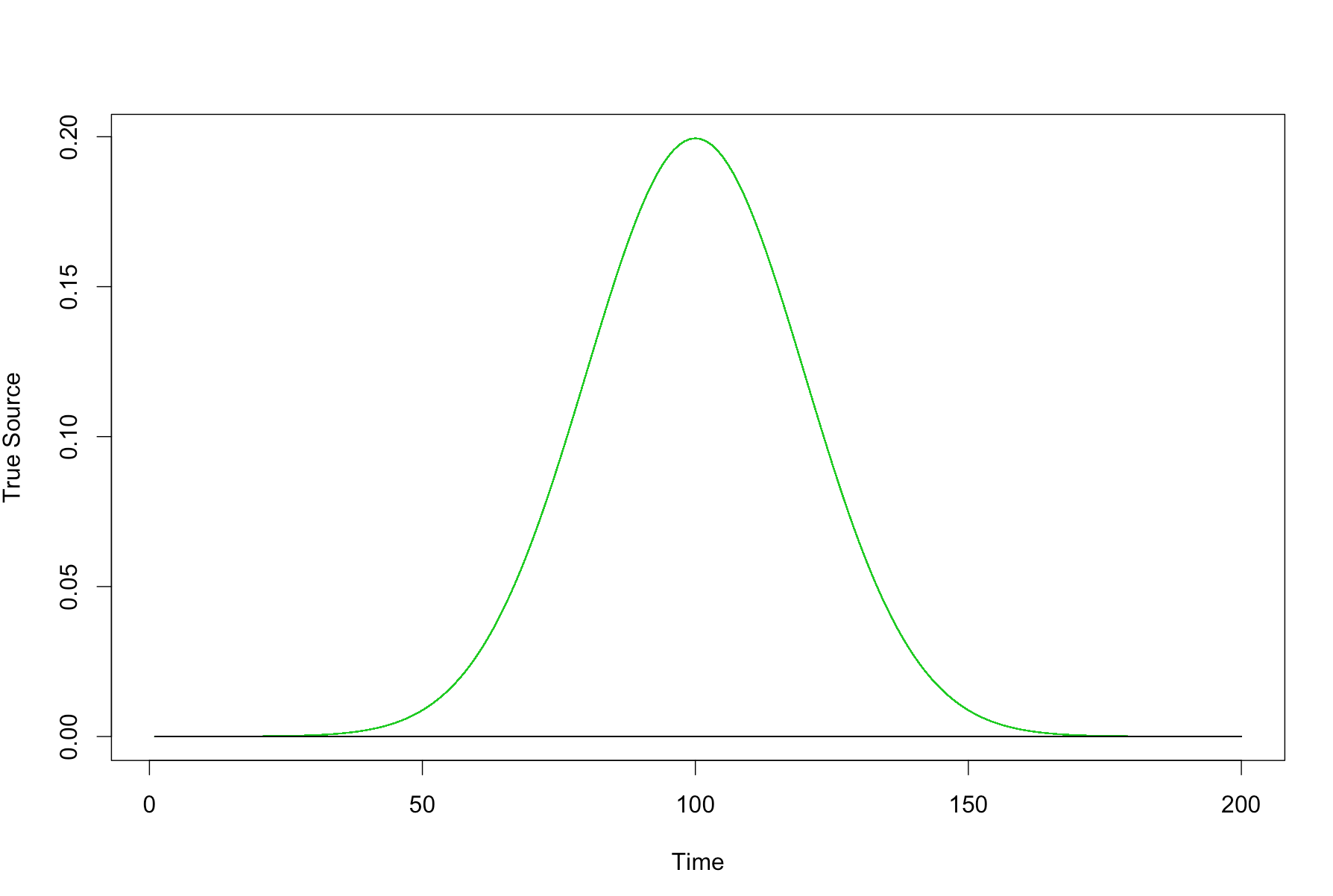} &
\includegraphics[scale=0.25]{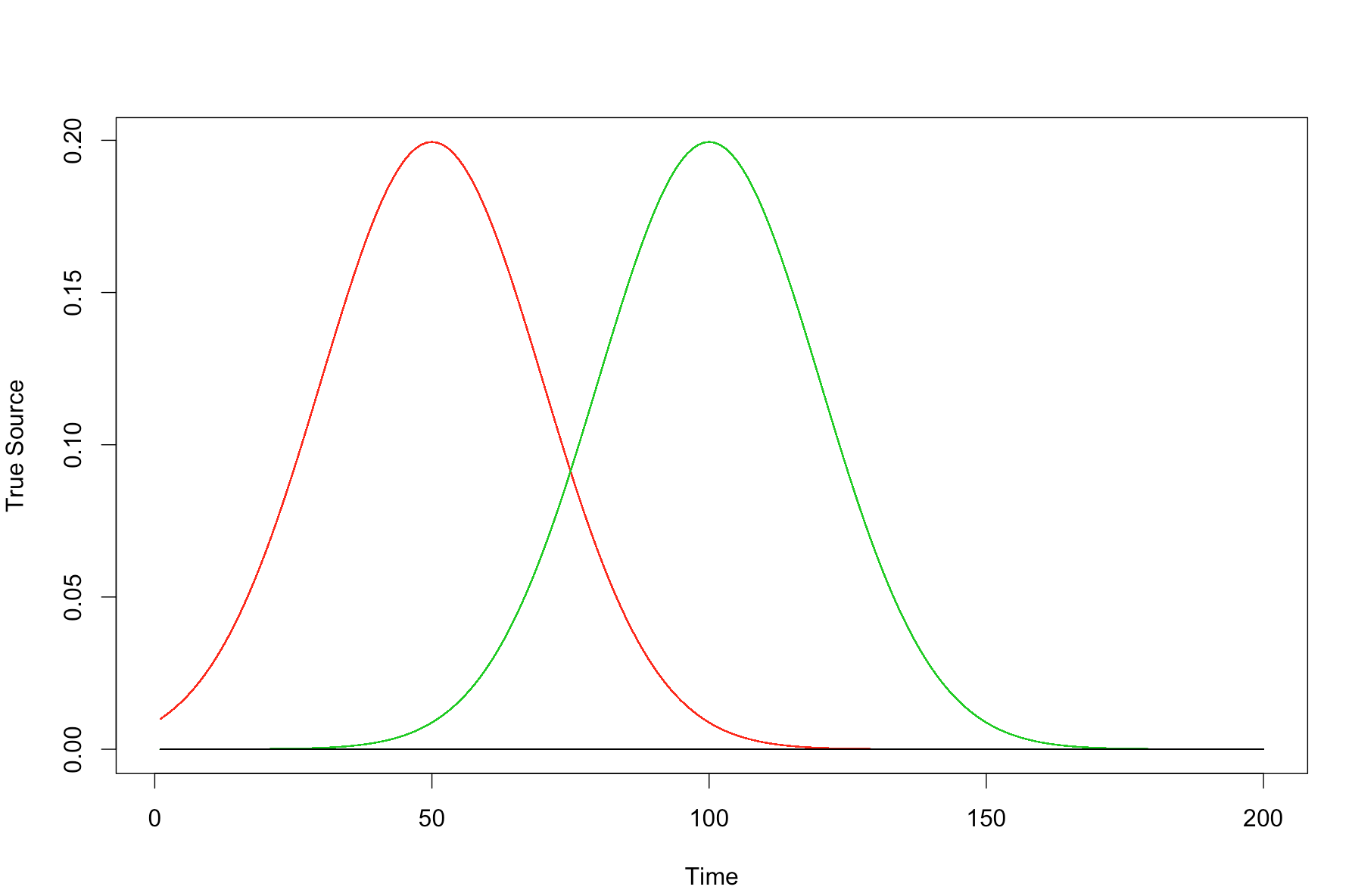}  \\
(a) & (b) \\ 
\includegraphics[scale=0.25]{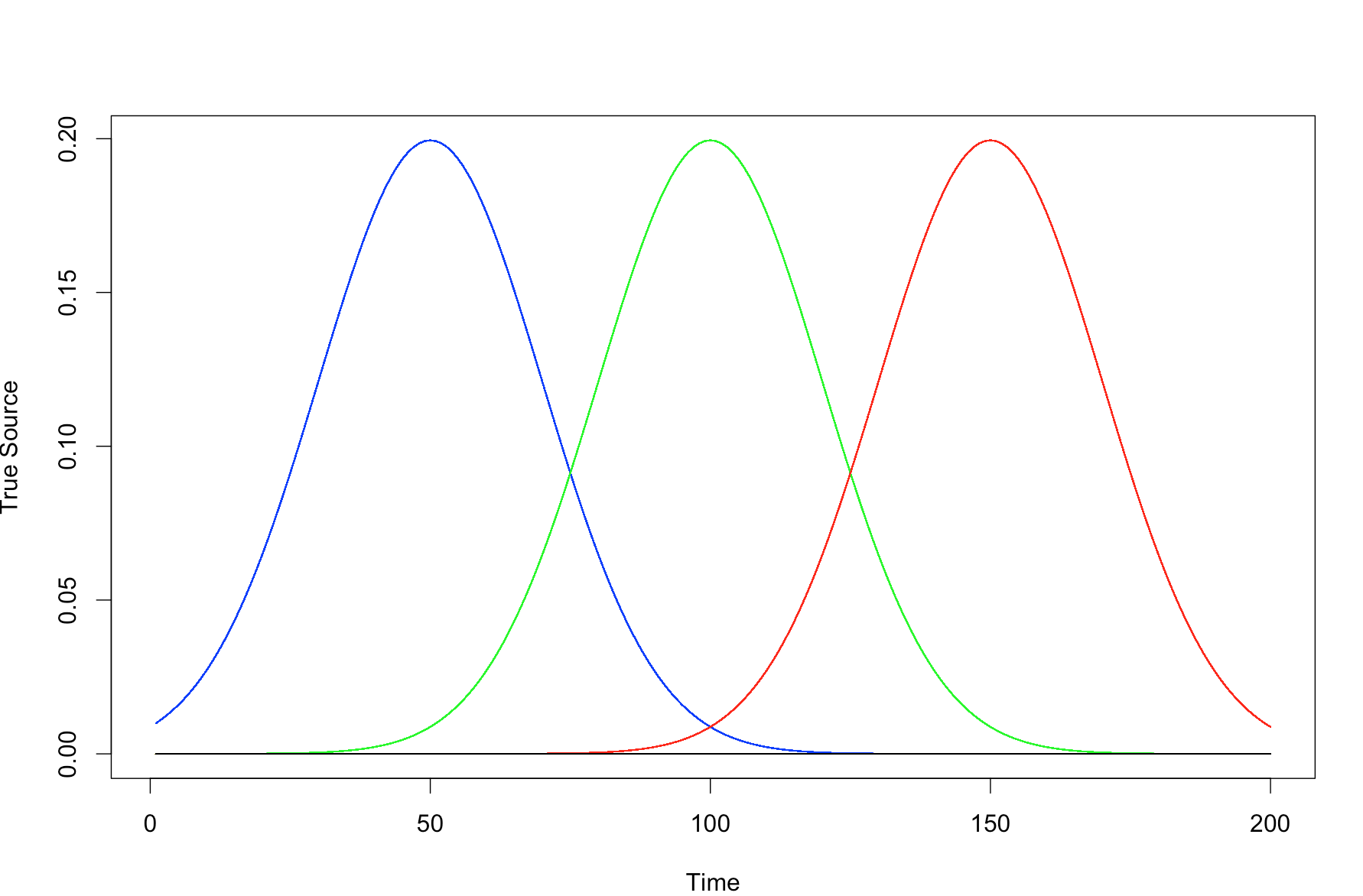}&
\includegraphics[scale=0.25]{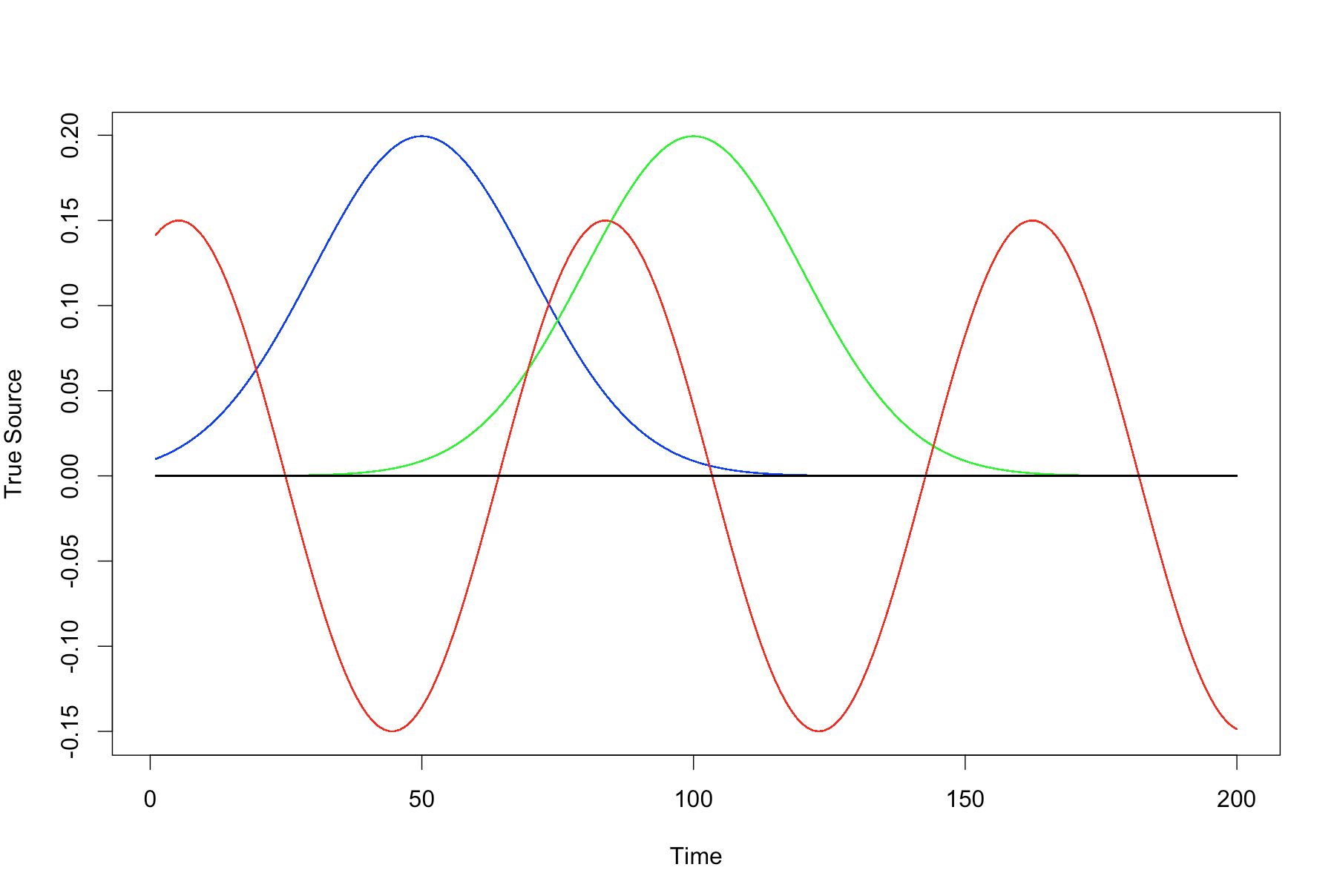} \\ 
(c) & (d) \\
\end{tabular}
\caption{The true signal $S_{j}(t)$ used in in each of the distinct active and inactive regions in the simulation study of Section 5.2, in the second part of the study where the mixture components are less well separated.}
\end{figure}

\section{Derivations for ICM algorithm}

The ICM algorithm requires the full conditional distribution for each model parameter.  The mode of this distribution is then used in the update step for that parameter. Here we derive the required full conditional distributions. 

The  full conditional distributions for $\sigma^2_M$ and $\sigma^2_E$ are obtained as follows:
\begin{align*}
	P(\sigma^2_M |Rest) & \propto \prod_{t=1}^{T}[P(M(t) | S(t), \sigma^2_M)] \times P(\sigma^2_M) \\
	& \propto \prod_{t=1}^{T}\bigg[|\sigma^2_M H_M| ^{-1/2} \exp\Big \{  -\frac{1}{2} \big(M(t) - X_M S(t)\big)^T (\sigma^2_M H_M) ^{-1} \big(M(t) - X_MS(t)\big)\Big\}\bigg] \\
	& \times (\sigma^2_M)^{-(a_M+1)}\exp(-b_M/\sigma^2_M)\\
	&  \propto  (\sigma^2_M)^{-TN_M/2 } \exp\Big \{  \sum_{t=1}^{T} -\frac{1}{2} \big(M(t) - X_MS(t)\big)^T (\sigma^2_M H_M) ^{-1} \big(M(t) - X_MS(t)\big)\Big\} \\
	& \times (\sigma^2_M)^{-(a_M+1)}\exp(-b_M/\sigma^2_M)\\
	& \propto (\sigma^2_M)^{-(a_M+\frac{TN_M}{2}+1)} \exp\Big \{  \sum_{t=1}^{T} -\frac{1}{2\sigma^2_M} \big(M(t) - X_MS(t)\big)^T H_M ^{-1} \big(M(t) - X_MS(t)\big)  - \frac{b_M}{\sigma^2_M}\Big\} \\
	& \propto (\sigma^2_M)^{-(a_M+\frac{TN_M}{2}+1)} \exp\Big \{  \frac{-1}{\sigma^2_M}\Big( \sum_{t=1}^{T} \frac{1}{2} \big(M(t) - X_MS(t)\big)^T H_M ^{-1} \big(M(t) - X_MS(t)\big)  +b_M\Big) \Big\} 
	\end{align*}
Therefore, the full conditional for $\sigma^2_M$ is an Inverse-Gamma distribution with
	\begin{align*}
	a^*_M &= a_M+\frac{TN_M}{2}\\
	b^*_M  &=  \sum_{t=1}^{T} \frac{1}{2} \big(M(t) - X_MS(t)\big)^T H_M ^{-1} \big(M(t) - X_MS(t)\big)  +b_M 
	\end{align*}
Similarly, we can get the full conditional for $\sigma^2_E$ as:
\begin{align*}
	P(\sigma^2_E |Rest) & \propto \prod_{t=1}^{T}[P(E(t) | S(t), \sigma^2_E)] \times P(\sigma^2_E) \\
	& \propto \prod_{t=1}^{T}\bigg[|\sigma^2_E H_E| ^{-1/2} \exp\Big \{  -\frac{1}{2} \big(E(t) - X_E S(t)\big)^T (\sigma^2_E H_E) ^{-1} \big(E(t) - X_ES(t)\big)\Big\}\bigg] \\
	& \times (\sigma^2_E)^{-(a_E+1)}\exp(-b_E/\sigma^2_E)\\
	&  \propto  (\sigma^2_E)^{-TN_E/2 } \exp\Big \{  \sum_{t=1}^{T} -\frac{1}{2} \big(E(t) - X_ES(t)\big)^T (\sigma^2_E H_E) ^{-1} \big(E(t) - X_ES(t)\big)\Big\} \\
	& \times (\sigma^2_E)^{-(a_E+1)}\exp(-b_E/\sigma^2_E)\\
	& \propto (\sigma^2_E)^{-(a_E+\frac{TN_E}{2}+1)} \exp\Big \{  \sum_{t=1}^{T} -\frac{1}{2\sigma^2_E} \big(E(t) - X_ES(t)\big)^T H_E ^{-1} \big(E(t) - X_ES(t)\big)  - \frac{b_E}{\sigma^2_E}\Big\} \\
	& \propto (\sigma^2_E)^{-(a_E+\frac{TN_E}{2}+1)} \exp\Big \{  \frac{-1}{\sigma^2_E}\Big( \sum_{t=1}^{T} \frac{1}{2} \big(E(t) - X_ES(t)\big)^T H_E ^{-1} \big(E(t) - X_ES(t)\big)  +b_E\Big) \Big\} 
	\end{align*}
Therefore, the full conditional for $\sigma^2_E$ is an Inverse-Gamma distribution with
	\begin{align*}
	a^*_E &= a_E+\frac{TN_E}{2}\\
	b^*_E  &=  \sum_{t=1}^{T} \frac{1}{2} \big(E(t) - X_ES(t)\big)^T H_E ^{-1} \big(E(t) - X_ES(t)\big)  +b_E 
	\end{align*}
The full conditional  for $\sigma^2_a$ is obtained as follows:
	\begin{align*}
	P(\sigma^2_a |Rest) &\propto  \prod_{t=2}^{T} \Big[ P(\pmb \mu^A(t) |\pmb \mu^A(t-1), \pmb A, \sigma^2_a ) \Big] \times P(\sigma^2_a) \\
	& \propto \prod_{t=2}^{T} \Big[ |\sigma^2_a \pmb I |^{-1/2}   \exp \big\{  -\frac{1}{2} (\pmb \mu^A(t) - \pmb {A\mu}^A(t-1)) ^T (\sigma^2_a \pmb I) ^{-1}   (\pmb \mu^A(t) - \pmb {A\mu}^A(t-1))   \big\}    \Big]  \\
	&\times (\sigma^2_a)^{-(a_a+1)} \exp(-b_a/\sigma^2_a) \\
	& \propto (\sigma^2_a)^{-\frac{(T-1)(K-1)}{ 2}}\exp \Big\{ \sum_{t=2}^{T}  -\frac{1}{2}   (\pmb \mu^A(t) - \pmb {A\mu}^A(t-1)) ^T (\sigma^2_a \pmb I) ^{-1}   (\pmb \mu^A(t) - \pmb {A\mu}^A(t-1))    \Big\} \\
	&\times (\sigma^2_a)^{-(a_a+1)} \exp(-b_a/\sigma^2_a) \\
	& \propto (\sigma^2_a)^{-\frac{(T-1)(K-1)}{ 2}} \exp \Big\{  \frac{1}{\sigma^2_a}\Big( -\frac{1}{2} \sum_{t=2}^{T}    (\pmb \mu^A(t) - \pmb {A\mu}^A(t-1)) ^T (\pmb \mu^A(t) - \pmb {A\mu}^A(t-1))   \Big) \Big\} \\
	&\times (\sigma^2_a)^{-(a_a+1)} \exp(-b_a/\sigma^2_a) \\
	& \propto (\sigma^2_a)^{-(a_a + {\frac{(T-1)(K-1)}{ 2}}+1)}  \\
	& \times \exp \Big\{  \frac{-1}{\sigma^2_a}\Big( \sum_{t=2}^{T}  \frac{1}{2}   (\pmb \mu^A(t) - \pmb {A\mu}^A(t-1)) ^T \pmb I ^{-1} (\pmb \mu^A(t) - \pmb {A\mu}^A(t-1))  + b_a\Big)  \Big\} 
	\end{align*}
Therefore, the full conditional distribution for $\sigma^2_a$ is still Inverse-Gamma distribution with new parameters as:
	\begin{align*}
	a^*_a &=  a_a + \frac{(T-1)(K-1)}{2} \\
	b^*_a &= \sum_{t=2}^{T}  \frac{1}{2}  (\pmb \mu^A(t) - \pmb {A\mu}^A(t-1)) ^T (\pmb \mu^A(t) - \pmb {A\mu}^A(t-1))  + b_a
	\end{align*}
For the matrix $\pmb A$, which describes the connectivity between states, we will  transform it into a vector via $vec(A)$, we have:
	\begin{align*}
	\pmb {A\mu}^A(t-1) &= vec(\pmb {A\mu}^A(t-1)) \\
	& = \Big(\pmb \mu^A(t-1) ^T\otimes \pmb I_{k-1} \Big) vec(\pmb A)\\
	& = \pmb {Kr_t} \times vec(\pmb A)\\
	& \text{where } \pmb {Kr_t} = \Big(\pmb \mu^A(t-1) ^T\otimes \pmb I_{k-1} \Big)
	\end{align*}
Then the full conditional could be obtained as:
	\begin{align*}
	P(vec(\pmb A) |Rest) &\propto \prod_{i=1}^{k-1}\prod_{j=1}^{k-1}P(A_{ij} |\sigma^2_A) \times \prod_{t=2}^{T} \Big[ P(\pmb \mu^A(t) |\pmb \mu^A(t-1), \pmb A, \sigma^2_a ) \Big] \\
	& \propto  MVN_{(k-1)^2} (vec(\pmb A); \pmb 0 , \sigma^2_A \pmb I_{(k-1)^2}) \\
	& \times \prod_{t=2}^{T} \exp\Bigg(   -\frac{1}{2\sigma^2_a}  \Big(  \pmb \mu^A(t) - \big(\pmb \mu^A(t-1)^T \otimes \pmb I_{k-1} \big) vec(\pmb A)\Big) ^T \\
	& \Big(  \pmb \mu^A(t) - \big(\pmb \mu^A(t-1)^T \otimes \pmb I_{k-1} \big) vec(\pmb A)\Big)  \Bigg)\\
	& \propto MVN_{(k-1)^2} (vec(\pmb A); \pmb 0 , \sigma^2_A \pmb I_{(k-1)^2}) \times \prod_{t=2}^{T} \exp\Bigg(   -\frac{1}{2\sigma^2_a}  \Big(  \pmb \mu^A(t) - \pmb {Kr_t} \times vec(\pmb A)\Big) ^T \\
	&  \Big(  \pmb \mu^A(t) - \pmb {Kr_t} \times vec(\pmb A)\Big)  \Bigg)\\
	& \propto \exp \Bigg(   - \frac{1}{2\sigma^2_A}vec(\pmb A)^Tvec(\pmb A)  + \frac{1}{\sigma^2_a} \bigg( \sum_{t=2}^{T} \pmb \mu^A(t)^T \pmb {Kr_t} \bigg)\times vec(\pmb A)   \\
	& - \frac{1}{2\sigma^2_a} vec(\pmb A) ^T  \bigg(   \sum_{t=2}^{T} \pmb{Kr_t}^T\pmb{Kr_t}\bigg)vec(\pmb A)\Bigg)\\
	& \propto  \exp\bigg( -\frac{1}{2} \big(vec(\pmb A) - \pmb V_1\big)^T \pmb C_1 \big(vec(\pmb A) - \pmb V_1\big)\bigg)
	\end{align*}
Therefore full conditional distribution for $vec(\pmb A)$ is $MVN_{(K-1)^2}(\pmb V_1,  \pmb C^{-1}_1)$, where
	\begin{align*}
	\pmb C_1 & = \frac{1}{\sigma^2_A} \pmb I_{(k-1)^2} + \frac{1}{\sigma^2_a} \bigg( \sum_{t=2}^{T} \pmb{Kr_t}^T\pmb{Kr_t}\bigg)\\
	\pmb V_1^T \pmb C_1 & =  \frac{1}{\sigma^2_a} \bigg( \sum_{t=2}^{T} \pmb \mu^A(t)^T \pmb {Kr_t} \bigg)\ \\
	\pmb V_1 &= \bigg( \frac{1}{\sigma^2_a} \Big( \sum_{t=2}^{T} \pmb \mu^A(t)^T \pmb {Kr_t} \Big) \times \pmb C^{-1}_1\bigg)^T
	\end{align*}

For all the variance components $\alpha_l$, the full conditional could be obtained together as:
	\begin{align*}
	P(\alpha_1 ,\alpha_2, . . . ,\alpha_k| \text{Rest}) & \propto [\prod_{j=1}^{p}\prod_{t=1}^{T}P(S_j(t)) |\pmb \mu(t), \pmb{\alpha, Z_{v(j)}}] \times  \prod_{l= 1}^{k}P(\alpha_l |a_\alpha,b_\alpha) \\
	& \propto [\prod_{j=1}^{p}\prod_{t=1}^{T}\prod_{l =1}^{k}N(S_j(t);\mu_l(t), \alpha_l)^{Z_{v(j)l}}] \times  \prod_{l= 1}^{k}IG(\alpha_l ;a_\alpha,b_\alpha) \\
	& \propto \prod_{l= 1}^{k}\Big[  \prod_{j=1}^{p}\prod_{t=1}^{T}\big(N(S_j(t);\mu_l(t), \alpha_l)^{Z_{v(j)l}}\big)\Big] \times  IG(\alpha_l ; a_\alpha,b_\alpha) \\
		& \propto \prod_{l= 1}^{k}\Bigg[  \prod_{j=1}^{p}\prod_{t=1}^{T}\bigg(\alpha_l^{-\frac{1}{2}} \exp(-\frac{(S_j(t) - \mu_l(t))^2}{2\alpha_l})\bigg) ^{Z_{v(j)l}}\times  (\alpha_l)^{-(a_\alpha+1)} \exp(-b_\alpha/\alpha_l )\Bigg]\\
		& \propto \prod_{l= 1}^{k}\Bigg[  \prod_{j=1}^{p}\prod_{t=1}^{T}\bigg(\alpha_l^{-\frac{1}{2}} \exp(-\frac{(S_j(t) - \mu_l(t))^2}{2\alpha_l})\bigg) ^{Z_{v(j)l}}\times  (\alpha_l)^{-(a_\alpha+1)} \exp(-b_\alpha/\alpha_l )\Bigg]\\
		& \propto \prod_{l= 1}^{k}\Bigg[  \prod_{j=1}^{p}\prod_{t=1}^{T}\bigg(\alpha_l^{-\frac{Z_{v(j)l}}{2}} \exp(-\frac{ Z_{v(j)l} (S_j(t) - \mu_l(t))^2}{2\alpha_l})\bigg)\\
		&\times  (\alpha_l)^{-(a_\alpha+1)} \exp(-b_\alpha/\alpha_l )\Bigg]\\
		& \propto \prod_{l= 1}^{k}\Bigg[  \bigg(\alpha_l^{-\frac{T\sum_{j=1}^{p}Z_{v(j)l}}{2}} \exp(-\frac{ \sum_{j=1}^{p}\sum_{t=1}^{T}Z_{v(j)l} (S_j(t) - \mu_l(t))^2}{2\alpha_l})\bigg)\\
		&\times  (\alpha_l)^{-(a_\alpha+1)} \exp(-b_\alpha/\alpha_l )\Bigg]\\
		& \propto \prod_{l= 1}^{k}\Bigg[  \alpha_l^{-\big(\frac{T\sum\limits_{j=1}^{p}Z_{v(j)l}}{2} +a_\alpha+1\big)} \exp \bigg(- \frac{1}{\alpha_l} \Big(\frac{ \sum\limits_{j=1}^{p}\sum\limits_{t=1}^{T}Z_{v(j)l} (S_j(t) - \mu_l(t))^2}{2 }  + b_\alpha\Big)\bigg) \Bigg]\\
	\end{align*}
Therefore, we can see that for each $\alpha_l$, the individual full conditional is still a Inverse-Gamma distribution with new parameters as:
	\begin{align*}
	a^*_{\alpha_l} & = \frac{T\sum\limits_{j=1}^{p}Z_{v(j)l}}{2} +a_\alpha\\
	b^*_{\alpha_l} & = \frac{ \sum\limits_{j=1}^{p}\sum\limits_{t=1}^{T}Z_{v(j)l} (S_j(t) - \mu_l(t))^2}{2 }  + b_\alpha
	\end{align*}
Also, full conditional for $\pmb\mu (t)$ for when $t = 1$: 
	\begin{align*}
	P(\pmb \mu(1) | \text{Rest}) & \propto \prod_{j=1}^{p}P(S_j(1)) | \pmb \mu^A(1), \pmb{\alpha, Z_{v(j)}}) \times P(\pmb \mu^A(2) |\pmb \mu^A(1), \pmb A, \sigma^2_a )\times P(\pmb \mu^A(1)|\sigma^2_{\mu_1}) \\
	& \propto [\prod_{j=1}^{p}\prod_{l =1}^{k}N(S_j(1);\mu_l(1), \alpha_l)^{Z_{v(j)l}}] \times [MVN_{k-1}(\pmb \mu^A(2);\pmb A\mu^A(1),   \sigma^2_a\pmb I)] \\
	&\times MVN_{k-1}(\pmb \mu^A(1); \pmb 0 , \sigma^2_{\mu_1} \pmb I)\\
	& \propto \bigg[ \prod_{j=1}^{p}\prod_{l=1}^{k}\exp\Big(-\frac{ Z_{v(j)l} (S_j(1) - \mu_l(1))^2}{2\alpha_l}\Big)\bigg] \times \exp\Big(  -\frac{1}{2\sigma^2_a} (\pmb\mu^A(2) - \pmb A \pmb\mu^A(1))^T \\
	&(\pmb\mu^A(2) - \pmb A \pmb\mu^A(1))\Big) \times \exp\Big( - \frac{1}{2\sigma^2_{\mu_1}} (\pmb \mu^A(1))^T (\pmb \mu^A(1))\Big) \\
	& \propto \exp \Bigg(-\frac{1}{2} \sum_{j=1}^{p}\sum_{l =1}^{k} \frac{ Z_{v(j)l}}{\alpha_l} (S_j(1) - \mu_l(1))^2 -\frac{1}{2\sigma^2_a} (\pmb\mu^A(2) - \pmb A \pmb\mu^A(1))^T (\pmb\mu^A(2) - \pmb A \pmb\mu^A(1)) \\
	&- \frac{1}{2\sigma^2_{\mu_1}} (\pmb \mu^A(1))^T (\pmb \mu^A(1))\Bigg) \\
	& \propto \exp \Bigg(  -\frac{1}{2} \sum_{j=1}^{p} \big(  S_j(1)\vec{I}_{k-1}  - \pmb \mu^A(1)\big)^T \text{Diag}(\frac{ Z_{v(j)l}}{\alpha_l}, l = 2, ... ,k)\big(  S_j(1)\vec{I}_{k-1}  - \pmb \mu^A(1)\big) \\
	& -\frac{1}{2\sigma^2_a} (\pmb\mu^A(2) - \pmb A \pmb\mu^A(1))^T (\pmb\mu^A(2) - \pmb A \pmb\mu^A(1)) - \frac{1}{2\sigma^2_{\mu_1}} (\pmb \mu^A(1))^T (\pmb \mu^A(1))\Bigg)\\
	&  \text{where}\;  \vec{I}_{k-1}  \;\text{is a all ones vector with length k-1}. \; \pmb D_j =\text{Diag}(\frac{ Z_{v(j)l}}{\alpha_l}, l = 2, ... ,k)\\
	P(\pmb \mu(1) | \text{Rest})  &\propto \exp\Bigg( \sum_{j=1}^{p}\Big( (S_j(1)\vec{I}_{k-1})^T \pmb D_j \pmb\mu^A(1)  -  \frac{1}{2}(\pmb\mu^A(1))^T \pmb D_j \pmb\mu^A(1) \Big )  +\frac{1}{\sigma^2_a} (\pmb\mu^A(2))^T \pmb A\pmb\mu^A(1)) \\
	& - \frac{1}{2\sigma^2_a} (\pmb\mu^A(1))^T \pmb A^T \pmb A (\pmb\mu^A(1)) - \frac{1}{2\sigma^2_{\mu_1}} (\pmb \mu^A(1))^T (\pmb \mu^A(1))\Bigg)\\
	& \propto  \exp\Bigg( \Big( \sum_{j=1}^{p}(S_j(1)\vec{I}_{k-1})^T \pmb D_j  +\frac{1}{\sigma^2_a} (\pmb\mu^A(2))^T \pmb A\Big)\pmb\mu^A(1)   \\
	& - \frac{1}{2}   \pmb\mu^A(1)^T \big\{   \sum_{j=1}^{p} D_j  + \frac{1}{\sigma^2_a} \pmb{A^TA}  + \frac{1}{\sigma^2_{\mu_1}}\pmb I_{k-1}  \big\} \pmb\mu^A(1)\Bigg)\\
	& \propto \exp\bigg(-\frac{1}{2} \Big(\pmb\mu^A(1) - \pmb M_1\Big)^T \pmb B_1 \Big(\pmb\mu^A(1) - \pmb M_1\Big)\bigg)
	\end{align*}
Therefore, the full conditional for $\pmb\mu(1)$ is multivariate normal distribution $MVN_{k-1}(\pmb M_1, \pmb B^{-1}_1)$with parameters as:	
	\begin{align*}
	\pmb B_1 & = \sum_{j=1}^{p} \pmb D_j  + \frac{1}{\sigma^2_a} \pmb{A^TA}  + \frac{1}{\sigma^2_{\mu_1}}\pmb I_{k-1} \\
	\pmb M_1^T \pmb B_1 &=  \sum_{j=1}^{p}(S_j(1)\vec{I}_{k-1})^T \pmb D_j  +\frac{1}{\sigma^2_a} (\pmb\mu^A(2))^T \pmb A\\
	\pmb M_1 & = \bigg(  \Big(\sum_{j=1}^{p}(S_j(1)\vec{I}_{k-1})^T \pmb D_j  +\frac{1}{\sigma^2_a} (\pmb\mu^A(2))^T \pmb A\Big) \times \pmb B^{-1}_1
\bigg) ^T
	\end{align*}
When $1 < t < T$, the full condition is:
	\begin{align*}
	P(\pmb \mu(t) | \text{Rest}) & \propto \prod_{j=1}^{p}P(S_j(t)) | 0,\pmb \mu^A(t), \pmb{\alpha, Z_{v(j)}}) \times P(\pmb \mu^A(t+1) |\pmb \mu^A(t), \pmb A, \sigma^2_a )\\
	& \times P(\pmb \mu^A(t) |\pmb \mu^A(t-1), \pmb A, \sigma^2_a )\\
	& \propto [\prod_{j=1}^{p}\prod_{l =1}^{k}N(S_j(t);\mu_l(t), \alpha_l)^{Z_{v(j)l}}] \times [MVN_{k-1}(\pmb \mu^A(t+1) ;\pmb A\mu^A(t) ,  \sigma^2_a\pmb I)] \\
	& \times  [MVN_{k-1}(\pmb \mu^A(t) ;\pmb A\mu^A(t-1),  \sigma^2_a\pmb I)] \\
	& \propto \bigg[ \prod_{j=1}^{p}\prod_{l=1}^{k}\exp\Big(-\frac{ Z_{v(j)l} (S_j(t) - \mu_l(t))^2}{2\alpha_l}\Big)\bigg] \times \exp\Big(  -\frac{1}{2\sigma^2_a} (\pmb\mu^A(t+1) - \pmb A \pmb\mu^A(t))^T \\
	& (\pmb\mu^A(t+1) - \pmb A \pmb\mu^A(t))\Big)  \times \exp\Big(  -\frac{1}{2\sigma^2_a} (\pmb\mu^A(t) - \pmb A \pmb\mu^A(t-1))^T  (\pmb\mu^A(t) - \pmb A \pmb\mu^A(t-1))\Big) \\
	& \propto \exp \Bigg(-\frac{1}{2} \sum_{j=1}^{p}\sum_{l =1}^{k} \frac{ Z_{v(j)l}}{\alpha_l} (S_j(t) - \mu_l(t))^2 -\frac{1}{2\sigma^2_a} \big(\pmb\mu^A(t+1) - \pmb A \pmb\mu^A(t)\big)^T \big(\pmb\mu^A(t+1) \\
	& - \pmb A \pmb\mu^A(t)\big)  -\frac{1}{2\sigma^2_a} \big(\pmb\mu^A(t) - \pmb A \pmb\mu^A(t-1)\big)^T \big(\pmb\mu^A(t) - \pmb A \pmb\mu^A(t-1)\big) \Bigg) \\
	& \propto \exp \Bigg(-\frac{1}{2} \sum_{j=1}^{p} \big(  S_j(t)\vec{I}_{k-1}  - \pmb \mu^A(t)\big)^T \text{Diag}(\frac{ Z_{v(j)l}}{\alpha_l}, l = 2, ... ,k)\big(  S_j(t)\vec{I}_{k-1}  - \pmb \mu^A(t)\big)\\
	& - \frac{1}{2\sigma^2_a} \big(\pmb\mu^A(t+1) - \pmb A \pmb\mu^A(t)\big)^T \big(\pmb\mu^A(t+1) - \pmb A \pmb\mu^A(t)\big)  \\
	& - \frac{1}{2\sigma^2_a} \big(\pmb\mu^A(t) - \pmb A \pmb\mu^A(t-1)\big)^T \big(\pmb\mu^A(t) - \pmb A \pmb\mu^A(t-1)\big)\Bigg)\\
		&  \text{where}\;  \vec{I}_{k-1}  \;\text{is a all ones vector with length k-1}. \; \pmb D_j =\text{Diag}(\frac{ Z_{v(j)l}}{\alpha_l}, l = 2, ... ,k)\\
	P(\pmb \mu(t) | \text{Rest}) &\propto \exp\Bigg( \sum_{j=1}^{p}\Big( (S_j(t)\vec{I}_{k-1})^T \pmb D_j \pmb\mu^A(t)  -  \frac{1}{2}(\pmb\mu^A(t))^T \pmb D_j \pmb\mu^A(t) \Big )  +\frac{1}{\sigma^2_a}(\pmb\mu^A(t+1)^T\pmb A )\pmb\mu^A(t)\\
	& - \frac{1}{2\sigma^2_a}\pmb\mu^A(t)^T\pmb A^T \pmb A \pmb\mu^A(t) - \frac{1}{2\sigma^2_a} \pmb\mu^A(t)^T\pmb\mu^A(t) +\frac{1}{\sigma^2_a}(\pmb\mu^A(t-1)^T\pmb A^T )\pmb\mu^A(t)\Bigg)\\
	& \propto \exp\bigg(-\frac{1}{2} \Big(\pmb\mu^A(t) - \pmb M_2\Big)^T \pmb B_2 \Big(\pmb\mu^A(t) - \pmb M_2\Big)\bigg)
	\end{align*}
Then, the full conditional distribution is a $MVN_{k-1}(\pmb M_2, \pmb B_2^{-1})$ as:
	\begin{align*}
	 \pmb B_2& = \sum_{j=1}^{p} \pmb D_j  + \frac{1}{\sigma^2_a} (\pmb{A^TA}  + \pmb I_{k-1})\\
	\pmb M_2^T \pmb B_2 &=  \sum_{j=1}^{p}(S_j(t)\vec{I}_{k-1})^T \pmb D_j +\frac{1}{\sigma^2_a} (\pmb\mu^A(t+1))^T \pmb A + \frac{1}{\sigma^2_a}(\pmb\mu^A(t-1)^T\pmb A^T )\\
	\pmb M_2 & = \Bigg( \bigg( \sum_{j=1}^{p}(S_j(t)\vec{I}_{k-1})^T \pmb D_j +\frac{1}{\sigma^2_a} (\pmb\mu^A(t+1))^T \pmb A + \frac{1}{\sigma^2_a}(\pmb\mu^A(t-1)^T\pmb A^T )\bigg) \times \pmb B^{-1}_2\Bigg)^T
	\end{align*}
When $t = T$, the full conditional is:
	\begin{align*}
	P(\pmb \mu(T) | \text{Rest}) & \propto \prod_{j=1}^{p}P(S_j(T)) |\pmb \mu^A(T), \pmb{\alpha, Z_{v(j)}}) \times P(\pmb \mu^A(T) |\pmb \mu^A(T-1), \pmb A, \sigma^2_a )\\
	& \propto [\prod_{j=1}^{p}\prod_{l =1}^{k}N(S_j(T);\mu_l(T), \alpha_l)^{Z_{v(j)l}}] \times [MVN_{k-1}(\pmb \mu^A(T) ;\pmb A\mu^A(T-1),  \sigma^2_a\pmb I)] \\
	& \propto \bigg[ \prod_{j=1}^{p}\prod_{l=1}^{k}\exp\Big(-\frac{ Z_{v(j)l} (S_j(T) - \mu_l(T))^2}{2\alpha_l}\Big)\bigg] \times \exp\Big(  -\frac{1}{2\sigma^2_a} (\pmb\mu^A(T) - \pmb A \pmb\mu^A(T-1))^T \\
	&(\pmb\mu^A(T) - \pmb A \pmb\mu^A(T-1))\Big) \\
	&\propto \exp \Bigg(-\frac{1}{2} \sum_{j=1}^{p}\sum_{l =1}^{k} \frac{ Z_{v(j)l}}{\alpha_l} (S_j(T) - \mu_l(T))^2 -\frac{1}{2\sigma^2_a} \big(\pmb\mu^A(T) - \pmb A \pmb\mu^A(T-1)\big)^T \\
	& \big(\pmb\mu^A(T) -\pmb A \pmb\mu^A(T-1)\big)  \Bigg)\\
	& \propto \exp \Bigg(-\frac{1}{2} \sum_{j=1}^{p} \big(  S_j(T)\vec{I}_{k-1}  - \pmb \mu^A(T)\big)^T \text{Diag}(\frac{ Z_{v(j)l}}{\alpha_l}, l = 2, ... ,k)\big(  S_j(T)\vec{I}_{k-1}  - \pmb \mu^A(T)\big)\\
	& -\frac{1}{2\sigma^2_a} \big(\pmb\mu^A(T) - \pmb A \pmb\mu^A(T-1)\big)^T \big(\pmb\mu^A(T) -\pmb A \pmb\mu^A(T-1)\big)  \Bigg)\\
	&  \text{where} \; \pmb D_j =\text{Diag}(\frac{ Z_{v(j)l}}{\alpha_l}, l = 2, ... ,k)\\
	P(\pmb \mu(T) | \text{Rest}) &\propto \exp\Bigg( \sum_{j=1}^{p}\Big( (S_j(T)\vec{I}_{k-1})^T \pmb D_j \pmb\mu^A(T)  -  \frac{1}{2}(\pmb\mu^A(T))^T \pmb D_j \pmb\mu^A(T) \Big )  \\
	&+\frac{1}{\sigma^2_a}(\pmb\mu^A(T-1)^T\pmb A^T )\pmb\mu^A(T)
	 - \frac{1}{2\sigma^2_a} \pmb\mu^A(T)^T\pmb\mu^A(T) \Bigg)\\
	& \propto \exp\bigg(-\frac{1}{2} \Big(\pmb\mu^A(T) - \pmb M_3\Big)^T \pmb B_3\Big(\pmb\mu^A(T) - \pmb M_3\Big)\bigg)
	\end{align*}

Therefore, full condition distribution when $t =T$ is a $MVN_{K-1}(\pmb M_3, \pmb B_3^{-1})$ with:
	\begin{align*}
	\pmb B_3 & = \sum_{j=1}^{p} \pmb D_j  + \frac{1}{\sigma^2_a}  \pmb I_{k-1}\\
	\pmb M_3^T \pmb B_3 &=  \sum_{j=1}^{p}(S_j(T)\vec{I}_{k-1})^T \pmb D_j +\frac{1}{\sigma^2_a}(\pmb\mu^A(T-1)^T\pmb A^T )\\
	\pmb M_3 &=\Bigg( \bigg( \sum_{j=1}^{p}(S_j(T)\vec{I}_{k-1})^T \pmb D_j  + \frac{1}{\sigma^2_a}(\pmb\mu^A(T-1)^T\pmb A^T )\bigg) \times \pmb B^{-1}_3\Bigg)^T
	\end{align*}

Regarding to $S_j(t)$ for $t= 1,2,..., T$, the full conditional distribution could be obtained as:
	\begin{align*}
	P(S_j(t)| \text{Rest})&\propto\prod_{t=1}^{T}[P(E(t) | S(t), \sigma^2_E)P(M(t) | S(t), \sigma^2_M)] \times [\prod_{t=1}^{T}P(S_j(t)) |\pmb \mu^A(t), \pmb{\alpha, Z_{v(j)}}] \\
	& \propto \prod_{t=1}^{T}\Bigg[\exp\Big \{  -\frac{1}{2} \big(M(t) - X_M S(t)\big)^T(\sigma^2_M H_M) ^{-1} \big(M(t) - X_MS(t)\big)\Big\} \\
	& \times  \exp\Big \{  -\frac{1}{2} \big(E(t) - X_E S(t)\big)^T(\sigma^2_E H_E) ^{-1} \big(E(t) - X_ES(t)\big)\Big\} \\
	& \times \prod_{l=1}^{k}\exp\Big(-\frac{ Z_{v(j)l} (S_j(t) - \mu_l(t))^2}{2\alpha_l}\Big)\Bigg] \\
	& \propto \prod_{t=1}^{T} \exp\Bigg[   -\frac{1}{2} \big(M(t) - X_M S(t)\big)^T(\sigma^2_M H_M) ^{-1} \big(M(t) - X_MS(t)\big) \\
	& -\frac{1}{2} \big(E(t) - X_E S(t)\big)^T(\sigma^2_E H_E) ^{-1} \big(E(t) - X_ES(t)\big)  -\frac{1}{2} \sum_{l=1}^{k} \frac{ Z_{v(j)l} (S_j(t) - \mu_l(t))^2}{\alpha_l}\Bigg]\\
	& \propto \prod_{t=1}^{T} \exp\Bigg[  -\frac{1}{2\sigma^2_M} \big(  M(t)^T  H_M^{-1} M(t)  - 2M(t)^T H_M^{-1}X_M S(t) + (X_MS(t))^TH_M^{-1}X_M S(t) \big)   \\
	&-\frac{1}{2\sigma^2_E} \big(  E(t)^T  H_E^{-1} E(t)  - 2E(t)^T H_E^{-1}X_E S(t) + (X_ES(t))^TH_E^{-1}X_ES(t) \big) \\
	& -\frac{1}{2} \sum_{l=1}^{k} \frac{ Z_{v(j)l} (S_j(t)^2 - 2\mu_l(t)S_j(t) + \mu_l(t)^2)}{\alpha_l} \Bigg]	\\
 \text{Let}\; X_M[,v] \; &\text {denote the}\; v\text{th}\; \text{column in the matrix}\; \text{Then, we can rewrite} \; X_MS(t) \; \text{as:} \\
  X_MS(t) &= \sum_{v=1}^{p}X_M[,v] S_v(t). \;\text{Then:}\\
	P(S_j(t)| \text{Rest}) & \propto \prod_{t=1}^{T} \exp\Bigg[  -\frac{1}{2\sigma^2_M} \Big(    - 2M(t)^T H_M^{-1} (\sum_{v=1}^{p}X_M[,v] S_v(t)) + \\
	& (\sum_{v=1}^{p}X_M[,v] S_v(t))^TH_M^{-1}(\sum_{v=1}^{p}X_M[,v] S_v(t)) \Big)  - \frac{1}{2\sigma^2_E} \Big( -2E(t)^T H_E^{-1} (\sum_{v=1}^{p}X_E[,v] S_v(t))  \\
	& + ((\sum_{v=1}^{p}X_E[,v] S_v(t)))^TH_E^{-1}(\sum_{v=1}^{p}X_E[,v] S_v(t)) \Big) -\frac{1}{2} \sum_{l=1}^{k} \frac{ Z_{v(j)l} (S_j(t)^2 - 2\mu_l(t)S_j(t) )}{\alpha_l} \Bigg]
		\end{align*}
We want to keep the terms that have $ S_j(t)$:
	\begin{align*}
	P(S_j(t)| \text{Rest}) &\propto \prod_{t=1}^{T} \exp\Bigg[  -\frac{1}{2\sigma^2_M} \Big( - 2M(t)^T H_M^{-1} X_M[,j] S_j(t))  \\ 
	& +(\sum_{v\neq j}^{}X_M[,v] S_v(t))^TH_M^{-1}(X_M[,j] S_j(t))  + (X_M[,j] S_j(t)) ^TH_M^{-1} (\sum_{v\neq j}^{}X_M[,v] S_v(t)) \\
	& + (X_M[,j] S_j(t)) ^TH_M^{-1}(X_M[,j] S_j(t))  \Big)   -\frac{1}{2\sigma^2_E} \Big( - 2E(t)^T H_E^{-1} X_E[,j] S_j(t))  \\ 
	& +(\sum_{v\neq j}^{}X_E[,v] S_v(t))^TH_E^{-1}(X_E[,j] S_j(t))  + (X_E[,j] S_j(t)) ^TH_E^{-1} (\sum_{v\neq j}^{}X_E[,v] S_v(t)) \\
	& + (X_E[,j] S_j(t)) ^TH_E^{-1}(X_E[,j] S_j(t))  \Big) -\frac{1}{2} \sum_{l=1}^{k} \frac{ Z_{v(j)l} (S_j(t)^2 - 2\mu_l(t)S_j(t) )}{\alpha_l}\Bigg] \\
	& \propto \prod_{t=1}^{T} \exp\Bigg[  -\frac{1}{2\sigma^2_M} \Big( - 2M(t)^T H_M^{-1} X_M[,j] S_j(t)  \\
	&  + 2(\sum_{v\neq j}^{}X_M[,v] S_v(t))^TH_M^{-1}(X_M[,j] S_j(t)) + (X_M[,j] S_j(t)) ^TH_M^{-1}(X_M[,j] S_j(t)) \Big)  \\
	&-\frac{1}{2\sigma^2_E} \Big( - 2E(t)^T H_E^{-1} X_E[,j] S_j(t)    + 2(\sum_{v\neq j}^{}X_E[,v] S_v(t))^TH_E^{-1}(X_E[,j] S_j(t)) \\
	&+ (X_E[,j] S_j(t)) ^TH_E^{-1}(X_E[,j] S_j(t)) \Big)-\frac{1}{2} \sum_{l=1}^{k} \frac{ Z_{v(j)l} (S_j(t)^2 - 2\mu_l(t)S_j(t) )}{\alpha_l}\Bigg]\\
	& \propto  \exp\sum_{t=1}^{T}\Bigg[  -\frac{1}{2\sigma^2_M} \Big( S_j(t)^2 \big( X_M[,j]  ^TH_M^{-1}X_M[,j]  \big)  +S_j(t) \big( - 2M(t)^T H_M^{-1} X_M[,j]   \\
	&  + 2(\sum_{v\neq j}^{}X_M[,v] S_v(t))^TH_M^{-1}X_M[,j] \big)  \Big)    -\frac{1}{2\sigma^2_E} \Big( S_j(t)^2 \big( X_E[,j]  ^TH_E^{-1}X_E[,j]  \big)  \\
	&+ S_j(t) \big( - 2E(t)^T H_E^{-1} X_E[,j]  + 2(\sum_{v\neq j}^{}X_E[,v] S_v(t))^TH_E^{-1}X_E[,j] \big)  \Big)\\
	& -\frac{1}{2} \sum_{l=1}^{k} \frac{ Z_{v(j)l} (S_j(t)^2 - 2\mu_l(t)S_j(t) )}{\alpha_l}\Bigg]\\
	&  \propto  \exp -\frac{1}{2}\sum_{t=1}^{T} \Bigg\{S_j(t)^2 \bigg[\frac{1}{\sigma^2_M} \Big( X_M[,j]  ^TH_M^{-1}X_M[,j] \big)+\frac{1}{\sigma^2_E} \big( X_E[,j]  ^TH_E^{-1}X_E[,j]  \Big) \\
	&+ \sum_{l=1}^{k}\frac{Z_{v(j)l}}{\alpha_l}\bigg]  +  S_j(t)\bigg[ \frac{1}{\sigma^2_M} \Big(  - 2M(t)^T H_M^{-1} X_M[,j] + 2(\sum_{v\neq j}^{}X_M[,v] S_v(t))^TH_M^{-1}X_M[,j]   \Big)  \\
	& + \frac{1}{\sigma^2_E} \Big(  - 2E(t)^T H_E^{-1} X_E[,j] + 2(\sum_{v\neq j}^{}X_E[,v] S_v(t))^TH_E^{-1}X_E[,j] \Big)  - 2 \sum_{l=1}^{k}\frac{\mu_l(t)}{\alpha_l}\bigg] \Bigg\}\\
	& \propto  \exp -\frac{1}{2}\sum_{t=1}^{T} \Bigg\{ S_j(t)^2 W_{1j} + S_j(t)W_{1j}\Bigg\}
	\end{align*}
Where:
	\begin{align*}
	W_{1j} &= \frac{1}{\sigma^2_M} \Big( X_M[,j]  ^TH_M^{-1}X_M[,j] \Big)+\frac{1}{\sigma^2_E} \Big( X_E[,j]  ^TH_E^{-1}X_E[,j]  \Big) + \sum_{l=1}^{k}\frac{Z_{v(j)l}}{\alpha_l}\\
	W_{2j}(t)&= \frac{1}{\sigma^2_M} \Big(  - 2M(t)^T H_M^{-1} X_M[,j] + 2(\sum_{v\neq j}^{}X_M[,v] S_v(t))^TH_M^{-1}X_M[,j]   \Big)\\
	& + \frac{1}{\sigma^2_E} \Big(  - 2E(t)^T H_E^{-1} X_E[,j] + 2(\sum_{v\neq j}^{}X_E[,v] S_v(t))^TH_E^{-1}X_E[,j] \Big)  - 2 \sum_{l=1}^{k}\frac{\mu_l(t)}{\alpha_l}
	\end{align*}	
Since we are interested in the full conditional distribution for $S_j(t)$ over all $t =1,2, . . ., T$, Then we can write as follows:
	\begin{align*}
	P(\pmb S_j |\text{Rest}) & \propto \exp\Big\{-\frac{1}{2} (\pmb S_j  - \pmb \mu_{S_j})^T \pmb \Sigma_{S_j}^{-1}(\pmb S_j  - \pmb \mu_{S_j})\Big\}
	\end{align*}
The precision matrix   $\pmb \Sigma^{-1}_{S_j} \text{ is a} \; T\times T \;\text{matrix where:}$
	\begin{align*}
	\pmb S_j^T \pmb \Sigma^{-1}_{S_j} \pmb S_j & = \sum_{t=1}^{T} S_j(t)^2W_{1j}\\
	\Rightarrow \pmb \Sigma^{-1}_{S_j}&  = Diag(W_{1j})\\
	 \text{Also:} \; \; -2 \pmb \mu_{S_j}^T \pmb \Sigma^{-1}_{S_j} \pmb S_j(t) & = \sum_{t=1}^{T} S_j(t)W_{2j}(t)\\
	 \Rightarrow -2\pmb \mu_{S_j}^T \pmb \Sigma^{-1}_{S_j}  \pmb S_j(t)&  = \pmb W_{2j} ^T \pmb S_j(t) \\
	 \text{where:}\;\; \pmb W_{2j}^T &=  (W_{2j}(1), W_{2j}(2), . . . , W_{2j}(T))\\
	 	 -2\pmb \mu_{S_j}^T \pmb \Sigma^{-1}_{S_j} & = \pmb W_{2j} ^T\\
	 \pmb \mu_{S_j} & =  \big(-\frac{1}{2}\pmb W_2 ^T\pmb \Sigma_{S_j} \big)^T\\
	 \pmb \mu_{S_j} & = -\frac{1}{2}\pmb \Sigma_{S_j} \pmb W_{2j}
	\end{align*}
	
	\begin{align*}	
	&\text{For}\; t=1,...., T\\
	& \pmb \Sigma^{-1}_{S_j}[t,t] = \frac{1}{\sigma^2_M} \Big( X_M[,j]  ^TH_M^{-1}X_M[,j] \Big)+\frac{1}{\sigma^2_E} \Big( X_E[,j]  ^TH_E^{-1}X_E[,j]  \Big) + \sum_{l=1}^{k}\frac{Z_{v(j)l}}{\alpha_l}\\
	&\text{For}\; t=1,2,..., T, \;\text{for}\; w=1,2,..., T, \text{and}\; t \neq w\\
	&  \pmb \Sigma^{-1}_{S_j}[t, w]  = \frac{1}{\sigma^2_M}(\sum_{v\neq j}^{}X_M[,v] S_v(t))^TH_M^{-1}X_M[,j]+ \frac{1}{\sigma^2_E}(\sum_{v\neq j}^{}X_E[,v] S_v(t))^TH_E^{-1}X_E[,j] 		
	\end{align*}
	

Now, for $\beta$, which is the spatial cohesion parameter for Potts model, the full condition distribution is :
	\begin{align*}
	 P(\beta| \text{Rest}) &\propto P(\pmb Z|\beta) \times P(\beta)\\
	 &\propto  \frac{\exp\{ \beta \sum_{h\sim j} \delta(\pmb{Z_v, Z_h})\} } {G(\beta)} \times \frac{1}{\beta_u}\\
	 & \text{where}\;\delta(\pmb{Z_v, Z_h})  = 2 \pmb{Z_v^{'}Z_h} - 1\\
	 & \text{or the approximation based on pseudolikelihood:}\\
	 & \propto P_{PL}(\pmb Z | \beta) \times P(\beta)
	\end{align*}
Noticing that the parameterization we used for Potts model, which is the same parameterization as McGrory et al (2009), is:
	\begin{align*}
	\delta(\pmb{Z_v, Z_h})  = 2 \pmb{Z_v^{'}Z_h} - 1 =\left\{
  \begin{array}{@{}ll@{}}
    -1, &  \pmb{Z_v, Z_h} \; \text{ are not in the same state}  \\
    1, & \pmb{Z_v, Z_h} \; \text{ are  in the same state}
  \end{array}\right.
	\end{align*}
Let $N_{np}$ denote the total number of neighbours,  $N_{ss}$ denote the total number of neighbours share the same state and $N_{ns}$ is the total number of neighbours that do not share the same state. Then we can rewrite the Potts model as:
	\begin{align*}
	P(\pmb{Z}|\beta_1)  &= \frac{\exp\{ \beta_1\sum_{h\sim j} \delta(\pmb{Z_v, Z_h})\} } {G(\beta_1)}	\\
	 &=\frac{1}{G(\beta_1)} \exp\big(\beta_1( N_{ss} - N_{ns})\big)\\
	 & =\frac{1}{G(\beta_1)} \exp\big(\beta_1( N_{ss} - (N_{np} - N_{ss}))\big)\\
	 & =\frac{1}{G(\beta_1)} \exp\big(2\beta_1 N_{ss} - \beta_1 N_{np}\big)\\
	  & =\frac{\exp(- \beta_1 N_{np})}{G(\beta_1)} \exp\big(2\beta_1 N_{ss} \big)
	\end{align*}
However, the parameterization used in Moores et al (2009) is that, $\sum_{h\sim j} \delta(\pmb{Z_v, Z_h})$ counts  the neighbours that share the same states. which could be expressed as:
	\begin{align*}
	P(\pmb{Z}|\beta_2)  &= \frac{\exp\{ \beta_2\sum_{h\sim j} \delta(\pmb{Z_v, Z_h})\} } {C(\beta_2)}	\\
	P(\pmb{Z}|\beta_2)  &= \frac{1 } {C(\beta_2)}\exp(\beta_2 N_{ss})
	\end{align*}
The reparameterization could be:
	\begin{align*}
	 2\beta_1 &= \beta_2\\
	 \frac{\exp(- \beta_1 N_{np})}{G(\beta_1)}  & = \frac{1 } {C(\beta_2)}\\
	 \Rightarrow C(\beta_2) \exp(- \beta_1 N_{np}) & = G(\beta_1)
	\end{align*}
The Pseudolikelihood for $\pmb Z$ is defined as:
	\begin{align*}
	P_{PL}(\pmb Z |\beta) & = \prod_{i=1}^{N_v}P(\pmb Z_i|\pmb Z_{-i}, \beta)\\
		& = \prod_{i=1}^{N_v}\frac{\exp \big( 2\beta \sum_{j=1}^{k}Z_{ij}\sum_{l\in\delta_i}Z_{lj}{}\big)}{\sum_{q=1}^{k} \exp (2\beta \sum_{l \in \delta_i}{}Z_{lq})}
	\end{align*}
For ICM update, and assuming a pseudolikelihood approximation, we want the value $\hat \beta$ that maximizes the following function over $[0, \beta_u]$
	\begin{align*}
 f(\beta) = P_{PL}(Z|\beta)P(\beta). 
	\end{align*}
For $f(\beta )$, 
	\begin{align*}
	f(\beta) & = \prod_{i=1}^{N_v}P(\pmb Z_i|\pmb Z_{-i}, \beta)\times P(\beta)\\
		& = \prod_{i=1}^{N_v}\frac{\exp \big( 2\beta \sum_{j=1}^{k}Z_{ij}\sum_{l\in\delta_i}Z_{lj}{}\big)}{\sum_{q=1}^{k} \exp (2\beta \sum_{l \in \delta_i}{}Z_{lq})} \times \pmb I (0 \leq \beta \leq \beta_u)\\
	 H(\beta) =  \log(f(\beta)) & = \log(I (0 \leq \beta \leq \beta_u)) +  \sum_{v=1}^{N_v}\Big[2\beta \sum_{j=1}^{k}Z_{ij}\sum_{l \in \delta_i}{}Z_{lj} -\log\big\{ \sum_{q=1}^{k} \exp(2\beta \sum_{l \in \delta_i}{}Z_{lq})\big\} \Big]\\
	& =  2\beta \sum_{i=1}^{N_v}\sum_{j=1}^{k}Z_{ij}\sum_{l \in \delta_i}{}Z_{lj} - \sum_{i=1}^{N_v}\log\{ \sum_{q=1}^{k} \exp(2\beta\sum_{l\in \delta_i}{} Z_{lq})\}
	\end{align*}
Taking the derivative	of $H(\beta)$ and assuming $\beta$ is in $[0, \beta_u]$, we have:
	\begin{align*}
	H'(\beta) & =  2 \sum_{i=1}^{N_v}\sum_{j=1}^{k}Z_{ij}\sum_{l \in \delta_i}{}Z_{lj}  -  \sum_{i=1}^{N_v} \{ \sum_{q=1}^{k} \exp(2\beta\sum_{l\in \delta_i}{} Z_{lq})\}^{-1} (\sum_{q=1}^{k}\exp(2\beta \sum_{l\in \delta_i}{} Z_{lq}))(2\sum_{l\in \delta_i}{}Z_{lq})
	\end{align*}
We will use the forms of $H(\beta)$ and $H'(\beta)$ to numerically maximize $H(\beta)$ at each ICM iteration. This constrained 1-dimensional maximization is easily carried out using numerical routines in the R programming language.

The full conditional distribution of $\pmb Z$ is:
	\begin{align*}
	P(\pmb Z |Rest ) & \propto [\prod_{j=1}^{p}\prod_{t=1}^{T}P(S_j(t)) | \pmb \mu^A(t), \pmb{\alpha, Z_{v(j)}}] \times P(\pmb Z| \beta) 
	\end{align*}
For now, let's focus on the label for $r$th voxel, which is $\pmb Z_r$.  Let $N_{jr}$ denote the number of points such that being mapped into $r$th voxel as $j | v(j) = r$. Then:
	\begin{align*}
	 P(\pmb Z_r |Rest) & \propto  \Big[ \prod_{j|v(j) = r}{} \prod_{l=1}^{k} \alpha_l^{-T Z_{v(j)l}/2} \exp \big( -\sum_{t=1}^{T}\frac{Z_{v(j)l}(S_j(t) - \mu_l(t))^2}{2\alpha_l}\big)\Big]  \times P_{PL}(\pmb Z_r|\beta)\\
	 & \propto \Big[ \prod_{j|v(j) = r}{} \prod_{l=1}^{k} \alpha_l^{-T Z_{v(j)l}/2} \exp \big( -\sum_{t=1}^{T}\frac{Z_{v(j)l}(S_j(t) - \mu_l(t))^2}{2\alpha_l}\big)\Big]  \\
	 & \times P(\pmb Z_r|Z_{\delta_r},\beta)\prod_{i=\delta_r}{}P(Z_i |Z_{\delta_i},\beta)\\
	 & \propto \Big[ \prod_{j|v(j) = r}{} \prod_{l=1}^{k} \alpha_l^{-T Z_{v(j)l}/2} \exp \big( -\sum_{t=1}^{T}\frac{Z_{v(j)l}(S_j(t) - \mu_l(t))^2}{2\alpha_l}\big)\Big] \times \exp(2\beta\sum_{j=1}^{k}\sum_{l\in \delta_r}{}Z_{rj}Z_{lj})\\
	 & \times  \prod_{i \in \delta_r}\frac{\exp(2\beta\sum_{j=1}^{k} Z_{ij}Z_{rj})}{\sum_{q=1}^{k} \exp (2\beta \sum_{l \in \delta_i}{}Z_{lq})}\\
	&  \propto \Big[ \prod_{j|v(j) = r}{} \prod_{l=1}^{k} \alpha_l^{-T Z_{v(j)l}/2} \exp \big( -\sum_{t=1}^{T}\frac{Z_{v(j)l}(S_j(t) - \mu_l(t))^2}{2\alpha_l}\big)\Big] \times \exp(2\beta\sum_{j=1}^{k}\sum_{l\in \delta_r}{}Z_{rj}Z_{lj})\\
	& \times \frac{\exp(2\beta\sum_{i \in \delta_i}{}\sum_{j=1}^{k} Z_{ij}Z_{rj})}{ \prod _{i \in \delta_i}(\sum_{q=1}^{k} \exp (2\beta \sum_{l \in \delta_i}{}Z_{lq}))}\\
	& \propto   \Big[ \prod_{j|v(j) = r}{} \prod_{l=1}^{k} \alpha_l^{-T Z_{v(j)l}/2} \exp \big( -\sum_{t=1}^{T}\frac{Z_{v(j)l}(S_j(t) - \mu_l(t))^2}{2\alpha_l}\big)\Big] \\
	& \times \frac{\exp\big(2\beta\sum_{j=1}^{k}\sum_{l\in \delta_r}{}Z_{rj}Z_{lj} + 2\beta\sum_{i \in \delta_i}{}\sum_{j=1}^{k} Z_{ij}Z_{rj}) + \big)}{\prod _{i \in \delta_i}(\sum_{q=1}^{k} \exp (2\beta \sum_{l \in \delta_i}{}Z_{lq}))}\\
	& \propto  \Big[ \prod_{j|v(j) = r}{} \prod_{l=1}^{k} \alpha_l^{-T Z_{v(j)l}/2} \exp \big( -\sum_{t=1}^{T}\frac{Z_{v(j)l}(S_j(t) - \mu_l(t))^2}{2\alpha_l}\big)\Big] \\
	& \times  \frac{\exp\big(4\beta\sum_{j=1}^{k}\sum_{l\in \delta_r}{}Z_{rj}Z_{lj} \big)}{\prod _{i \in \delta_r}(\sum_{q=1}^{k} \exp (2\beta \sum_{l \in \delta_i}{}Z_{lq}))}\\
	& \propto \Big[  \prod_{l=1}^{k} \alpha^{-T N_{jr}Z_{rl}/2}  \Big] \times \exp\Big[ -\sum_{l=1}^{k}Z_{rl}\sum_{t=1}^{T}\sum_{j|v(j) =r}{}\frac{(S_j(t) - \mu_l(t))^2}{2\alpha_l} + 4\beta\sum_{j=1}^{k}\sum_{l\in \delta_r}{}Z_{rj}Z_{lj}\Big] \\
	& \times  \frac{1}{\prod _{i \in \delta_r}(\sum_{q=1}^{k} \exp (2\beta \sum_{l \in \delta_i}{}Z_{lq}))}
	\end{align*}

When $Z_{rh} = 1$, We could find that:
	\begin{align*}
	 P(\pmb Z_{rh} = 1 |Rest) & \propto\alpha^{-T N_{jr}/2} \times \exp \Big[- \sum_{t=1}^{T}\sum_{j|v(j) =r}{}\frac{(S_j(t) - \mu_l(t))^2}{2\alpha_l} + 4\beta\sum_{l\in \delta_r}{}Z_{rj}Z_{lj}\Big]\\
	 & \times \frac{1}{ \prod _{i \in \delta_r}\sum_{q=1}^{k} \exp \big(2\beta (\pmb I(q = h) + \sum_{l \in \delta_i, l \neq r}{}Z_{lq})\big)}
	\end{align*}
Then, the probability of $Z_{rh} = 1$ give the rest could be computed as:
	\begin{align*}
	P(Z_{rh} = 1 | Rest) = \frac { \frac{\alpha_h^{-T N_{jr}/2} \times \exp \Big[ -\sum_{t=1}^{T}\sum_{j|v(j) =r}{}\frac{(S_j(t) - \mu_h(t))^2}{2\alpha_h} + 4\beta\sum_{l\in \delta_r}{}Z_{rh}Z_{lh}\Big]}{ \prod _{i \in \delta_r}\sum_{q=1}^{k} \exp \big(2\beta (\pmb I(q = h) + \sum_{l \in \delta_i, l \neq r}{}Z_{lq})\big)}}{\sum_{h=1}^{k}\frac{\alpha_h^{-T N_{jr}/2} \times \exp \Big[ -\sum_{t=1}^{T}\sum_{j|v(j) =r}{}\frac{(S_j(t) - \mu_h(t))^2}{2\alpha_h} + 4\beta\sum_{l\in \delta_r}{}Z_{rh}Z_{lh}\Big]}{ \prod _{i \in \delta_r}\sum_{q=1}^{k} \exp \big(2\beta (\pmb I(q = h) + \sum_{l \in \delta_i, l \neq r}{}Z_{lq})\big)}}
	\end{align*}
For block updating, let's define the block setting as:
	\begin{align*}
	B &=  \{\text{Black Square Index}\}\\
	W &=  \{\text{White Square Index}\}
	\end{align*}
Then, the labelling for  white square or black square could be expressed as:
	\begin{align*}
	\pmb Z_B &= \{Z_i | \;i \in B\}\\
	\pmb  Z_W &= \{Z_i | \; i \in W\}
	\end{align*}
We know that (we are using Moores et al.):
	\begin{align*}
	P(\pmb  Z_B | \pmb  Z_W, \beta) & \propto P(\pmb Z|\beta_2)\\
	  & \propto \prod _{l\in \pmb B}{} f_l(\pmb Z_l|\pmb Z_W, \beta_2)\\
	  & \propto \prod_{l \in B}{} \prod_{j \in \delta_l}{} \exp\big(\beta_2\times \delta (\pmb Z_l , \pmb Z_j)\big)\\
	  & \propto \prod_{l \in B}{} \exp\big(\beta_2 \sum_{j \in \delta_l }{} \delta(\pmb Z_l, \pmb Z_j)\big)\\
	  & \propto \prod_{l\in B}{}\exp\big(2\beta \pmb Z_l^{T} \sum_{j \in \delta_l }{} \pmb Z_j)\big)
	\end{align*}
By symmetry, we could also get the white square as:
	\begin{align*}
	P(\pmb  Z_W | \pmb  Z_B, \beta)	  & \propto \prod_{l\in W}{}\exp\big(2\beta \pmb Z_l^{T} \sum_{j \in \delta_l }{} \pmb Z_j)\big)
	\end{align*}
With normalizing constant, we can get $P(\pmb  Z_B | \pmb  Z_W, \beta) $ as:
	\begin{align*}
	P(\pmb Z_B | \pmb Z_W, \beta)  = & \prod_{l\in B}{} \frac{\exp\big(2\beta \pmb  Z_l^{T} \sum_{j \in \delta_l }{} \pmb Z_j)\big)}{\sum_{q=1 }{k}\exp\big(2\beta \pmb e_q^{T} \sum_{j \in \delta_l }{}\pmb  Z_j)\big)}
	\end{align*}
where $\pmb e_q ^{T} = (0, 0, 0 , 0, 0, 0, 1, 0, 0, ....)$ is a coordinate vector that $q$th location is 1 and 0 otherwise. Now, the full conditional distribution for $Z_B$ is:
	\begin{align*}
	P(\pmb Z_B|Rest) & \propto  \Big[\prod_{j=1}^{p}\prod_{t=1}^{T}P(S_j(t)) | \pmb \mu^A(t), \pmb{\alpha, Z_{v(j)}}\Big] \times P(\pmb Z| \beta) \\
	&  \propto  \Big[\prod_{j=1}^{p}\prod_{t=1}^{T}P(S_j(t)) | \pmb \mu^A(t), \pmb{\alpha, Z_{v(j)}}\Big] \times P(\pmb Z_B| \pmb Z_W, \beta)\\
	& \propto  \Big[  \prod_{l \in B}{}\prod_{j|v(j) = l}{}\prod_{t=1}^{T}P(S_j(t)) | \pmb \mu^A(t), \pmb{\alpha, Z_{v(j)}}\Big] \times P(\pmb Z_B| \pmb Z_W, \beta)\\
	& \propto \prod_{l \in B}{} \Big( \big[ \prod_{j|v(j) = l}{} \prod_{t=1}^{T}P(S_j(t) | \pmb \mu^A(t), \pmb{\alpha, Z_{v(j)}})\big] \times \exp(2\beta \pmb Z_l^{T} \sum_{j \in \delta_l }{} \pmb Z_j) \Big)\\
	& \propto  \prod_{l \in B}{} \Big(  \big[ \prod_{j|v(j) = l}{} \prod_{t=1}^{T} \prod_{\kappa=1}^{k} \alpha_{\kappa} ^{-Z_{v(j)\kappa}/ 2} \exp(-\frac{Z_{v(j)\kappa}(S_j(t) - \mu_{\kappa}(t))^2}{2\alpha_{\kappa}})\big]\times\exp(2\beta \pmb Z_l^{T} \sum_{j \in \delta_l }{} \pmb Z_j)  \Big)\\
	& \propto \prod_{l \in B}{} \big[ \prod_{\kappa=1}^{k}  \prod_{j|v(j) = l}{}  \alpha_{\kappa}^{-TZ_v(j)\kappa / 2} \times \exp(- \sum_{t=1}^{T}\frac{Z_{v(j)\kappa}(S_j(t) - \mu_{\kappa}(t))^2}{2\alpha_{\kappa}})\big] \times\exp(2\beta \pmb Z_l^{T} \sum_{j \in \delta_l }{} \pmb Z_j)\\
	& \propto \prod_{l \in B}{}  [\prod_{\kappa=1}^{k} \alpha_{\kappa} ^{-TZ_{lk}}]^{N_{jr}/2}\times \exp \Big(-\frac{1}{2} \sum_{j|v{j} = l}{}\sum_{\kappa=1}^{k}\frac{Z_{l\kappa}}{\alpha_{\kappa}} \sum_{t=1}^{T}(S_j(t) - \mu_{\kappa}(t))^2 + 2\beta \sum_{\kappa=1}^{k} \sum_{j\in\delta_l}{}Z_{lk}Z_{jk}\Big)
	\end{align*}
Therefore, we can have that ($\kappa$ is the voxel and $h$ is the mixture component):
	\begin{align*}
	P(Z_{\kappa h} = 1| Rest) = \frac{ \alpha_{h}^{-TN_{j\kappa}/2 } \times \exp \big(  -\frac{1}{2} \sum_{j|v(j) = \kappa}{} \alpha_h^{-1}\sum_{t=1}^{T} ( S_j(t) - \mu_{h}(t))^2 +  2\beta\sum_{v \in \delta_\kappa}{}Z_{vh}\big) }{	\sum_{ l=1}^{K}\alpha_{l}^{-TN_{j\kappa}/2 } \times \exp \big(  -\frac{1}{2} \sum_{j|v(j) = \kappa}{} \alpha_l^{-1}\sum_{t=1}^{T} ( S_j(t) - \mu_{l}(t))^2 +  2\beta\sum_{v \in \delta_\kappa}{}Z_{vl}\big) 
}
	\end{align*}
	
		\begin{align*}
	\sum_{ l=1}^{K}\alpha_{l}^{-TN_{j\kappa}/2 } \times \exp \big(  -\frac{1}{2} \sum_{j|v(j) = \kappa}{} \alpha_l^{-1}\sum_{t=1}^{T} ( S_j(t) - \mu_{l}(t))^2 +  2\beta\sum_{v \in \delta_\kappa}{}Z_{vl}\big) 
	\end{align*}
		
The full conditional for $\pmb Z_B$ is:	
	\begin{equation*}
	P(\pmb Z_B |Rest)  =  \prod_{l\in B}{} \Bigg[ \frac{ \splitfrac{[\prod_{\kappa=1}^{k} \alpha_{\kappa} ^{-TZ_{lk}}]^{N_{jr}} \times \exp \Big(-\frac{1}{2} \sum_{j|v{j} = l}{}\sum_{\kappa=1}^{k}\frac{Z_{l\kappa}}{\alpha_{\kappa}} \sum_{t=1}^{T}(S_j(t) - \mu_{\kappa}(t))^2} {+ 2\beta \sum_{\kappa=1}^{k} \sum_{j\in\delta_l}{}Z_{lk}Z_{jk}\Big)
}} {\sum_{h=1}^{k} \alpha_{h}^{-TN_{jl}/2 } \times \exp \big(  -\frac{1}{2} \sum_{j|v(j) = l}{} \alpha_h^{-1}\sum_{t=1}^{T} ( S_j(t) - \mu_{\kappa}(t))^2 +  2\beta\sum_{v \in \delta_l}{}Z_vh\big) 
}	\Bigg]
	\end{equation*}
The full conditional for $\pmb Z_W$ is obtained in the same way and has the same form modulo minor change in notation.

\section{Analysis of Synthetic Data}
 In this section we evaluate our methodology on a number of test cases using synthetic data. We consider four cases each with different levels of complexity in the true scene. For simplicity, in each of our four examples we fix $K$  in our algorithm to be the true number of mixture components,  though,  in Section 4.2 of the main manuscript (MM) we present a simulation study designed to evaluate our estimator of the number of mixture components $\hat K_{ICM}$.
  
\subsection{Two Sources with Gaussian Signals}
We simulate brain activity on 8,196 locations on the cortex with two active subregions, the first containing  250 locations and the second containing 150 locations. The locations including those comprising the active regions are depicted in Figure 3, panel (a). The neural activity $S_j(t)$ for locations $j$ in each of the active subregions is based on the two Gaussian curves depicted in Figure 4, panel (a) while inactive locations have $S_j(t) = 0$. 

The neural activity $\vS(t)$ is projected onto the MEG and EEG sensor arrays using the forward operators $\vX_M$ and $\vX_E$. The simulated data are then obtained by adding Gaussian noise at each sensor, where the variance of the noise at each sensor is set to be 5\% of the temporal variance of the signal at that sensor. 

We run the ICM algorithm with $K=3$ with the cortical locations divided into $J = 250$ clusters over a 3D grid of $N_v=450$ voxels. The required computing time is roughly 400 seconds on a MacBook Pro with a 2.7 GHz Intel Core i5 processor and 8 GB memory. The estimated allocation variables $\hat\vZ$ are depicted in Figure 3, panel (b). The two regions of neural activity appear to be correctly localized spatially, while the estimated source time series are depicted in Figure 4, panel (b). The temporal patterns of the latter match the true signals depicted in Figure 3, panel (a) reasonably well with both temporal peaks being correctly identified, though there appears to be some over-estimation of the amplitudes in the first component. The overestimation in amplitude can be up to a factor of 2 and then, there is also some underestimation by as much as a factor of 4. Zooming in suggests that there are a few cases where the signal (in the first peak) would not have been detected because it would be lost in the noise.   In summary, both Figure 3 panel (b) and Figure 4, panel (b) indicate that the spatial and temporal localizations are adequate, in the sense that the broad scale features of the signal are recovered, and the required computing time is also reasonable given the complexity of the model and the dimension of parameter space. 
	
\begin{figure}[htbp]
\centering
\begin{tabular}{cc}
\vspace{-5em} \includegraphics[scale=0.35]{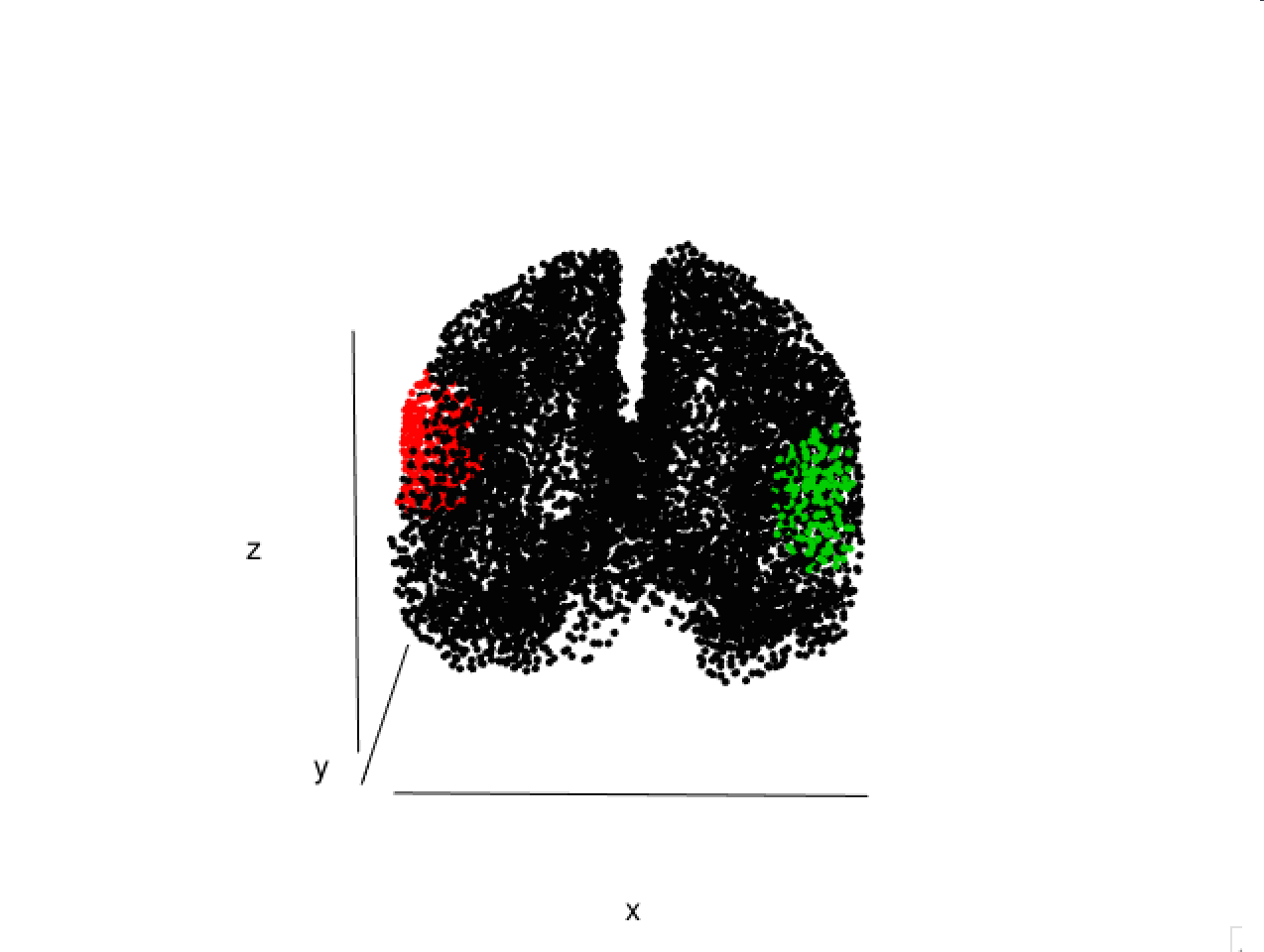} & \hspace{-8em}
\includegraphics[scale=0.35]{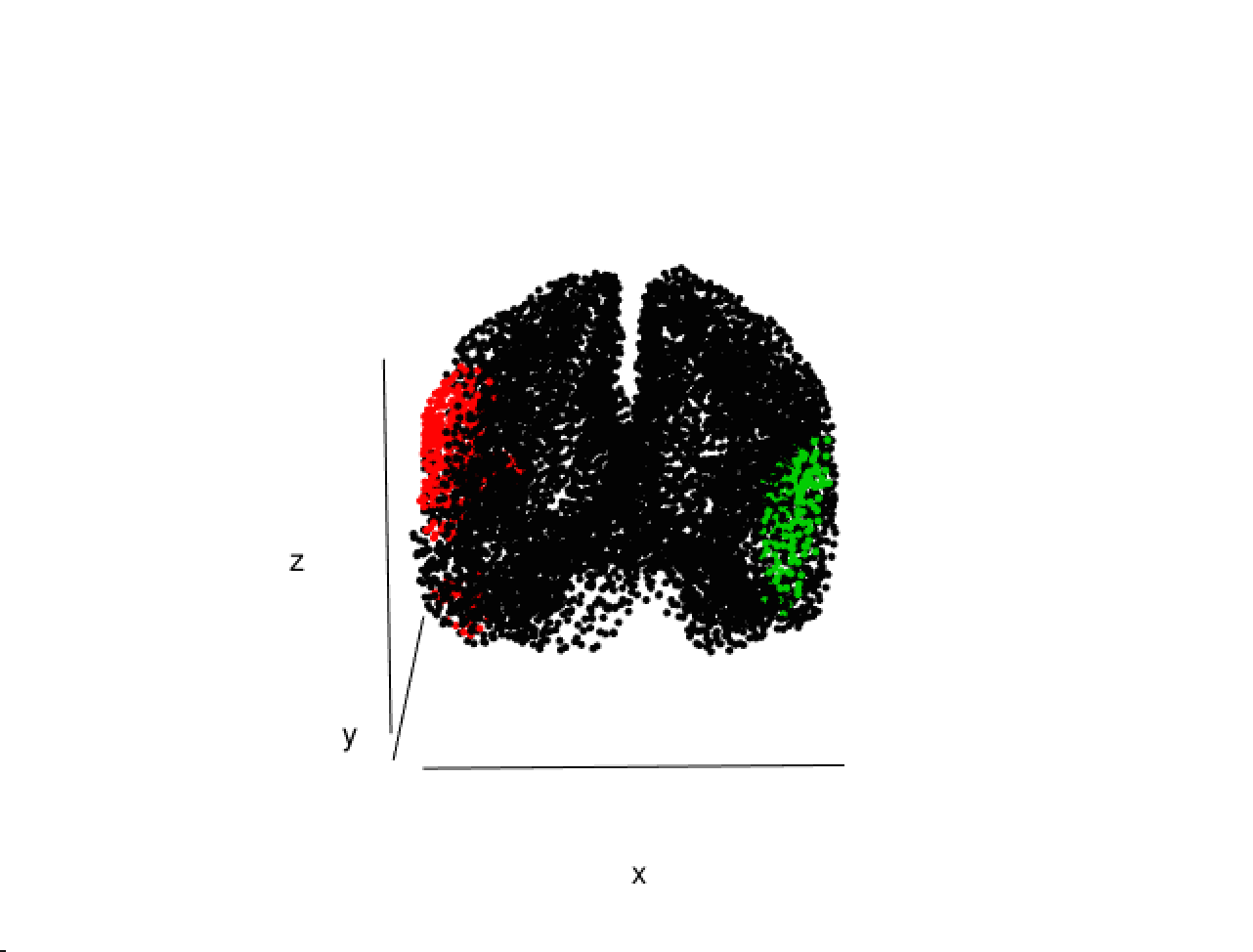}  \\ 
(a) & \hspace{-8em} (b)\\ \vspace{-2em}
 \includegraphics[scale=0.35]{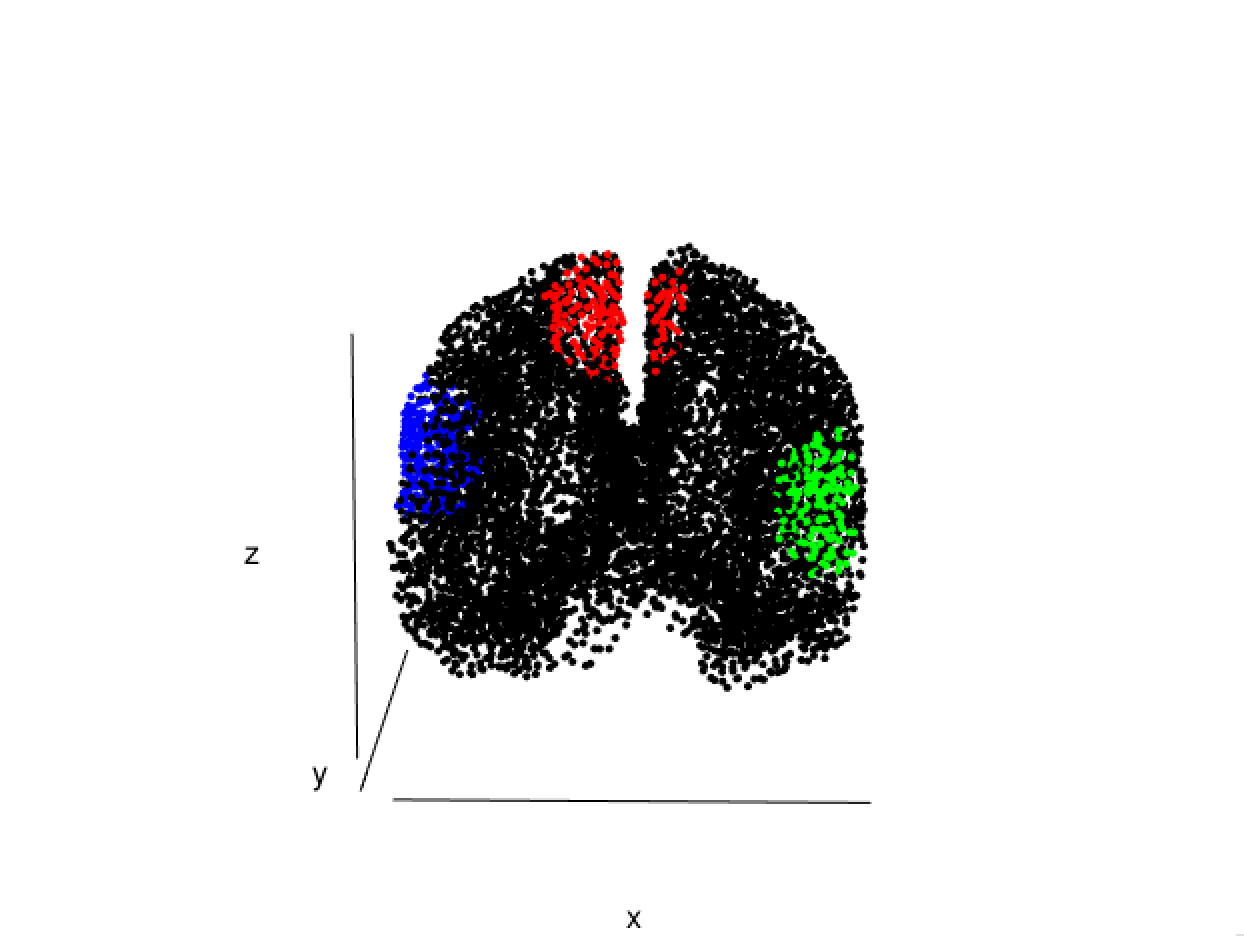} & \hspace{-8em}
\includegraphics[scale=0.35]{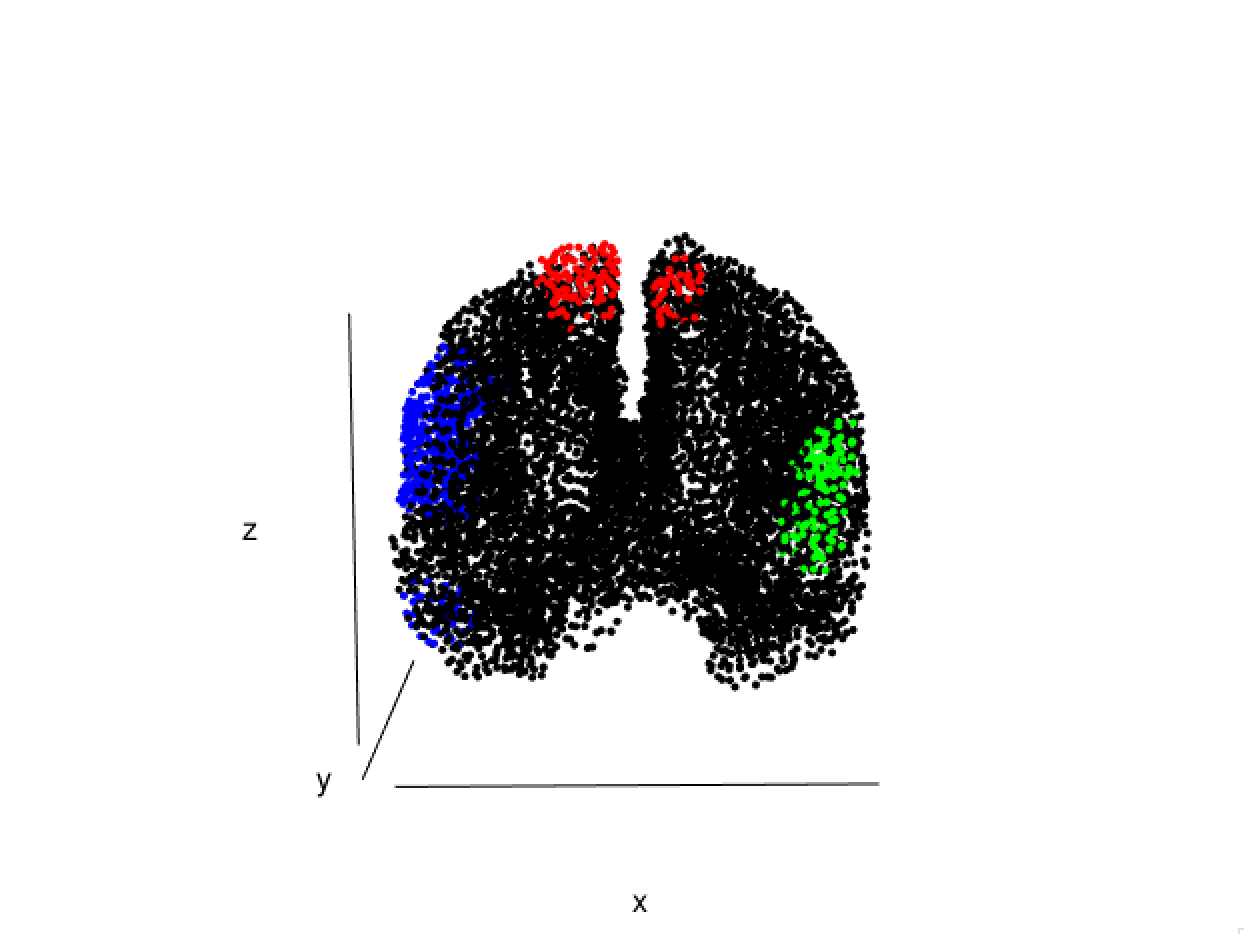} \\
(c) & \hspace{-8em} (d) \\
\vspace{-4em} \includegraphics[scale=0.35]{K_4_loc.png} & \hspace{-8em}
\includegraphics[scale=0.35]{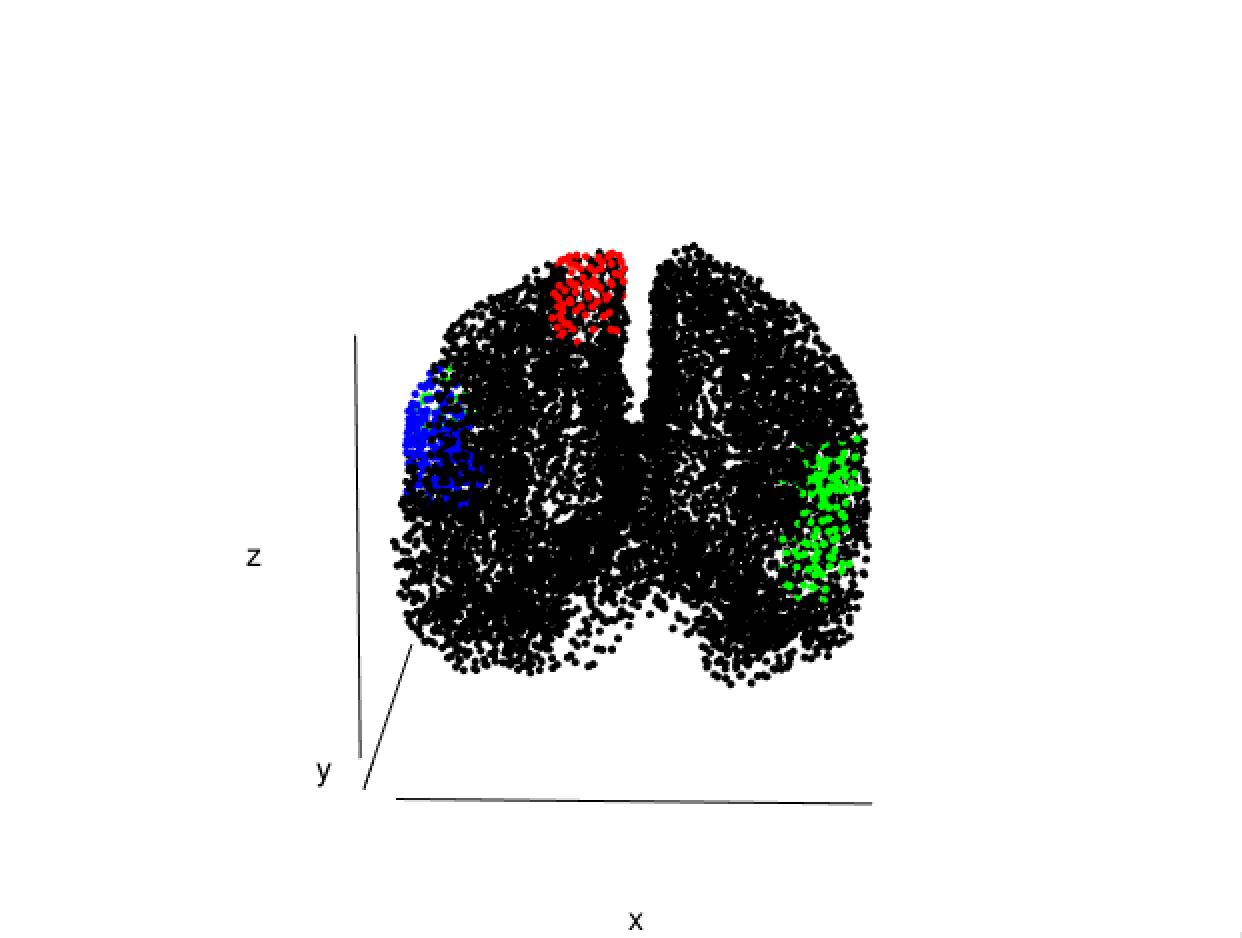}\\
(e) & \hspace{-8em} (f)\\
\end{tabular}
\caption{The true partition of the cortex into active and inactive states for the three examples of simulation studies of Section 4 (MM) (for $K=3$ and $K=4$) are depicted in the left column. The right column presents the corresponding estimated mixture allocation variables ($\hat{\vZ}$).}
\end{figure}

\begin{figure}[htbp]
\centering
\begin{tabular}{cc}
\includegraphics[scale=0.19]{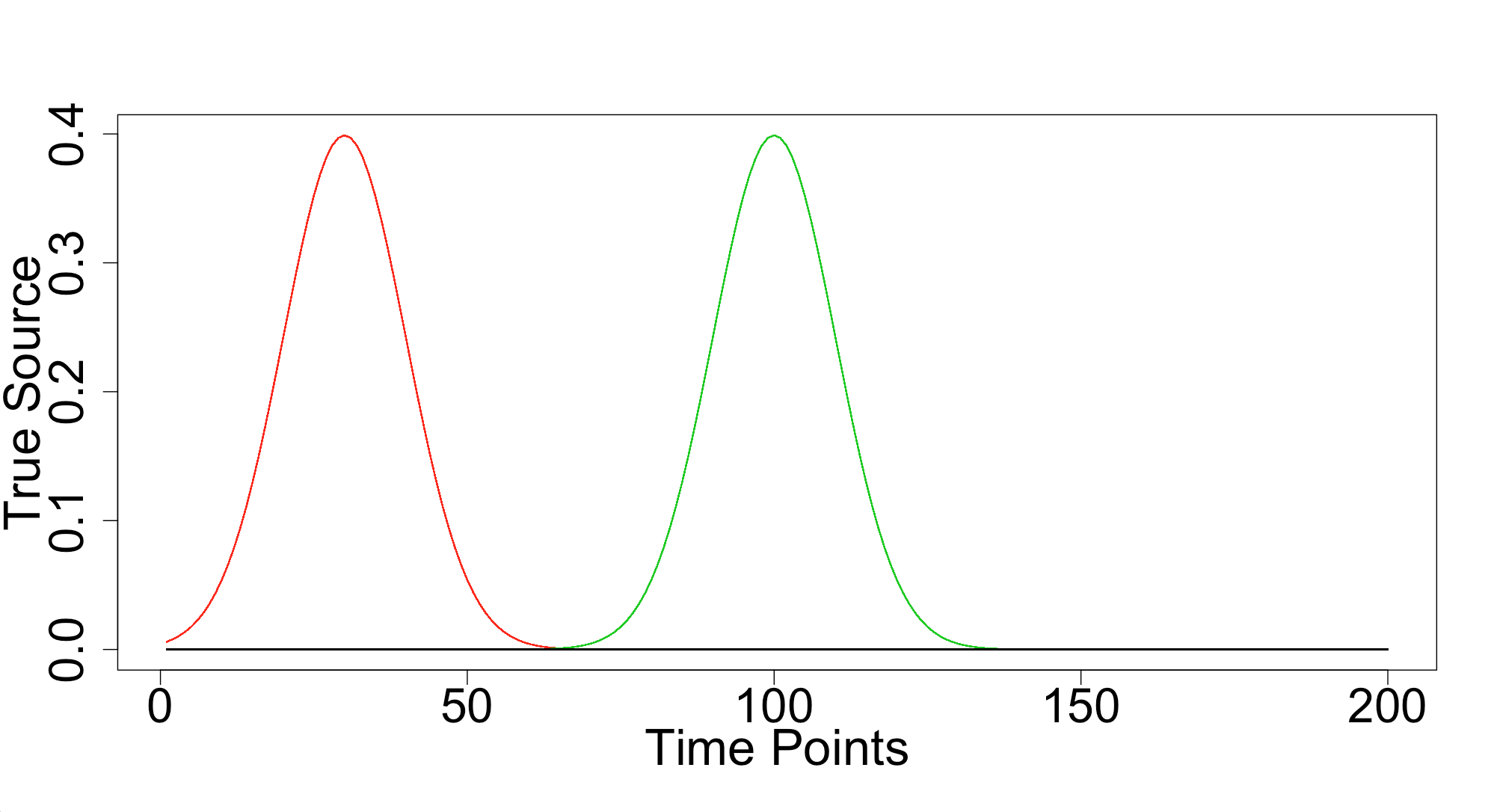} &
\includegraphics[scale=0.19]{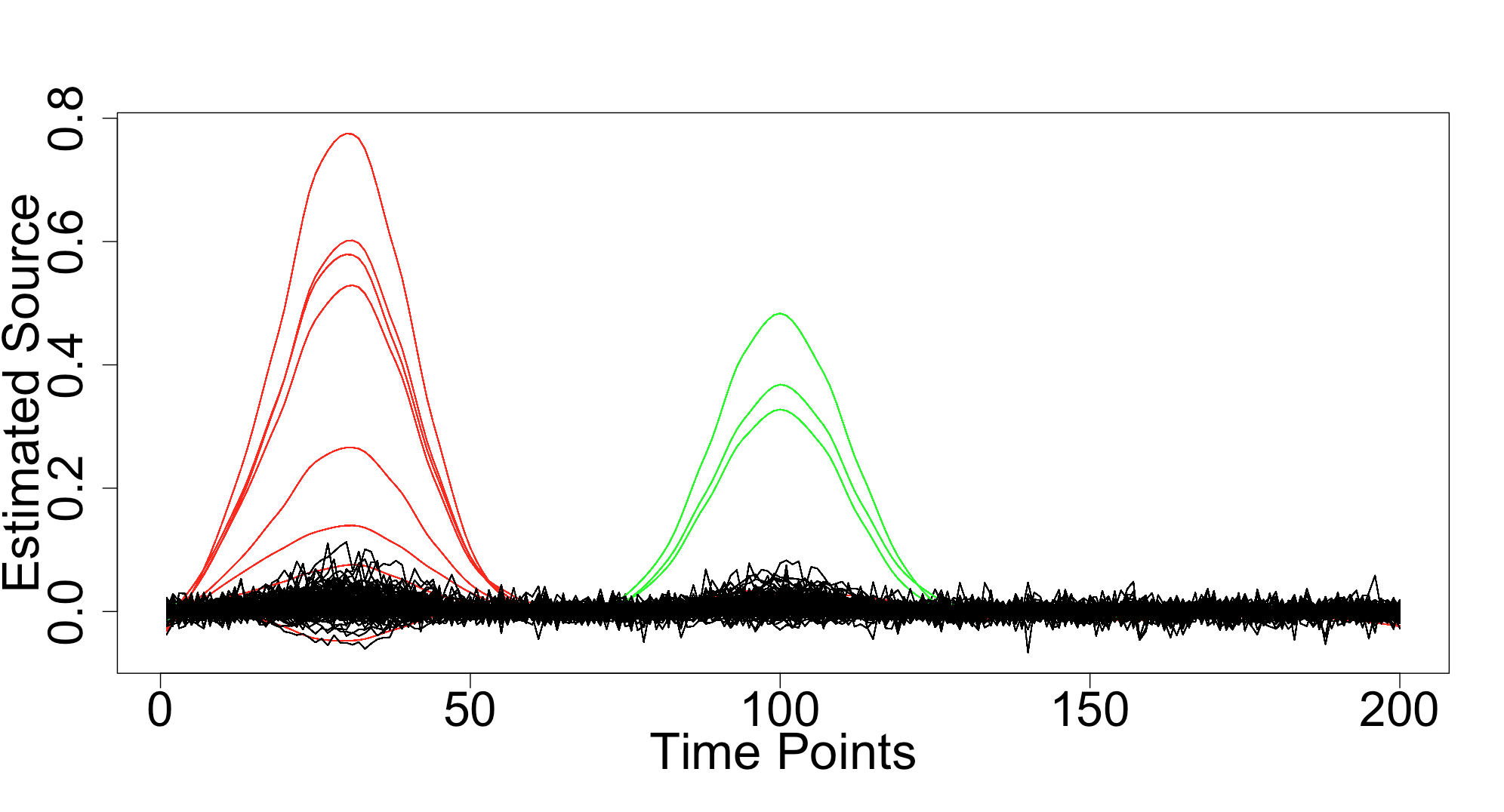}  \\ 
(a) & (b) \\
\includegraphics[scale=0.19]{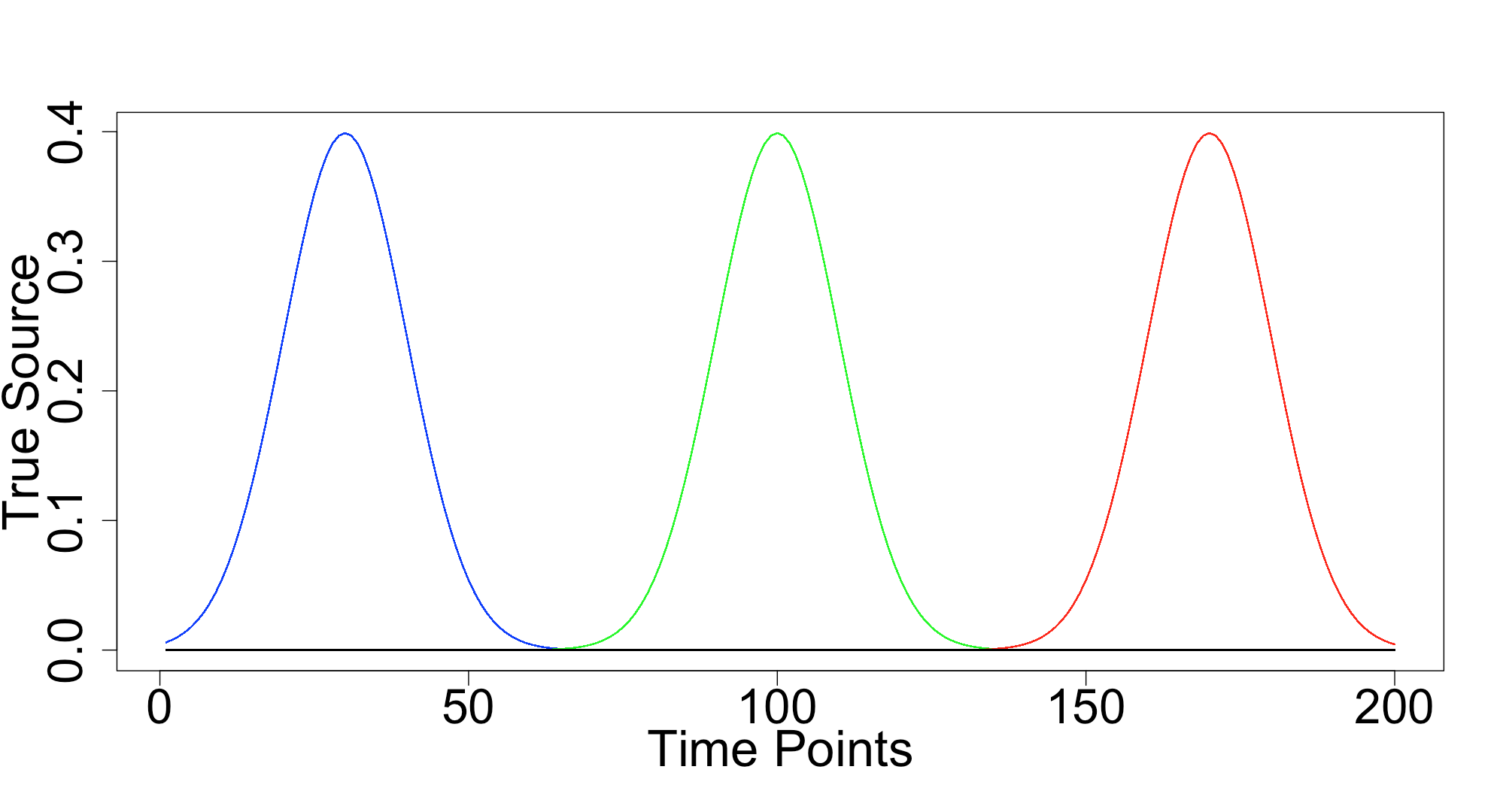} &
\includegraphics[scale=0.19]{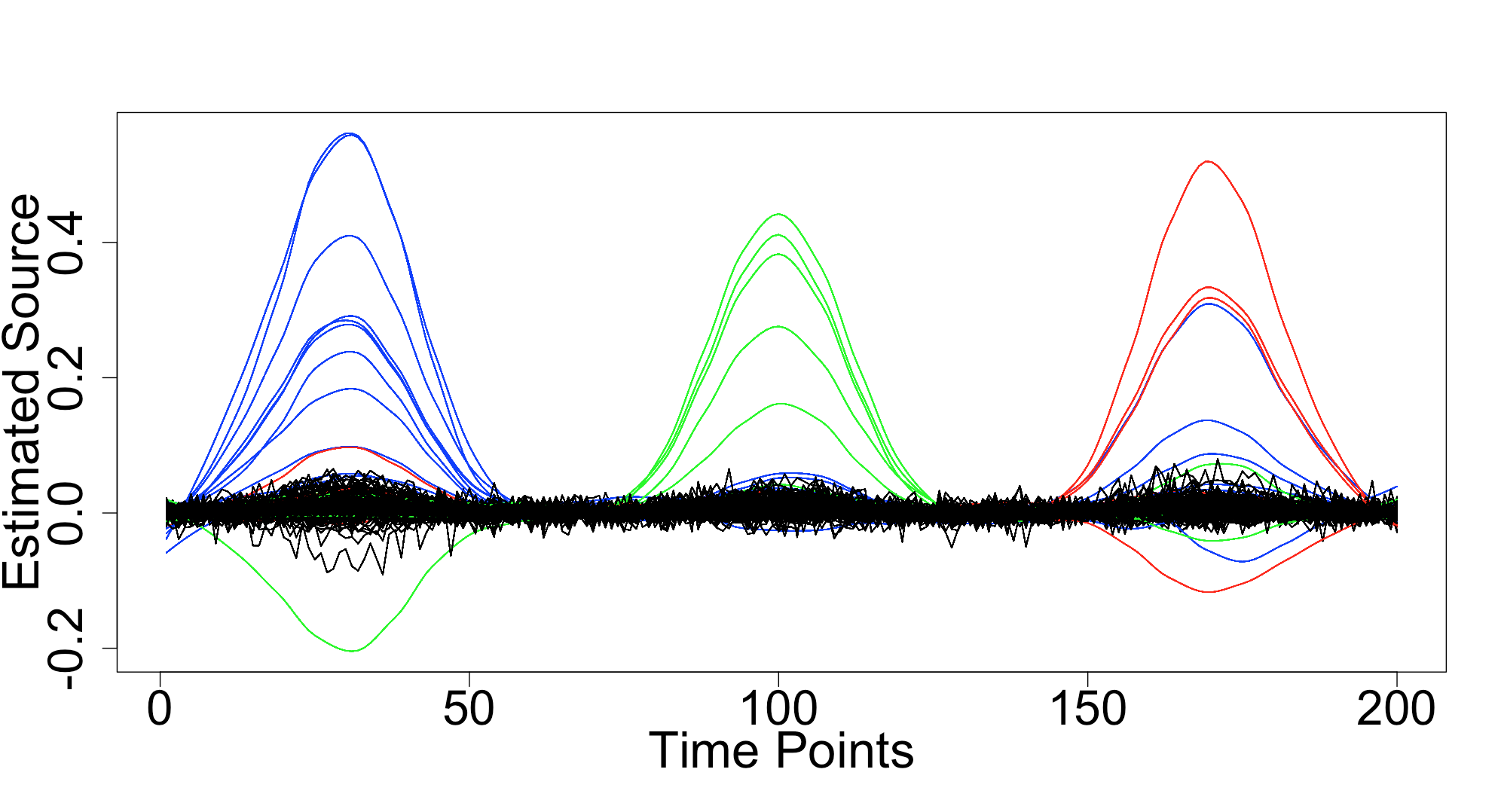} \\
(c) & (d)\\
\includegraphics[scale=0.19]{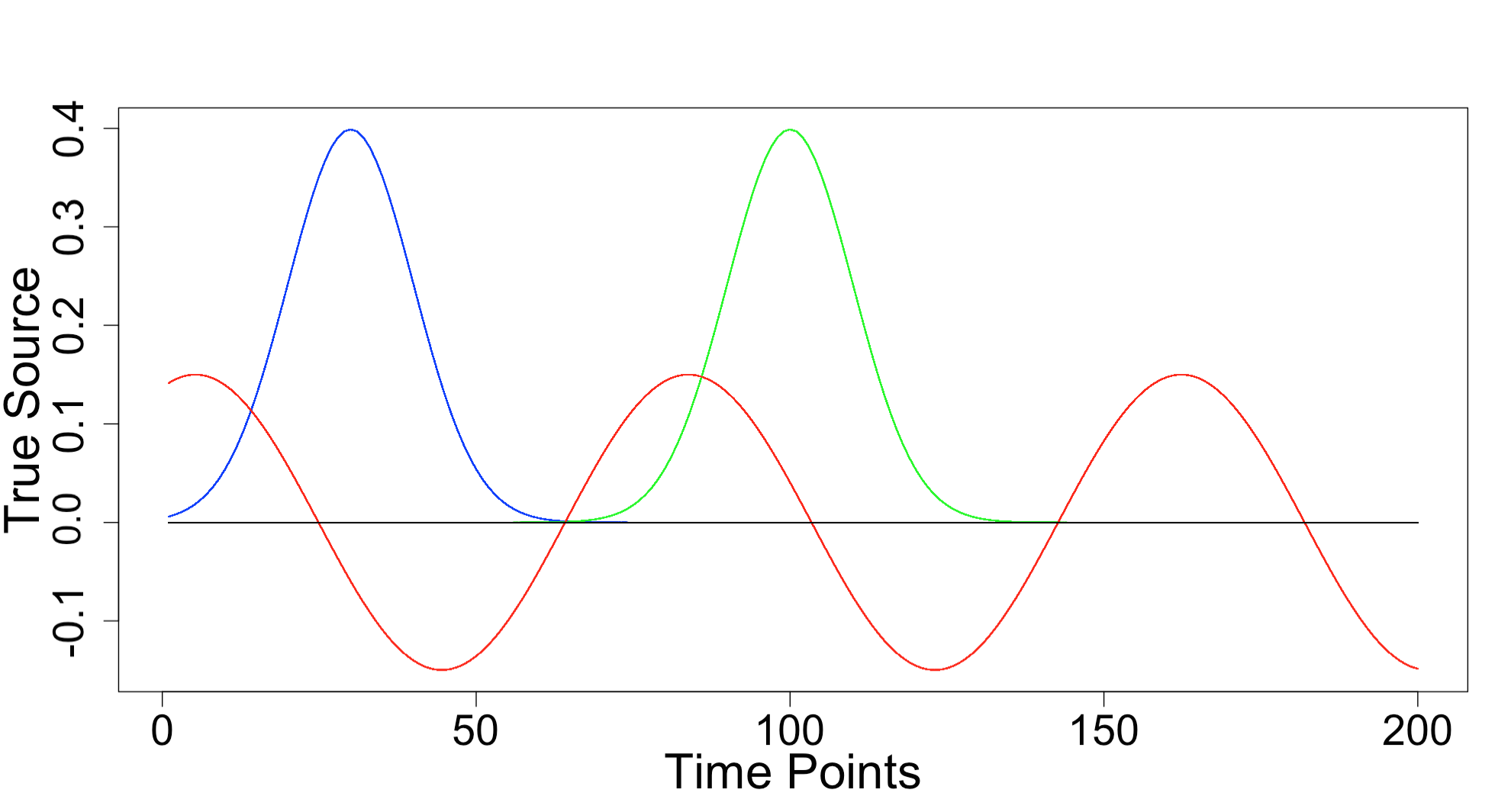} &
\includegraphics[scale=0.19]{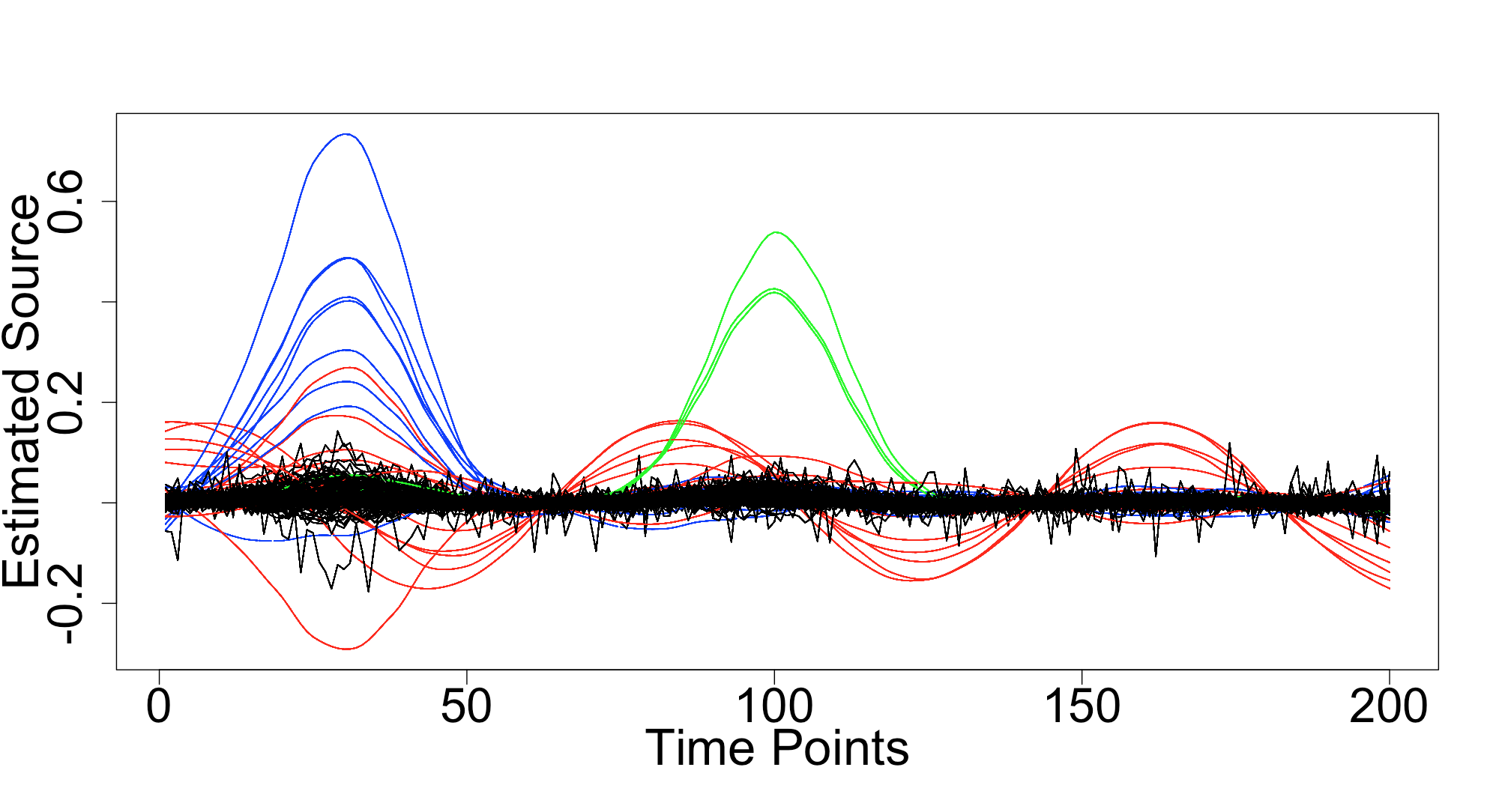}\\
(e) & (f)\\
\includegraphics[scale=0.19]{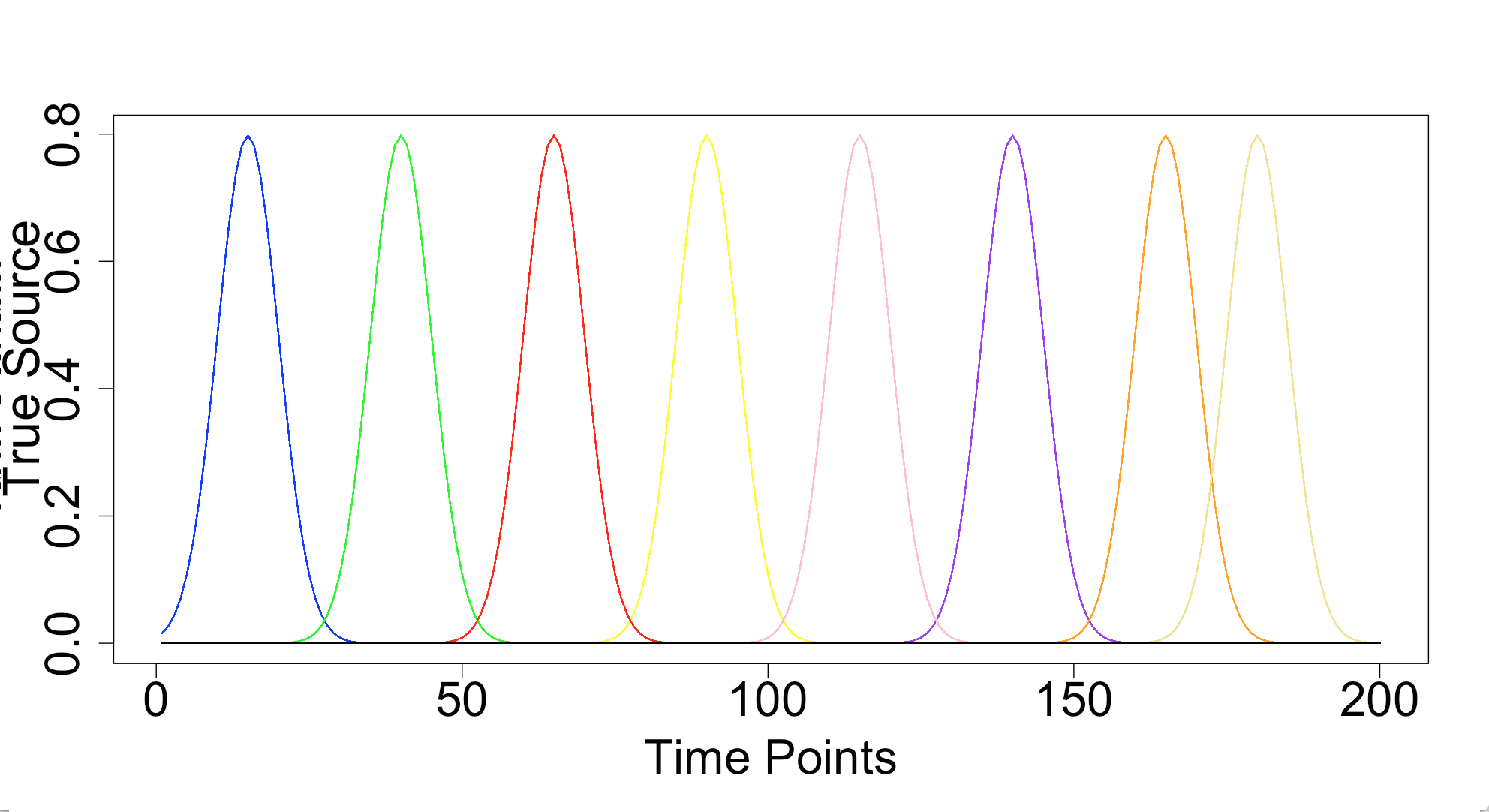} &
\includegraphics[scale=0.19]{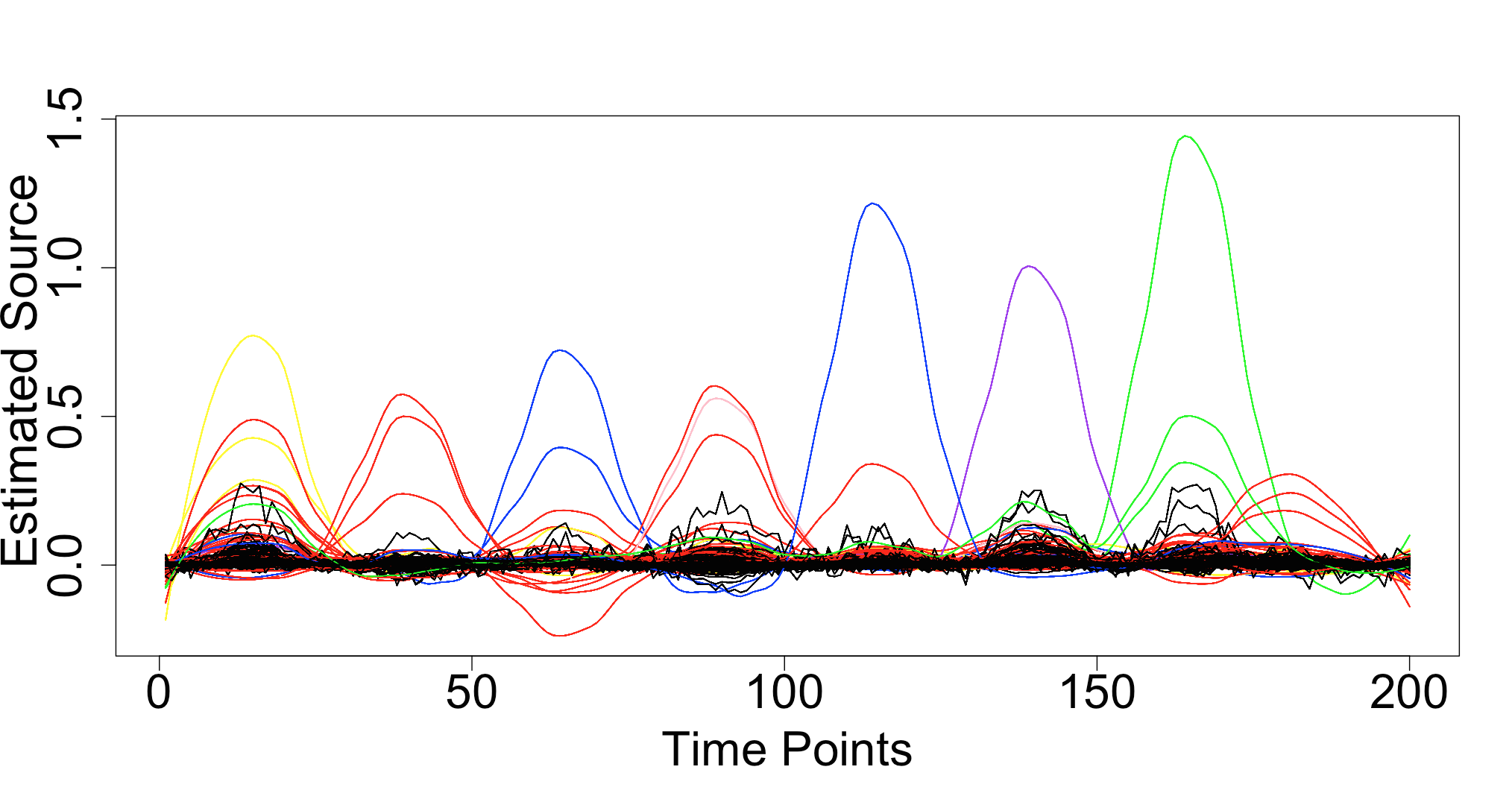}\\
(g) & (h)
\end{tabular}
\caption{The true signal $S_{j}(t)$ used in each of the distinct active and inactive regions in the simulation studies of Section 4 (MM) (for $K=3$, $K=4$ and $K=9$) are depicted in the left column. The right column presents the corresponding estimated sources $\hat{S}_{j}(t)$ at each location of the cortex.}
\end{figure}

\clearpage

\subsection{Three Sources with Gaussian or Sinusoidal Signals}
In our second example we consider three subregions of activity as depicted in Figure 3, panel (c). The neural activity $S_j(t)$ is based on the three Gaussian curves depicted in Figure 4, panel (c). The data are otherwise simulated in the same manner as described in the previous section. We run the ICM algorithm with $K=4$ and other settings as previously described and the running time is again approximately 400 seconds. The estimated allocation variables are depicted in Figure 3, panel (d), and when compared to the true allocations in Figure 3, panel (c), we see that the three regions of neural activity have been identified reasonably well spatially. Comparing the true signals in Figure 4, panel (c), to the estimated sources in Figure 4, panel (d), we see that the three temporal peaks of activity have been correctly identified, though we note that the curves ($S_{j}(t)$) for a few locations have been estimated incorrectly with respect to their shape and some of these have estimated amplitudes that are negative. Still, the overall reconstruction of the neural activity appears reasonably accurate in this  case.

In our third example we again consider the three subregions of activity where the Gaussian signal in the third region is replaced with a sinusoid as depicted in Figure 4, panel (e). In this case the signal from the third activated region overlaps with both signals from the other active regions. The ICM algorithm requires the same computing time as in the previous examples. The estimated mixture allocation variables are depicted in Figure 3, panel (f), and comparing to the true scene our algorithm appears to have correctly spatially localized the three regions of neural activity, though part of the third region (the red coloured region in the third row of Figure 3) appears to have been incorrectly classified as 'inactive'. Examining the estimated sources $\hat{S}_{j}(t)$ in Figure 4, panel (f), in comparison to the true signals in Figure 4, panel (e), we see that the patterns of the temporal signals including the sinusoid appear to be mostly well estimated.

\subsection{Eight Sources with Gaussian Signals}
In our fourth and final example we consider a much more difficult setting where there are eight active regions, each with spatial extent roughly one-fourth that of the active regions considered in the previous three examples. The true spatial configuration of the nine states is depicted from various different angles in Figure 5. The temporal profile of the brain activation is represented with eight Gaussian signals depicted in Figure 4, panel (g). The running time for the algorithm is approximately 800 seconds and we obtain, in this case, $\hat{K}_{ICM} = 7$. The estimated signals are shown in Figure 4, panel (h), while the estimated state allocation variables are shown in Figure 6, with the panels of this figure corresponding to the panels of Figure 5 where the true states are depicted. Overall, the algorithm is able to capture roughly the broad spatiotemporal pattern of brain activation, though the quality of the estimates relative to the simpler examples is not as high. Nevertheless, it appears as though the approach can be applied successfully to reconstruct general patterns of brain activity in this more difficult case, while the true number of states in the brain is under-estimated by 2. 

\begin{figure}[htbp]
\centering
\begin{tabular}{cc}
\vspace{-4em} \includegraphics[scale=0.45]{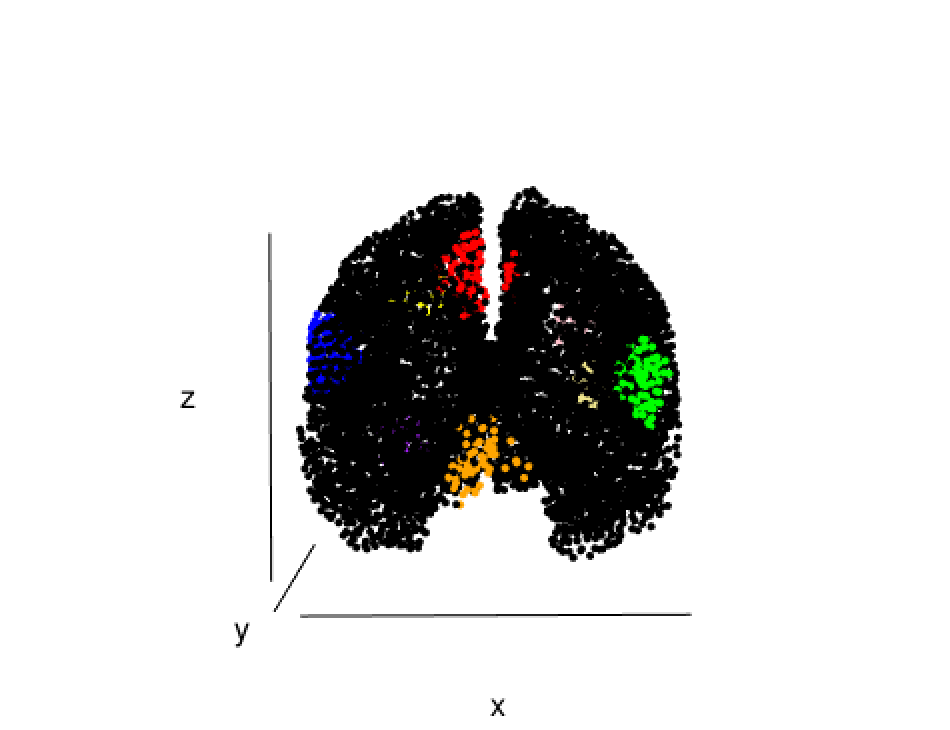} & \hspace{-7em}
\includegraphics[scale=0.45]{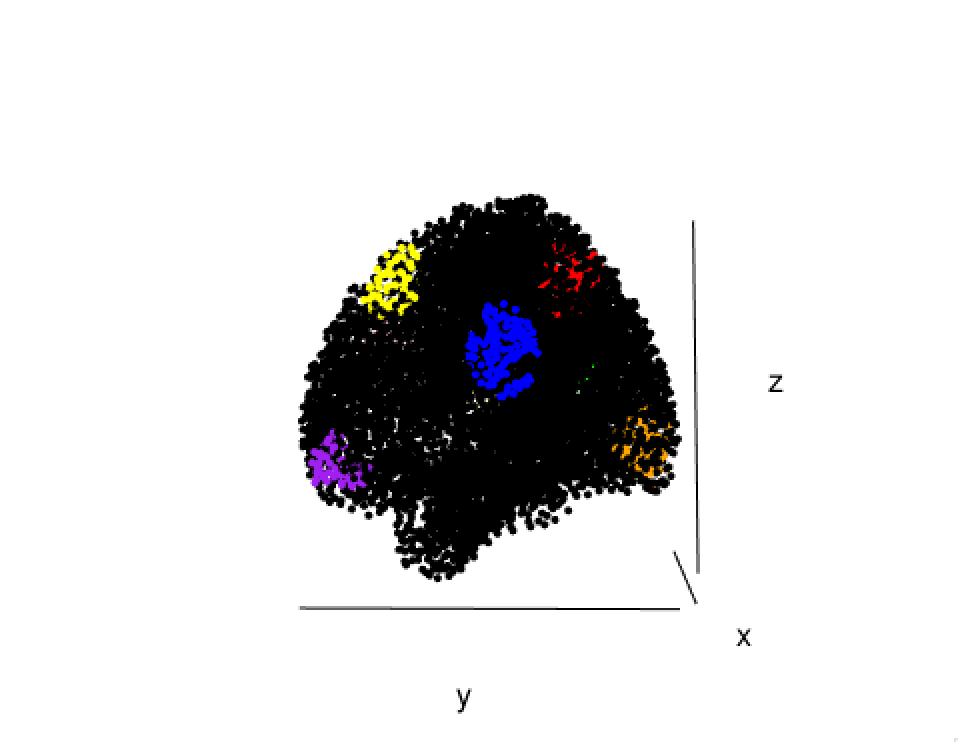}  \\ 
(a) & \hspace{-7em} (b)\\ \vspace{-3em}
 \includegraphics[scale=0.45]{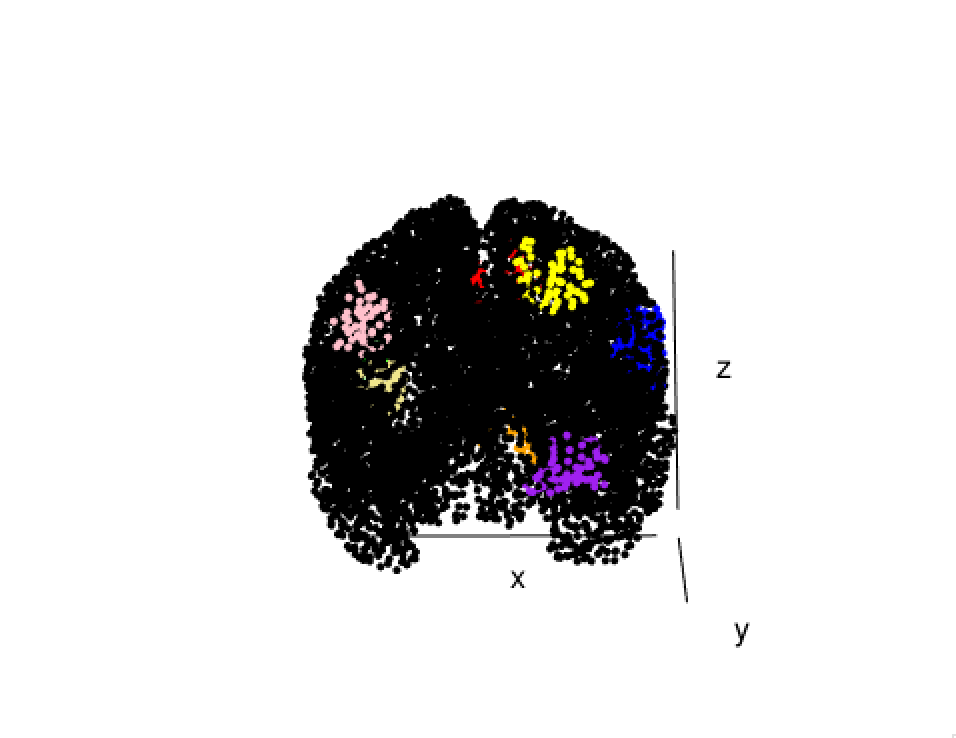} & \hspace{-7em}
\includegraphics[scale=0.45]{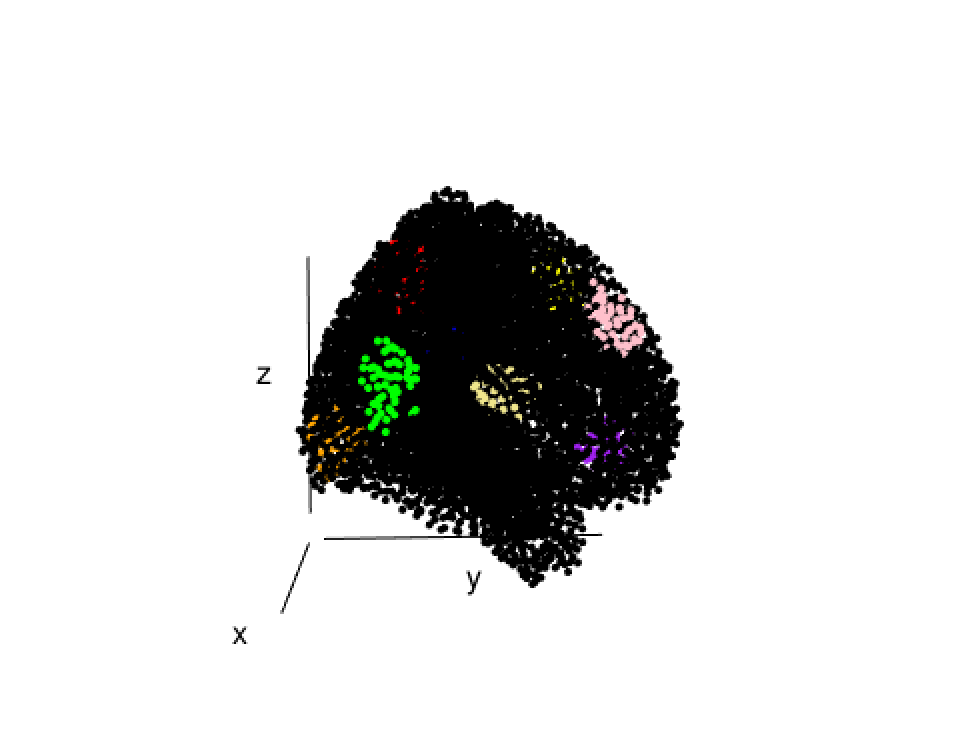} \\
(c) & \hspace{-7em} (d) \\
\vspace{-4em} \includegraphics[scale=0.45]{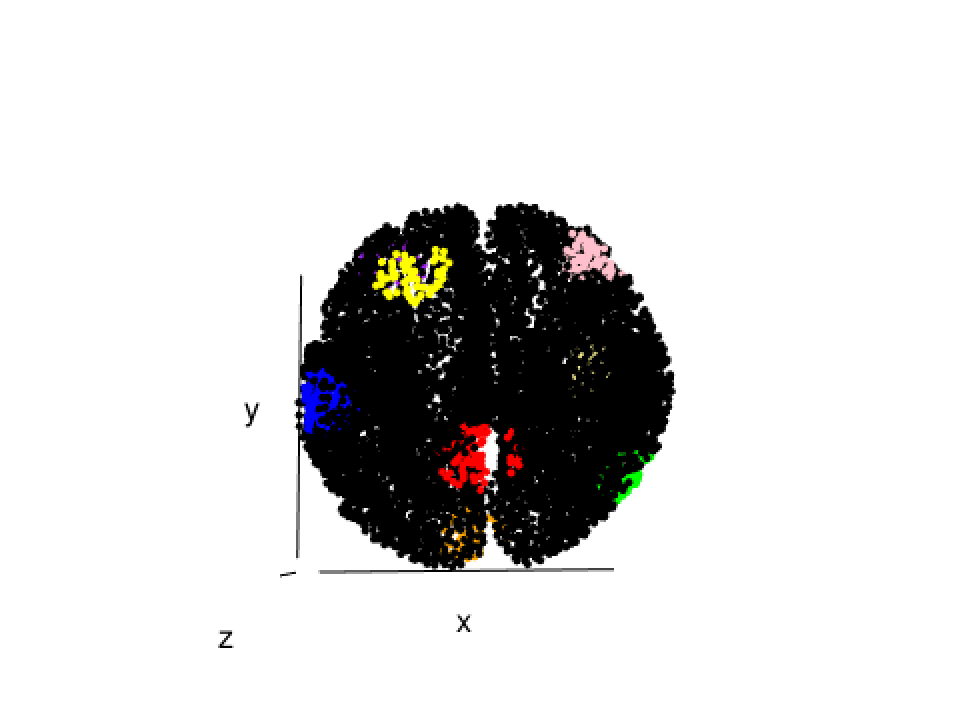} & \hspace{-8em}
\includegraphics[scale=0.45]{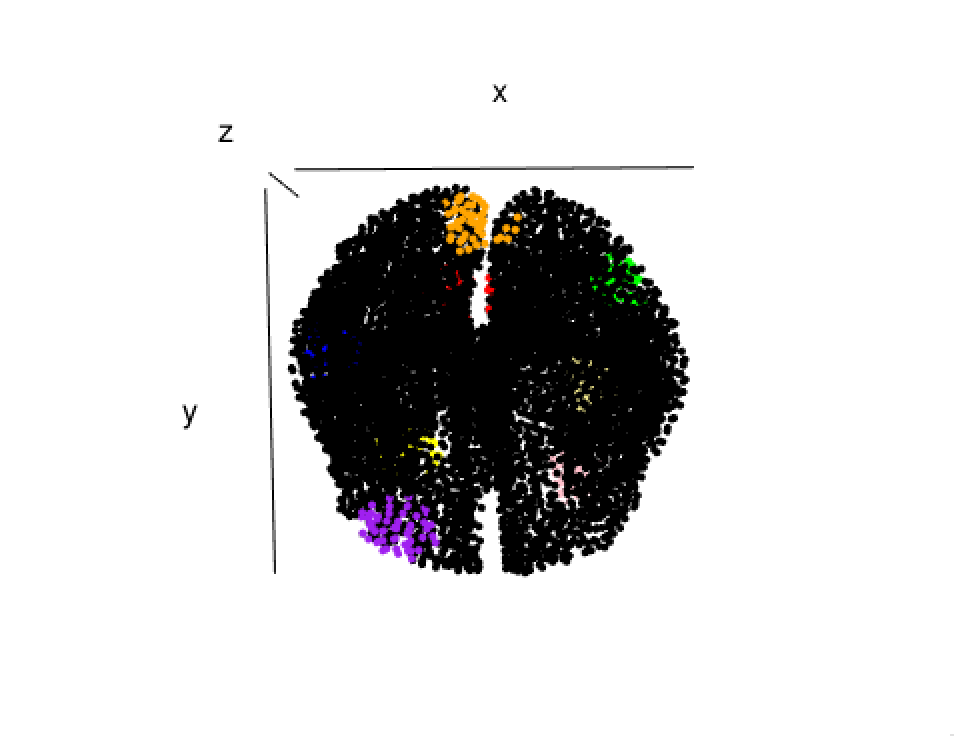}\\
(e) & \hspace{-7em} (f)\\
\end{tabular}
\caption{The true partition of the cortex into active and inactive states for the case of $K=9$ states.}
\end{figure}

\begin{figure}[htbp]
\centering
\begin{tabular}{cc}
\vspace{-4em} \includegraphics[scale=0.35]{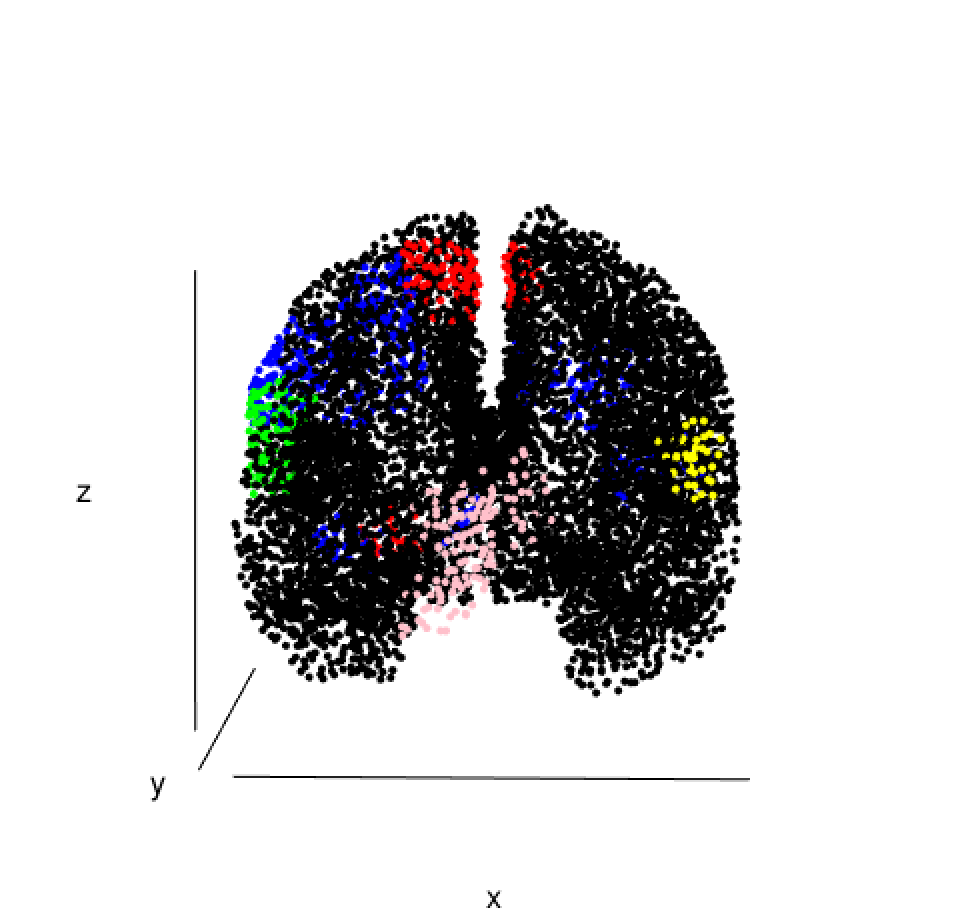} & 
\includegraphics[scale=0.35]{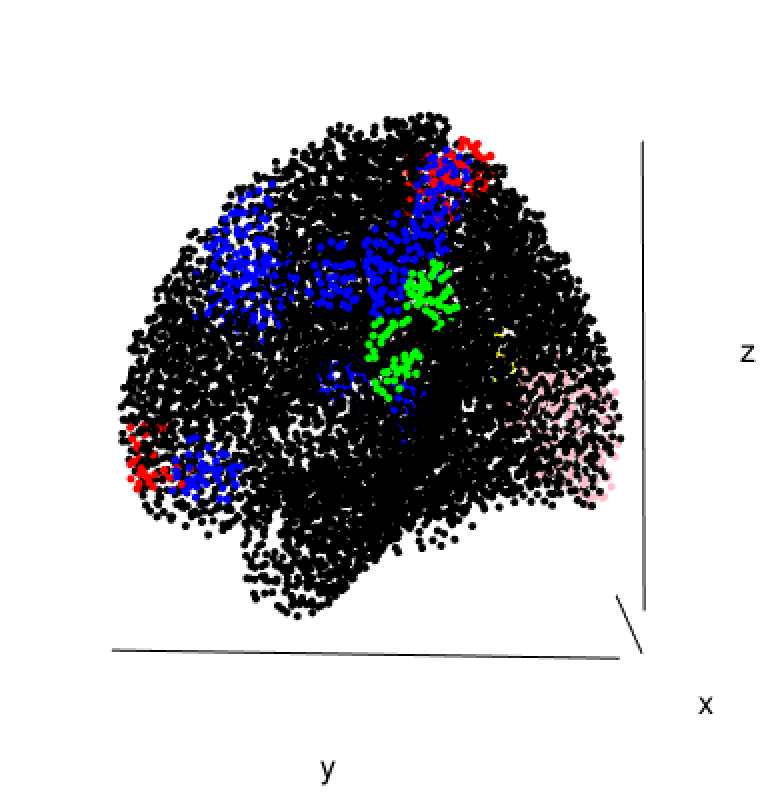}  \\ 
(a) & \hspace{-7em} (b)\\ \vspace{-3em}
 \includegraphics[scale=0.35]{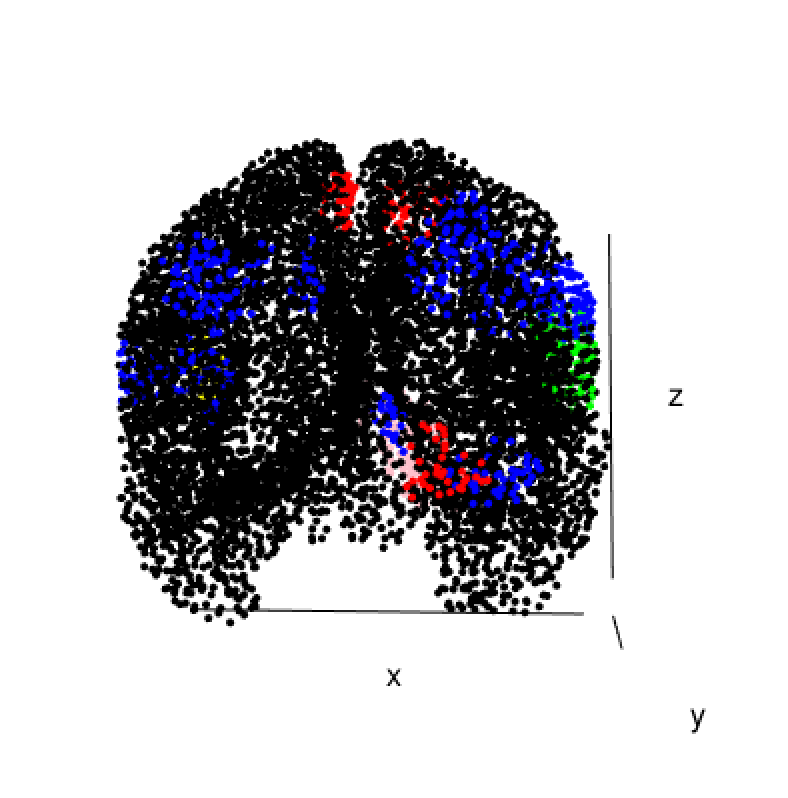} & 
\includegraphics[scale=0.35]{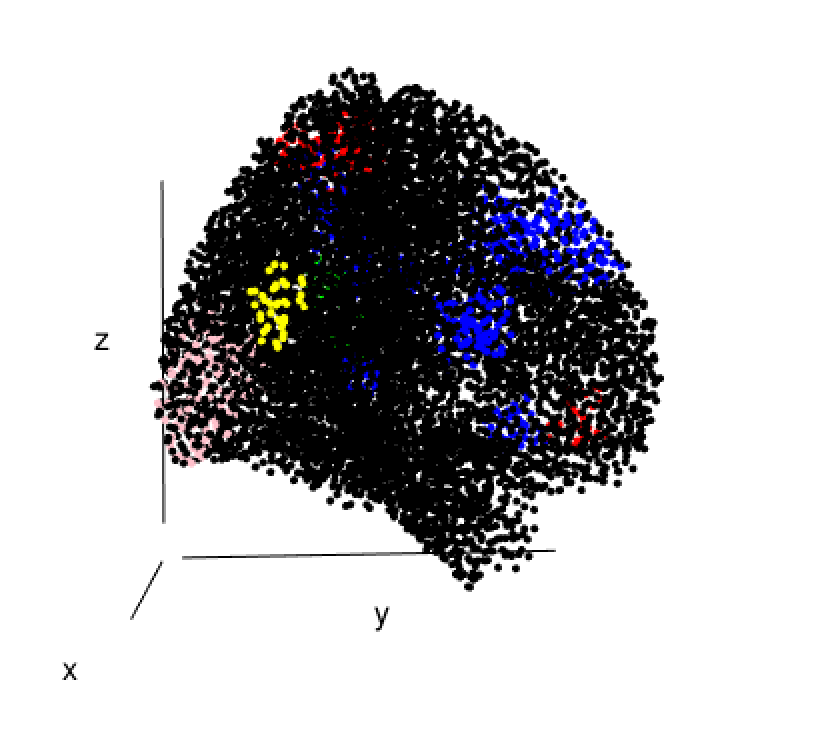} \\
(c) & \hspace{-7em} (d) \\
\vspace{-4em} \includegraphics[scale=0.35]{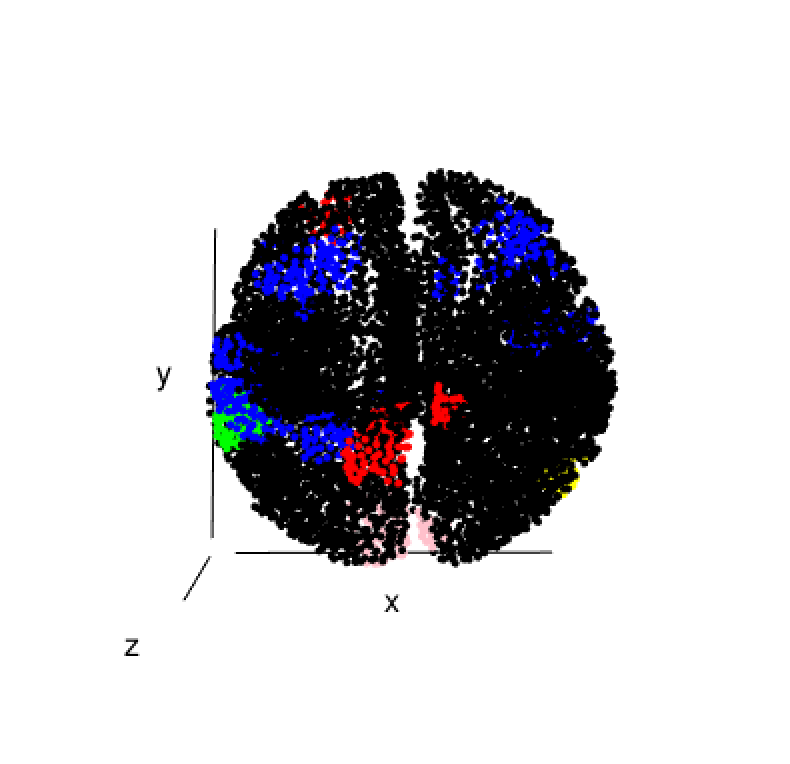} & 
\includegraphics[scale=0.35]{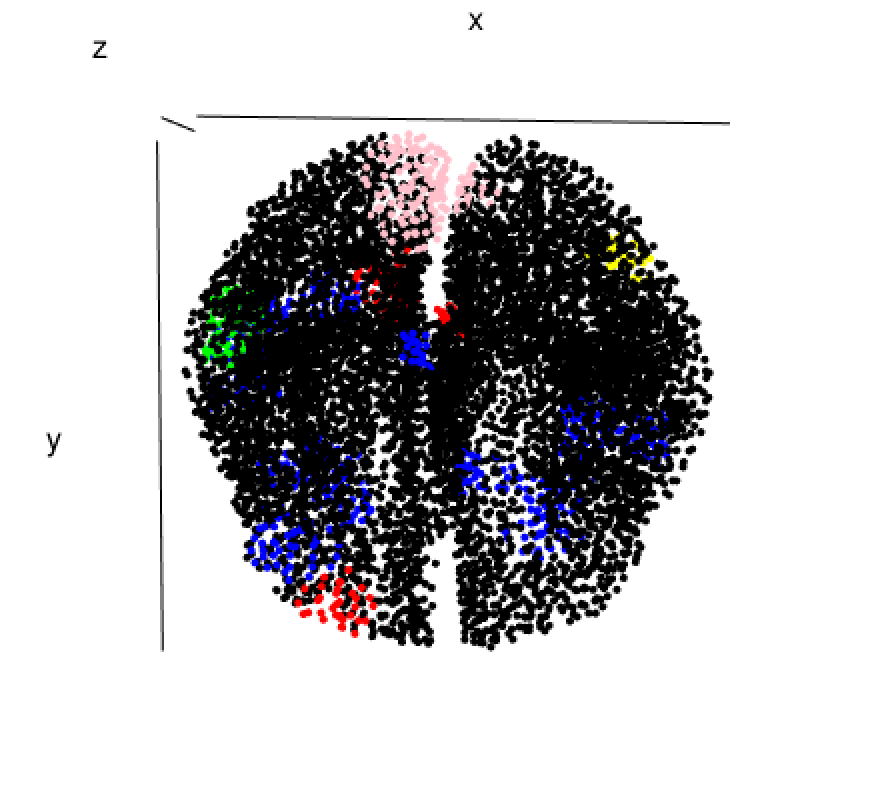}\\
(e) & \hspace{-7em} (f)\\
\end{tabular}
\caption{The estimated allocation of the cortex into active and inactive states for the case of $K=9$ true states. In this case $\hat{K}_{ICM} = 7$. The panels in this figure correspond to the panels in Figure 5.}
\end{figure}

\end{document}